\documentclass{emulateapj}
\usepackage{natbib}
\usepackage{graphicx}
\usepackage{url}
\usepackage{color}

\begin{document}

\title{Photometric Redshifts with the LSST: Evaluating Survey Observing Strategies}

\author{Melissa L. Graham\altaffilmark{1}}
\author{Andrew J. Connolly\altaffilmark{1}}
\author{\v{Z}eljko Ivezi\'{c}\altaffilmark{1}}
\author{Samuel J. Schmidt\altaffilmark{2}}
\author{R. Lynne Jones\altaffilmark{1}}
\author{Mario Juri\'{c}\altaffilmark{1}}
\author{Scott F. Daniel\altaffilmark{1}}
\author{Peter Yoachim\altaffilmark{1}}
\altaffiltext{1}{Department of Astronomy, University of Washington, Box 351580, U.W., Seattle WA 98195}
\altaffiltext{2}{Department of Physics, UC Davis, One Shields Avenue, Davis CA 95616}

\begin{abstract}
In this paper we present and characterize a nearest-neighbors color-matching photometric redshift estimator that features a direct relationship between the precision and accuracy of the input magnitudes and the output photometric redshifts. This aspect makes our estimator an ideal tool for evaluating the impact of changes to LSST survey parameters that affect the measurement errors of the photometry, which is the main motivation of our work (i.e., it is not intended to provide the ``best" photometric redshifts for LSST data). We show how the photometric redshifts will improve with time over the 10-year LSST survey and confirm that the nominal distribution of visits per filter provides the most accurate photo-$z$ results. The LSST survey strategy naturally produces observations over a range of airmass, which offers the opportunity of using an SED- and $z$-dependent atmospheric affect on the observed photometry as a color-independent redshift indicator. We show that measuring this airmass effect and including it as a prior has the potential to improve the photometric redshifts and can ameliorate extreme outliers, but that it will only be adequately measured for the brightest galaxies, which limits its overall impact on LSST photometric redshifts. We furthermore demonstrate how this airmass effect can induce a bias in the photo-$z$ results, and caution against survey strategies that prioritize high-airmass observations for the purpose of improving this prior. Ultimately, we intend for this work to serve as a guide for the expectations and preparations of the LSST science community with regards to the minimum quality of photo-$z$ as the survey progresses.
\end{abstract}
\keywords{LSST; photometric redshifts}

\section{Introduction} \label{sec:intro}

Photometric redshifts (photo-$z$'s) are an essential part of every cosmological science goal of the Large Synoptic Survey Telescope (LSST), including studies of large scale structure, weak lensing, galaxy clusters, and supernova host galaxies \citep{2008arXiv0805.2366I}. Since the initial framework for using galaxy photometry to infer distance or redshift was presented by \cite{1962IAUS...15..390B}, the methodology has evolved over time. Techniques for estimating photometric redshifts were stimulated in particular by the era of large-scale surveys such as the Sloan Digital Sky Survey, and more recently by advances in machine learning technology. Since the LSST sample will contain billions of galaxies, the science impact of systematic biases will dominate over statistical errors. The next-generation of photo-$z$ estimators are in development, testing a variety of approaches for providing the most precise cosmological parameter measurements from LSST data (e.g., \citealt{2016MNRAS.461.3432S,2016PASP..128j4502S,2017ApJ...838....5L,2017arXiv170405988T}). In this paper we present and characterize a nearest-neighbors color-matching photo-$z$ method that is comparable to, for example, the photo-$z$ estimators recently presented by \cite{2008ApJ...683...12B} and \cite{2012ApJS..201...32S}. This estimator is not intended to provide the ``best" photometric redshifts for LSST data. Instead, it serves as an efficient tool to evaluate potential changes to LSST survey parameters that affect the quality of the photometry because, as we will show, the precision and accuracy of the output photo-$z$ are directly related to the magnitude errors of the input galaxy catalog. An optimal photo-$z$ estimator designed for precision cosmology would, among other attributes, deliver significantly improved scatter and bias estimates at $z_{\rm phot} > 1.5$ compared to the estimator that we use in this work.

One of the goals for LSST science is to provide photometry of high enough quality to derive precise photometric redshifts that will enable significant advances in cosmological studies. To express this photometric quality in terms of photometric redshift capabilities, the Science Requirements Document (SRD; \citealt{LPM-17}) defines some minimum target values\footnote{Note that there is a typo in the SRD at the time of this publication, to be corrected soon, where the $(1+z)^{-1}$ factor for the target values is not clearly stated.} for photometric redshifts for an $i<25$, magnitude-limited sample of $4\times10^9$ galaxies from $0.3<z<3.0$ as: (1) the root-mean-square error in photo-$z$ must be $< 0.02(1+z_{\rm phot})$; (2) the fraction of outliers must be $<10\%$; and (3) the average bias must be $< 0.003(1+z_{\rm phot})$. We emphasize that these targets are {\it minimum deliverables} for the LSST; leading-edge cosmological studies may require more stringent specifications and next-generation photo-$z$ estimators may deliver on that front -- see Section 3.8.1 of the LSST Science Book \citep{2009arXiv0912.0201L} and Section 3.7 of the LSST DESC white paper \citep{2012arXiv1211.0310L} for descriptions of photo-z performance with LSST. However, they are suitable reference targets for this work, and we explain in further detail how we calculate and evaluate these statistics in Section~\ref{ssec:stats}. These SRD targets apply to photometry from the nominal 10 year LSST wide-fast-deep survey, which means that the final stacked image depth and signal-to-noise -- and their uniformity across the sky -- in all six filter bandpasses is of primary concern. Although the details of how the LSST will reach full depth (i.e., cadence) are not as important for meeting the SRD's targets for photo-$z$ from LSST, they will affect the rate of progress towards the main cosmological science goals. In this work, we use the photo-$z$ estimator to investigate how the LSST photo-$z$ will generally improve as the magnitude errors decrease over time, and to estimate the minimum photo-$z$ quality in the first $1$--$2$ years as an indicator of the science that will be possible with the earliest data releases. We also use the estimator to simulate the impact that any changes to the LSST survey strategy which impacts the final depths, such as the relative fraction of visits allotted to the $u$-band filter, might have on the photometric redshifts. An early version of this work can also be found in the LSST Observing Strategy Whitepaper \citep{2017arXiv170804058L}.

One of the biggest challenges for photometric redshifts is minimizing the fraction of outliers, which are caused by a degeneracy between galaxies at low-$z$ with similar colors to those at high-$z$. This degeneracy is mainly driven by the Lyman break in a high-$z$ galaxy spectrum being redshifted to the position of the Balmer break in a low-$z$ galaxy spectrum. To break this degeneracy requires a way to gain information about the spectral energy distribution (SED) and/or redshift from observations that are independent of the photometric color. Examples include applying a prior based on apparent magnitude or size, because bright, large galaxies are more likely to be at low-$z$. In this work we will explore an atmospheric effect whereby the amount of extinction as a function of airmass is sensitive to the galaxy's observed SED, and therefore the redshift of the object. The potential to use such an effect is a relatively unique by-product of the wide-area survey style of LSST, as the $>800$ required visits of each field over ten years guarantees that at least some observations will be done at high airmass. We will investigate whether the slope of the magnitude-airmass relationship can be used as a redshift indicator, and discuss the prospects for measuring this subtle effect with LSST data, its impact on the photometric redshift results, and whether this should influence the LSST survey strategy.

In Section~\ref{sec:cat} we describe the simulated galaxy catalog that we use for this experiment. In Section~\ref{sec:method} we describe and characterize the photometric redshift estimator and the statistical measures that we use to analyze the results. In Section~\ref{sec:lsstparams} we explore how changes to the LSST observing strategy may effect the predicted photo-$z$ quality in the earliest data releases, and demonstrate the utility of our photo-$z$ method's direct relation between magnitude errors and the precision and accuracy of the photo-$z$ results. In Section~\ref{sec:air} we describe how an alteration of the effective filter transmission function at different airmass leads to a SED- and $z$-dependent photometric effect that can help to refine a galaxy's photo-$z$ estimate, and show how it could be used to improve the results from our photo-$z$ estimator. We summarize and conclude in Section~\ref{sec:conc}.

\section{Simulated Galaxy Catalog}\label{sec:cat}

\begin{figure*}
\begin{center}
\includegraphics[width=8.2cm,trim={1cm 4.5cm 1cm 4.5cm}, clip]{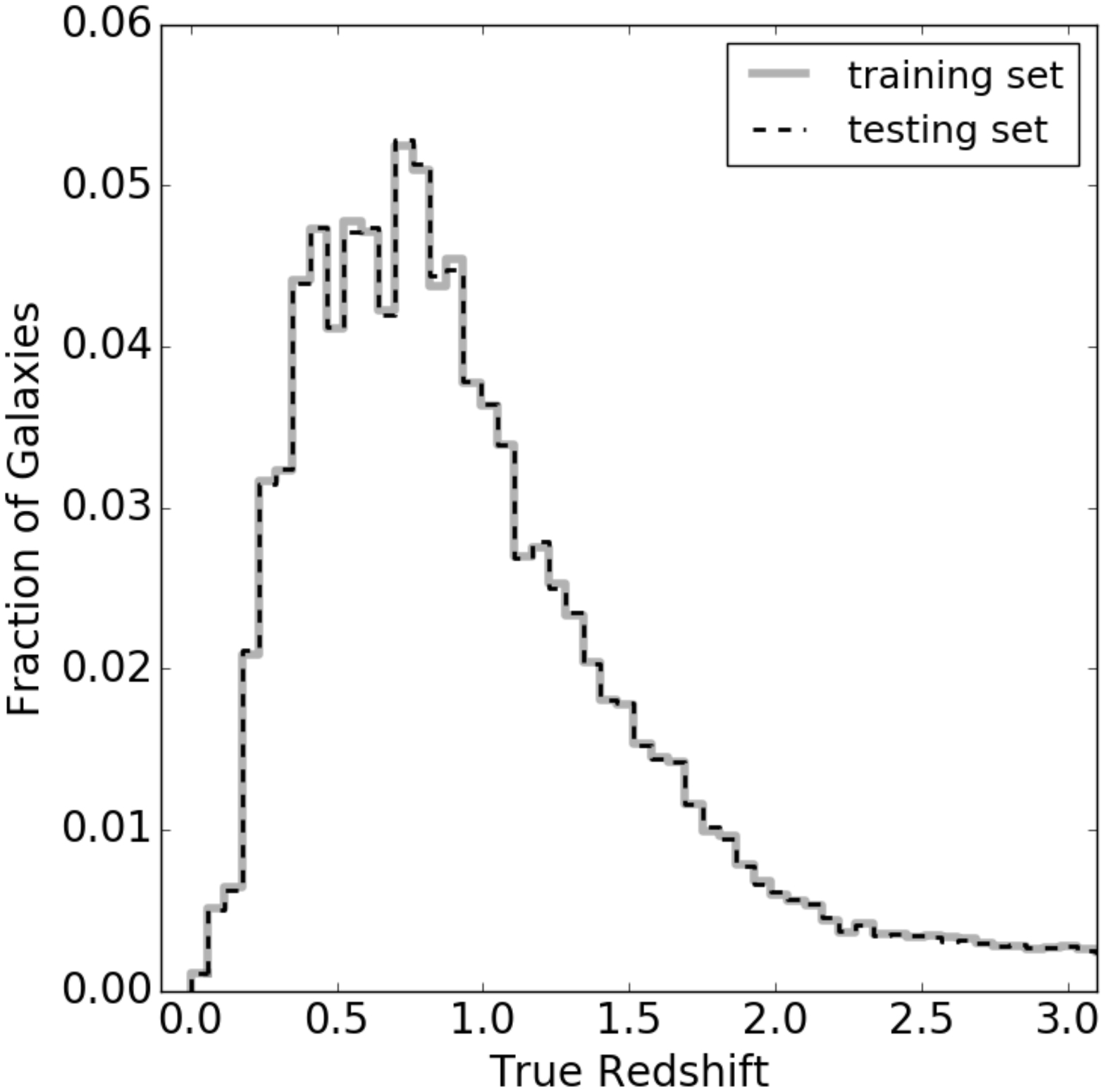}
\includegraphics[width=8.2cm,trim={1cm 4.5cm 1cm 4.5cm}, clip]{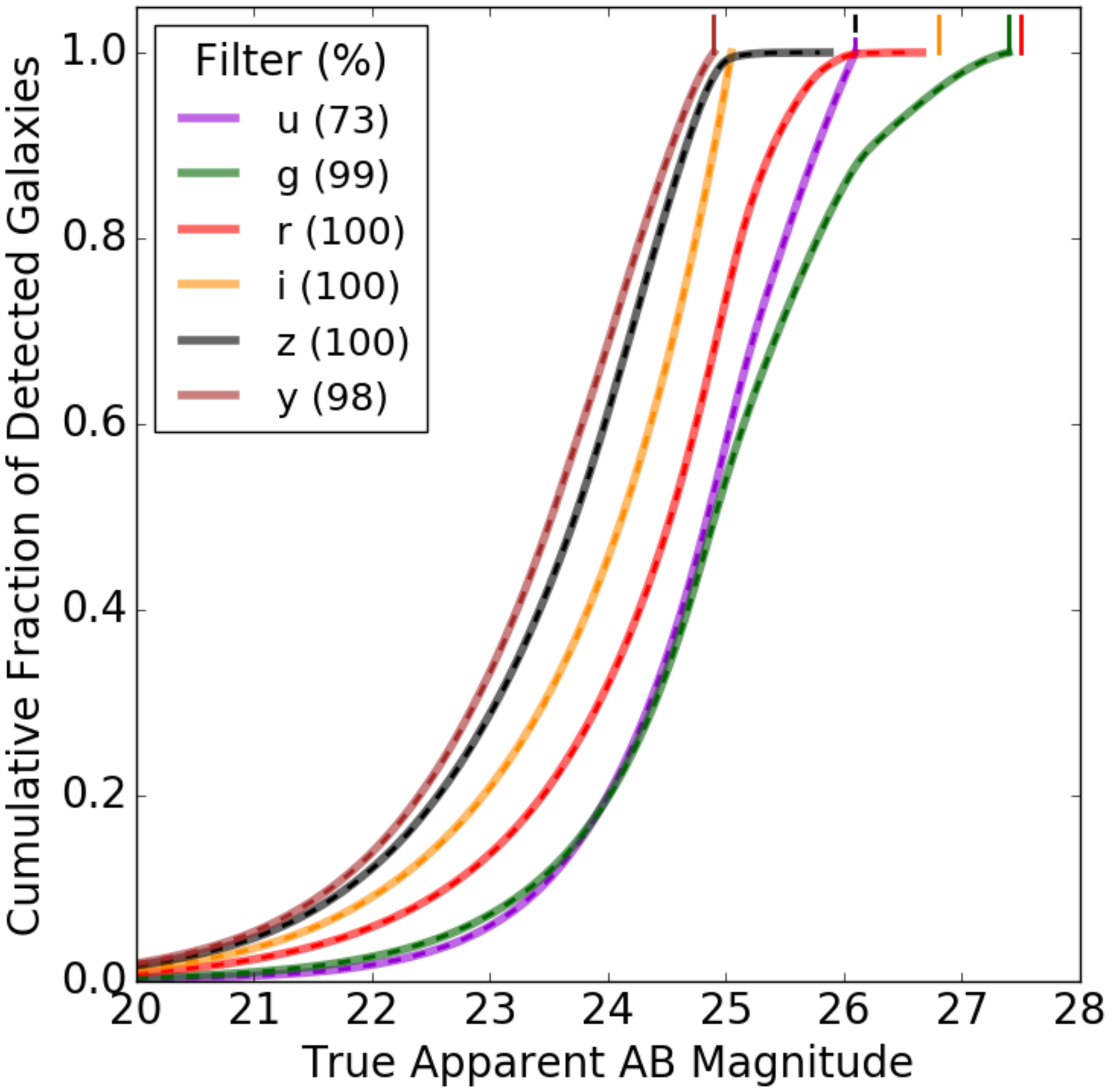}
\includegraphics[width=8.2cm,trim={1cm 4.5cm 1cm 4.5cm}, clip]{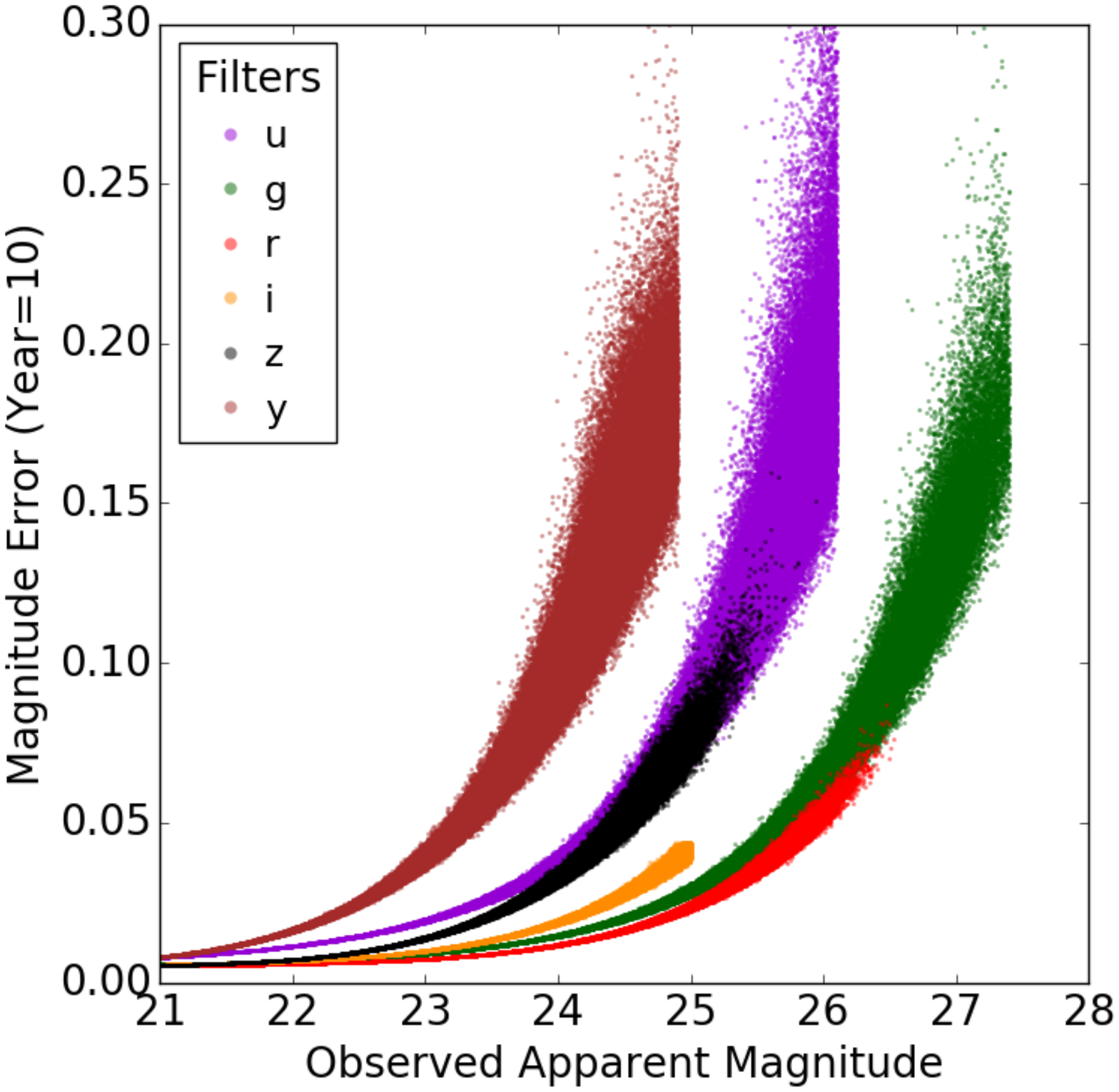}
\includegraphics[width=8.2cm,trim={1cm 4.5cm 1cm 4.5cm}, clip]{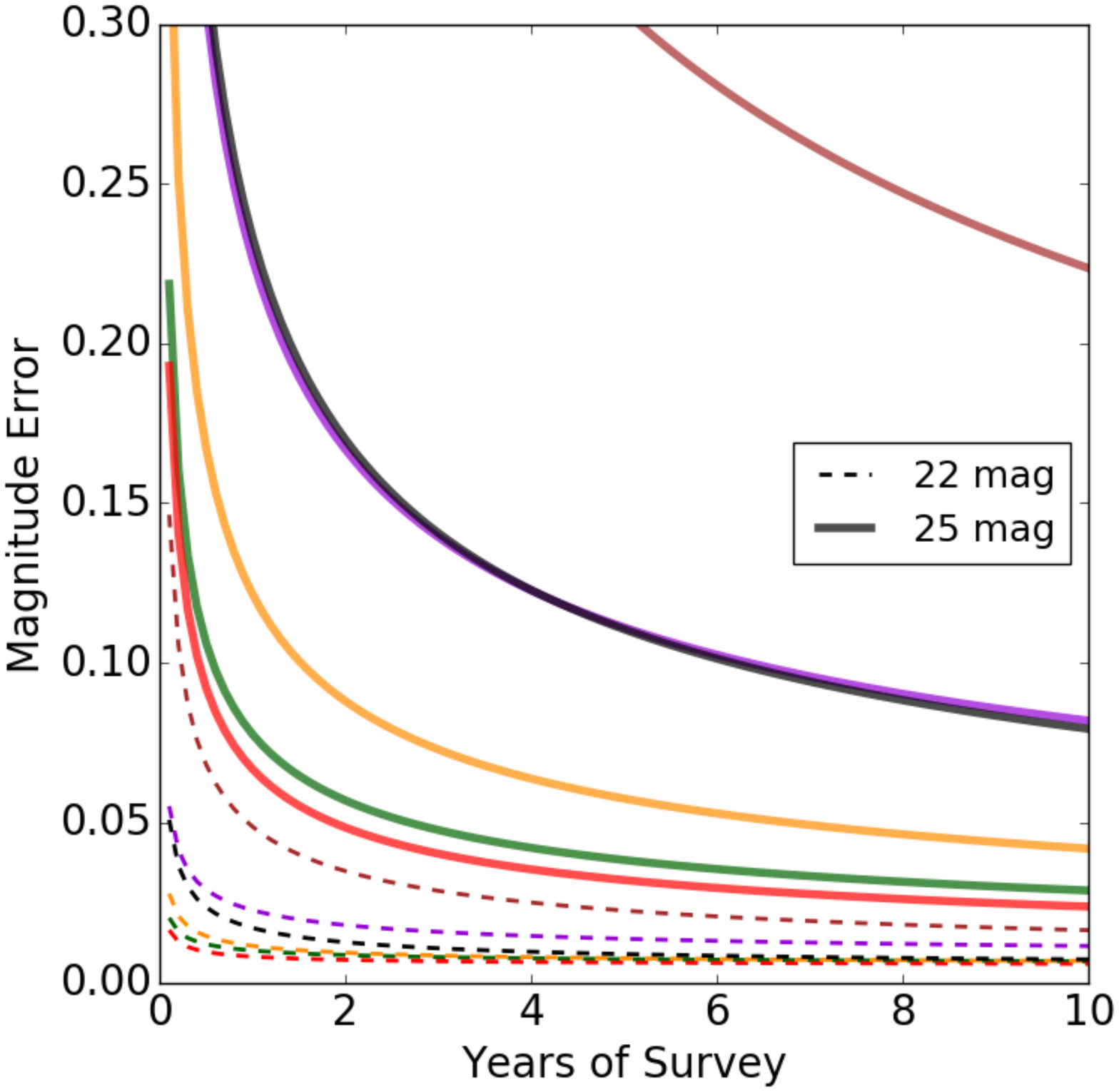}
\caption{ {\it Top left:} The redshift distribution of the training (solid) and test (dashed) sets of galaxies after a limit of $i<25$ mag has been applied. {\it Top right:} The distribution of true apparent magnitudes in the six LSST filters $ugrizy$ for the training (solid) and test (dashed) sets of galaxies after a limit of $25$ mag has been applied in the $i$-band and the predicted 10-year limiting magnitudes for LSST have been applied to the other five filters (shown as ticks in the upper right corner). The percentage of galaxies detected in each filter is given in parentheses in the legend. {\it Bottom left:} The magnitude errors $vs.$ the simulated observed apparent magnitudes, for galaxies with $i<25$. Cuts based on the predicted 10-year limiting magnitude for the LSST have also been applied in each filter. {\it Bottom right:} The magnitude error $vs.$ year of survey for galaxies of magnitude 22 (dashed) or 25 (solid) in each of the six LSST filters $ugrizy$, using the same line color scheme as the panel above. \label{fig:cats}}
\end{center}
\end{figure*}

For our experiments with LSST photometric redshifts, we start with a large simulated galaxy catalog containing ``true" apparent magnitudes and redshifts, and then generate observational errors and scatter the data to simulate ``observed" magnitudes using a prescription that is appropriate for the LSST photometric system. We use a galaxy catalog based on the Millennium simulation \citep{2005Natur.435..629S}, in particular a catalog based on the galaxy formation models of \cite{2014MNRAS.439..264G}, constructed using the lightcone construction techniques described by \cite{2013MNRAS.429..556M}\footnote{Documentation for this catalog can be found at \url{http://galaxy-catalogue.dur.ac.uk}}. This simulated galaxy catalog was designed to model the optical and near-infrared properties of galaxies, including emission lines, and with appropriate limits also serves as a realistic representation of future LSST catalogs. For this work, we start with the $\sim2.6\times10^6$ simulated galaxies with a true redshift of $z < 3.5$ and a true apparent $i$-band magnitude of $m_i < 25.05$. This is slightly deeper than the ranges in redshift and magnitude for which the SRD has defined the LSST photometric redshift goals, as described in Section~\ref{sec:intro}, to avoid any edge effects in our assessment statistics\footnote{In retrospect, we should have used a deeper true apparent $i$-band magnitude because the catalog is missing a small fraction of up-scattered galaxies in $i$-band, as indicated by the sharp upper limit on $i$-band magnitude error in the lower-left panel of Figure \ref{fig:cats}. This mistake has slightly depopulated the faint-end of our simulated galaxy luminosity function. We have rerun our photo-$z$ estimator using a true apparent magnitude limit of $i<25.5$ and found no statistically discernible impact on the overall results, and so conclude that this oversight has not influenced any of our conclusions.}.

The photo-$z$ estimator that we describe in Section~\ref{sec:method} requires ``training" and ``test" galaxy catalogs: the former is equivalent to a set of galaxies with ``known" or spectroscopic redshifts, and the latter to a sample of galaxies for which photo-$z$ will be estimated. To generate these subsets and ensure no overlap, we choose randomly without replacement from the greater catalog of $\sim2.6\times10^6$ simulated galaxies. The roles of the training and test sets in our experiment are described in further detail in Section~\ref{ssec:photoz}. In Figure \ref{fig:cats}, the top left panel shows the redshift distributions for these two subsets, and the top right panel shows the distributions in true apparent magnitude for each of the six LSST filters $ugrizy$. The distributions in the top right plot only include galaxies brighter than the predicted 10-year limiting magnitude in each filter: $u<26.1$, $g<27.4$, $r<27.5$, $z<26.1$, and $y<24.9$.

To incorporate an LSST-like observational uncertainty into our catalog we simulate observed apparent magnitudes from the true catalog magnitudes by adding a normal random scatter with a standard deviation equal to the predicted magnitude error for each galaxy. Predicted magnitude errors for the LSST, as described in Section 3.2.1 of \cite{2008arXiv0805.2366I}, depend on the galaxy's magnitude and the total survey exposure time elapsed in a given filter (for this work, we assume no additional components from e.g., deblending). We assume standard observing conditions, a mean airmass of $1.2$, and a uniform survey progression that accumulates a total of 56, 80, 184, 184, 160, and 160 visits in filters $ugrizy$ by year 10 (where each visit is 30 seconds of integration time). We use the LSST simulations software package described by \cite{2014SPIE.9150E..14C} to calculate the magnitude errors. In the lower left panel of Figure \ref{fig:cats} we plot the magnitude errors $vs.$ the simulated observed magnitudes for our test set of galaxies. It is clear from this plot that all galaxies with $i<25$ mag are also detected in the adjacent filters, $r$ and $z$, but that the 10-year detection limits in $u$, $g$, and $y$ cause a small fraction of our galaxies to be undetected in those filters. This plot furthermore shows that by first simulating the observed magnitudes and then applying the detection limits to an underlying catalog that extends fainter than $i=25$, we have realistically included galaxies that are scattered brighter than their true magnitudes. In the lower right panel of Figure \ref{fig:cats} we show how the magnitude error improves with year of survey in each filter, for galaxies of magnitude $22$ and $25$. While most of the improvement is gained in the first 2 years, faint galaxies experience significant improvement in the second half of the survey. 

In this experiment, whenever we consider a change to LSST parameters that would effect the measured magnitudes and uncertainties (e.g., the number of visits per field), we recalculate the errors and re-simulate the observed magnitudes as appropriate. The uncertainty on color is calculated as the root of the sum of the squares of the magnitude uncertainties in the two filters (i.e., the magnitude uncertainties from the two filters are added in quadrature, under the assumption of uncorrelated errors). In the next section, we describe how the training and test sets are used in our photometric redshift estimator.

\section{The Photometric Redshift Estimator}\label{sec:method}

In this section we describe the nearest-neighbors photo-$z$ estimator that we use for this work. This estimator uses the photometric colors and their errors to identify a color-matched subset of training galaxies for each simulated test galaxy (recall that the training set is analogous to a spectroscopic galaxy sample with LSST photometry). The stringency of what constitutes a color match, and thus the size of the color-matched subset of training galaxies, is controlled by a user-defined parameter. The photo-$z$ for the test galaxy is determined from this color-matched subset with one of three options (also defined by the user). These details are described further in Section~\ref{ssec:photoz}. One of the main reasons we are motivated to develop and characterize this method is that it removes any systematics introduced by assigning a SED or spectral template to each source in order to derive a redshift estimate. This intermediate step can blur the correlations between photometric uncertainties and photo-$z$ results, which we want to avoid, because exploring those correlations for a future LSST-like catalog is one of the main goals of this work. We want to be clear that we are not advocating that this method be deployed for the official LSST data products, and certainly not claiming that it provides the best photo-$z$ estimates, only that it is appropriate for our purposes in this experiment. 

In Section~\ref{ssec:photoz} we describe the method of our nearest-neighbors photo-$z$ estimator, and in Section~\ref{ssec:stats} we explain the statistical measures that we will use to analyze our results. In Section~\ref{ssec:ppfcsq} we evaluate how the application of initial queries in magnitude or color, the color-matching threshold level, and the color-matched selection method influence the photometric redshifts. In Section~\ref{ssec:N} we assess the number of galaxies to include in the test set in order to achieve statistically robust results, and discuss the training set in Section~\ref{ssec:train}. 

\subsection{Description of the Photo-$z$ Estimator}\label{ssec:photoz}

The first step in our photo-$z$ estimator is to use the color of a test galaxy to identify the nearest-neighbors in color-space from the training set of galaxies. To do this, for every galaxy in the test set we calculate the $\chi^2$ Mahalanobis distance, $D_M$, in color-space with respect to the set of training galaxies:

\begin{equation}\label{eq:DM}
D_M = \sum_{\rm 1}^{N_{\rm colors}} \frac{( c_{\rm train} - c_{\rm test} )^2}{ (\delta c_{\rm test})^2},
\end{equation}

\noindent where $c$ is color, $\delta c$ is measurement error in color, and $N_{\rm colors}$ is the total number of colors. Typically 5 colors are constructed with the 6 LSST magnitudes: $u-g$, $g-r$, $r-i$, $i-y$, $y-z$. Any magnitudes fainter than the predicted detection limits are considered non-detections\footnote{Specifically, their values are set to {\sc NaN} and we use {\sc numpy.nansum} for Equation \ref{eq:DM}.} The number of degrees of freedom (DoF) for the resulting $\chi^2$ distribution is equal to the number of colors used; non-detections in any given magnitude cause all associated colors to be excluded from $D_M$ and appropriately lowers the DoF. We identify the color-matched subset of training galaxies as those with $D_M$ less than a threshold value, which is defined by the percent point function (PPF): for example, for $N_{\rm dof} = 5$, PPF $=95\%$ of all training galaxies consistent with the test galaxy will have $D_M < 11.07$ (and PPF $=68\%$ have $D_M<5.86$). We consider the choice of PPF value in Section~\ref{ssec:ppfcsq}. Including forced photometry and/or upper limits in filters with non-detections, which will have larger and/or non-Gaussian measurement errors, is an option that we do not consider for this method at this time. 

From this color-matched subset, a single training galaxy is chosen by one of three methods: a random selection, the best color match, or a random selection weighted by $D_M^{-1}$. These three selection methods are compared in Section~\ref{ssec:ppfcsq}. The test galaxy's $z_{\rm phot}$ is assigned to be that selected training set galaxy's $z_{\rm true}$, and it's photo-$z$ uncertainty ($\delta z_{\rm phot}$) is assigned to be the standard deviation in the redshifts of the color-matched subset. This photo-$z$ uncertainty could be used to reject poorly-determined photo-$z$ and improve the overall results, but since this work is not concerned with producing photo-z for scientific use, we do not make any cuts to the test set of galaxies based on $\delta z_{\rm phot}$.

The final 10-year version of the spectroscopic sample is anticipated to contain millions of galaxies. Although the final 10-year catalog of LSST photometric redshifts will contain 4 billion galaxies, for this work we need only a subset large enough to properly assess the photometric redshift quality. The necessary size of the test subset is evaluated in Section~\ref{ssec:N}, and we discuss the size and depth of the training set in Section~\ref{ssec:train}. Calculating $D_M$ in five colors for a million training galaxies is computationally intensive -- and unnecessary, since most of them will be rejected from the color-matched subset. To save time, for each test galaxy we apply an initial query to the training set to identify galaxies that have similar $g-r$ and $r-i$ colors and $i$-band magnitudes. This step also serves as a kind of magnitude pseudo-prior on the photometric redshift. We discuss the implementation of, and acceptable scenarios for, these initial queries in Section~\ref{ssec:ppfcsq}.

\subsection{Statistical Measures and Outlier Characterization} \label{ssec:stats}

As described in Section~\ref{sec:intro}, the main goal of this work is to assess the impact on photo-$z$ from changes to the LSST photometric quality, not the overall performance of our photo-$z$ estimator with respect to any science requirements. However, we do need to ensure that we are using a photo-$z$ estimator that is at least good enough to produce results which meet the minimum expectations. Furthermore, we find it useful to have globally-defined target values to serve as static reference points for all our experiments, in which we are typically comparing the relative photo-$z$ results as we change parameters that affect the galaxy magnitude errors. Recall that in Section~\ref{sec:intro} we introduced the SRD's minimum target values for photometric redshifts as: (1) the root-mean-square error in photo-$z$ must be $< 0.02(1+z_{\rm phot})$; (2) the fraction of outliers must be $<10\%$; and (3) the average bias must be $< 0.003(1+z_{\rm phot})$. The SRD does not specify whether these values apply to tomographic bins in redshift, and only references the full range of $0.3 \leq z_{\rm phot} \leq 3.0$. For our purposes we adopt the broadest possible interpretation of the SRD's target values as applicable to the full range, but fully recognize that future precision cosmological studies might require more stringent constraints, especially in high-$z$ bins. 

In order to assess the photometric redshift results we define a set of statistical measures, where $z_{\rm true}$ is the ``true" catalog redshift, $z_{\rm phot}$ is the photometric redshift, and the difference is represented as $\Delta z = z_{\rm true} - z_{\rm phot}$. For the photo-$z$ error we use $\Delta z_{(1+z)} = (z_{\rm true} - z_{\rm phot})/(1+z_{\rm phot})$, where the denominator compensates for the larger uncertainty at high redshifts and allows for a meaningful comparison of photo-$z$ error across the redshift range of $0.3\leq z_{\rm phot} \leq 3.0$. For our measurement of the robust standard deviation in $\Delta z_{(1+z)}$ we use the interquartile range (IQR), identified by determining the upper and lower limits on $\Delta z_{(1+z)}$ that contain 50\% of the catalog galaxies. We divide the full-width of the IQR by $1.349$ to convert it to the equivalent of the standard deviation of a Gaussian distribution ($\sigma_{\rm IQR}$). We measure the robust bias as the mean value of $\Delta z_{(1+z)}$ for galaxies within the IQR: $\overline{\Delta z_{\rm(1+z), IQR}}$. We bootstrap the uncertainties on the robust standard deviation and bias by randomly drawing a subset with replacement and recalculating the statistic 1000 times.

Outlier galaxies are identified as those with $\Delta z_{(1+z)} > 3\sigma_{\rm IQR}$ or $\Delta z_{(1+z)} > 0.06$, whichever is {\it larger} (as defined by the SRD). The value of $\sigma_{\rm IQR}$ that we use to identify outliers is calculated based on all galaxies in the full range of photo-$z$, $0.3 \leq z_{\rm phot} \leq 3.0$ (i.e., we take the approach of identifying outliers in a global sense, not in redshift bins). Galaxies that become extreme outliers, with $|\Delta z| > 1$, have $z_{\rm true} < 0.5$ or $>2.0$ but are assigned $z_{\rm phot} >2.0$ or $<0.5$. The underlying physical origin of these photo-$z$ outliers is when the Lyman break in a high-$z$ galaxy spectrum is redshifted to the position of the Balmer break in a low-$z$ galaxy, which causes the photometric colors to be similar. In our experiment, we find that galaxies that become extreme outliers are typically fainter in $i$ and bluer in $u-g$ and $g-r$ than non-outlier galaxies, which is consistent with being caused by this degeneracy. We also find that the redshift distribution of the color-matched subset of training galaxies -- in other words, the probability density function from which the photo-$z$ is estimated -- is much more frequently bi-modal for extreme outliers than for galaxies with more accurate photo-$z$. Consequently, these extreme outlier galaxies usually have a large photo-$z$ uncertainty, $\delta z_{\rm phot}$, because we use the standard deviation in redshift of the color-matched subset of training galaxies as the $\delta z_{\rm phot}$ (Section~\ref{ssec:photoz}).

In the following experiments we will typically present our results in two ways. The first way is with plots of $z_{\rm true}$ $vs.$ $z_{\rm phot}$, with a domain of $0<z_{\rm phot}<3$ and a range of $0<z_{\rm true}<3.5$. Both the test and training sets of galaxies extend to a redshift of $3.5$ in order to avoid edge effects at the highest redshift that we include in our analysis, $z_{\rm phot}=3$. In these plots of $z_{\rm true}$ $vs.$ $z_{\rm phot}$ we will usually plot all galaxies with a semi-transparent black point in order to convey a sense of density in $z_{\rm true}$ $vs.$ $z_{\rm phot}$, and then color all galaxies classified as an outlier with more opaque red point. Since the definition of an outlier depends on the $\sigma_{\rm IQR}$, which is calculated only over $0.3 \leq z_{\rm phot} \leq 3.0$, we do not identify outliers at $z<0.3$. We use these plots of $z_{\rm true}$ $vs.$ $z_{\rm phot}$ to illustrate the quality of the photo-$z$ results for a single run of our estimator. 

The second way that we will typically present our results is with plots of the robust standard deviation or bias in $z_{\rm phot}$-bins across our analysis range, from $0.3 \leq z_{\rm phot} \leq 3.0$. We use these plots to compare the statistical results for multiple runs of our photo-$z$ estimator in which we have varied the input parameters. It is important to note that the classification of galaxies as outliers is based on the $\sigma_{\rm IQR}$ over the full range of $0.3 \leq z_{\rm phot} \leq 3.0$, but when we compute the robust standard deviation or bias in a redshift bin we use the IQR of galaxies in that bin. This means that the IQR of a single bin might contain globally-defined outliers. This is especially true at high redshifts where there is more scatter in $\Delta z_{(1+z)}$ and less galaxies per bin.

It is also important to note that when the robust standard deviation or bias in individual $z_{\rm phot}$-bins exceeds the SRD's targets, which we represent as a horizontal dashed lines in such plots, this does not indicate an absolute failure of the estimator. As described above, for our purposes the SRD's minimum target values apply to the full range of $0.3 \leq z_{\rm phot} \leq 3.0$, and so we also represent the robust standard deviation or bias across this range as a horizontal colored bar in such plots. A single run of our photo-$z$ estimator can be considered to have ``met the SRD's minimum targets" so long as the statistical measure from all galaxies in the full range of $0.3 \leq z_{\rm phot} \leq 3.0$ is below the targeted value.

\subsection{Photo-$z$ Method Parameters}\label{ssec:ppfcsq}

Here we compare the results of five runs with our photo-$z$ estimator in which we vary three parameters: the type of initial query imposed on the training set that reduces the computational time (described below); the PPF value that controls which training set galaxies are included in the color-matched set; and the method by which a galaxy is chosen from the color-matched set. For all five of these runs we use the same $10^6$ training set and $5\times10^4$ test set galaxies, and simulate magnitude errors for a full-depth, 10-year LSST survey. 

\begin{figure}
\begin{center}
\includegraphics[width=8.2cm,trim={1cm 4.5cm 1cm 4.5cm},clip]{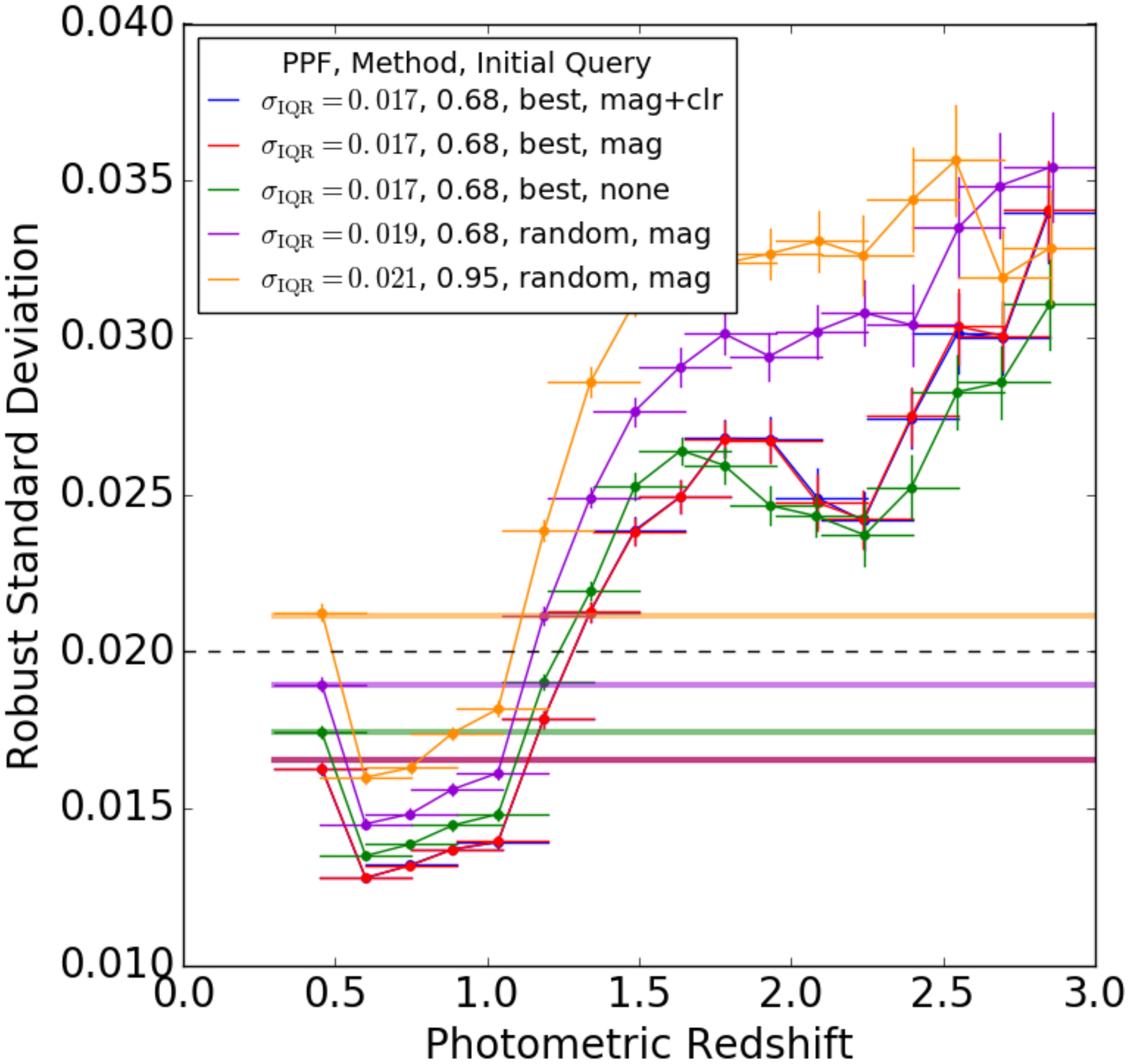}
\includegraphics[width=8.5cm,trim={0.5cm 4.5cm 0.5cm 4.5cm},clip]{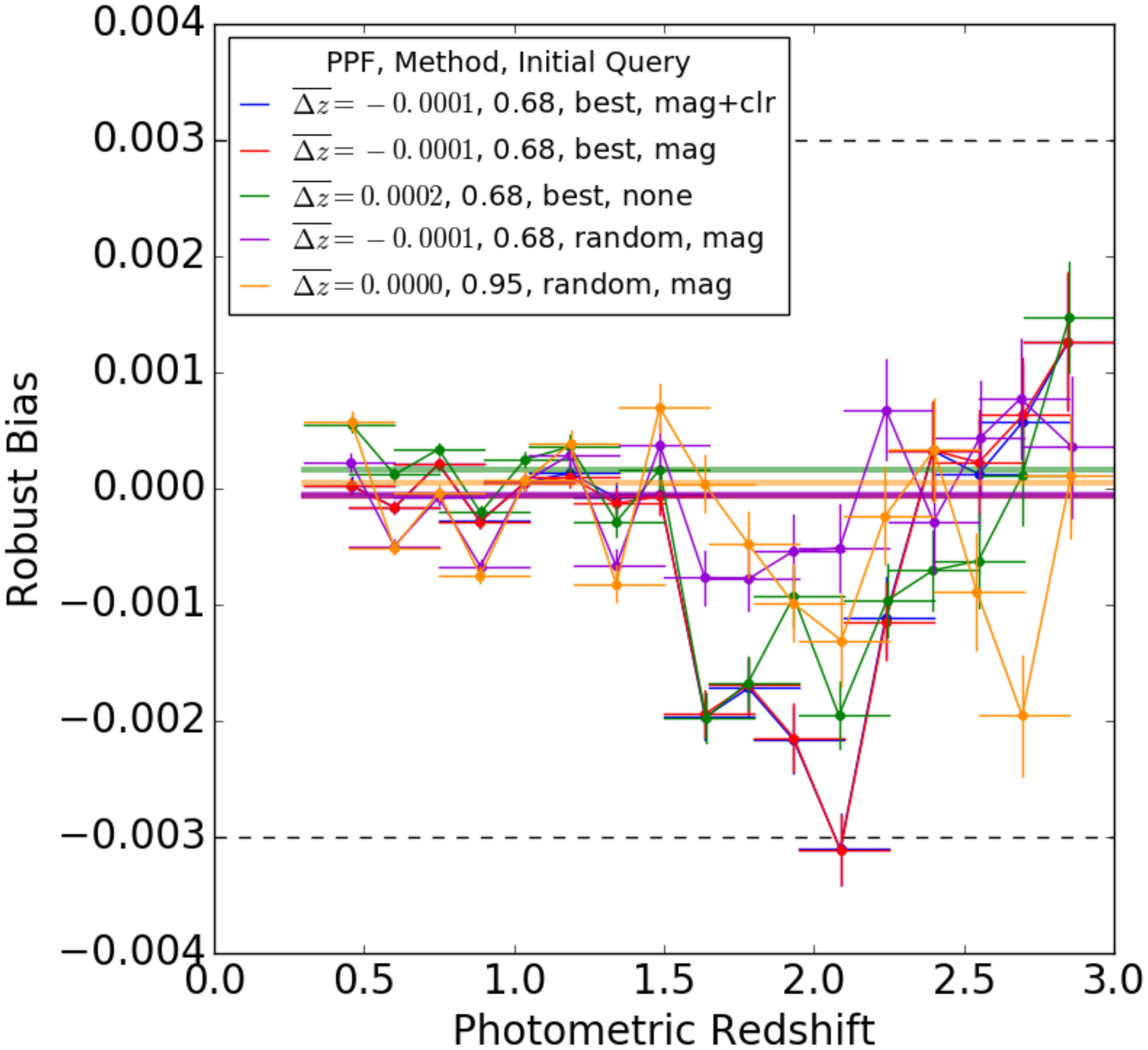}
\caption{The robust standard deviation ({\it top}) and bias ({\it bottom}) of $\Delta z_{(1+z)}$ in bins of $z_{\rm phot}$ for five runs in which we vary three options: (1) no initial query (green), an initial query based on apparent magnitude (red), or initial queries based on magnitude and color (blue; very similar to the red line); (2) the PPF value, $0.68$ (purple) or $0.95$ (orange); and (3) the method of choosing either the best (red) or a random (purple) training set galaxy from the color-matched set. Horizontal colored lines mark the value of the statistic over the full redshift range $0.3 \leq z_{\rm phot} \leq 3.0$, which are also listed in the legend. Horizontal dashed lines mark the SRD's target value for each statistic. \label{fig:ppfcsq}} 
\end{center}
\end{figure}

The results are presented in Figure \ref{fig:ppfcsq}, where we plot the statistical measures of robust standard deviation and bias as a function of $z_{\rm phot}$. We find that the SRD target value is met for the bias regardless of how these three parameters are set, but that the target for standard deviation is not met if we use $PPF=0.95$ and a random selection from the color-matched subset of training galaxies. We do not show it here, but the SRD's target for fraction of outliers is also always met. Each of the variable aspects of our photo-$z$ estimator are discussed in turn below.

\begin{figure}
\begin{center}
\includegraphics[width=8.5cm,trim={1cm 4.5cm 1cm 4.5cm},clip]{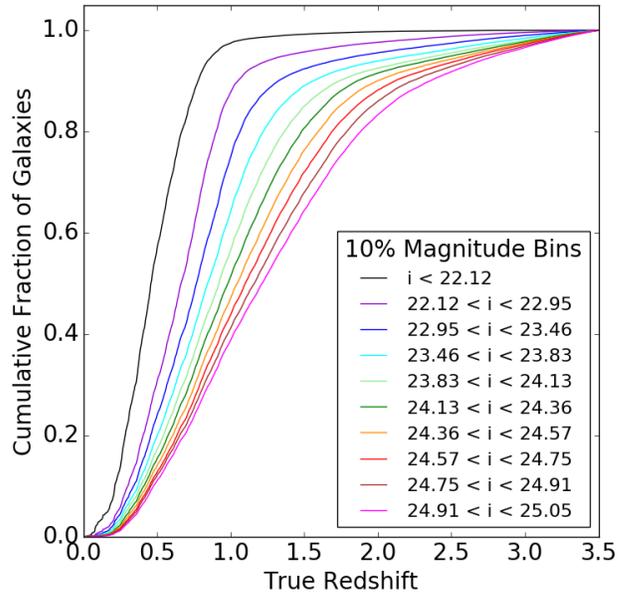}
\caption{Distributions of true catalog redshift in bins of true apparent $i$-band magnitude, with each bin containing 10\% of the catalog. In this case there are $2\times10^5$ galaxies per bin. \label{fig:mbins}} 
\end{center}
\end{figure}

\textbf{Initial queries --} We identify two queries that can be applied to the training set in order to limit the number of galaxies for which $D_M$ must be calculated, and save computational time: one is based on $i$-band magnitudes and the other on $g-r$ and $r-i$ colors. For the first query, we start with the training set's distribution of $i$-band magnitudes and identify the $\pm5\%$ of galaxies on either side of the test galaxy's $i$-band magnitude. The magnitude range of this $\pm5\%$ subset is always greater than the largest uncertainty on $i$-band magnitude, $\pm0.05$ (Figure \ref{fig:cats}). We then only calculate $D_M$ for this 10\% sample, a representative set of magnitude-matched training galaxies. Redshift distributions in 10\% bins in $i$-band magnitude are shown in Figure \ref{fig:mbins}. This query effectively functions as a magnitude prior, but we call it a pseudo-prior because it is not providing a magnitude-based probabilistic weight on the photo-$z$ selection, which is the more typical application of the word ``prior". This query proportionally lowers the computational time required by $\sim90\%$ (to $\sim3$ hours on a personal laptop for a training set of $10^6$ galaxies). We can further reduce this run time by another $\sim60\%$ by not calculating $D_M$ for training set galaxies that have no chance of being included in the color-matched subset\footnote{Computational run time could also be further reduced by the use of, e.g., k-d trees}. To do this we use the three filters with the lowest photometric errors, $g$, $r$, and $i$, and only calculate $D_M$ for training set galaxies with $g-r$ and $r-i$ colors within $\pm0.3$ of the test galaxy. This limit is set by taking the maximum error in color for the faintest detected galaxy with a 1-year LSST survey from the bottom right panel of Figure \ref{fig:cats} and rounding up. By choosing the three filters with the lowest magnitude errors we eliminate a larger number of training set galaxies with significantly disparate colors, but these are the two colors with the most weight in $D_M$ (Equation \ref{eq:DM}) and so we must ensure that our initial color query does not bias the statistical results. In Figure \ref{fig:ppfcsq} we plot the robust standard deviation and bias in bins of $z_{\rm phot}$ for runs in which we do not use an initial query in magnitude or color (green line), use an initial query in magnitude only (red line), or use an initial query in both magnitude and color (blue line). We find that the results are almost identical whether or not we make a preliminary cut based on color, which means that it is a safe way to cut down the computational time (i.e., the blue and red lines are coincident\footnote{Although we do not show it in Figure \ref{fig:ppfcsq}, the statistical results for the run in which we apply only an initial query on color is coincident with the results for no query}). We also find that the robust standard deviation and bias over the full redshift range of $0.3 \leq z_{\rm phot} \leq 3.0$ are closer to the SRD's targets when we do include the initial magnitude query (i.e., an $i$-band magnitude pseudo-prior), but that in the high-redshift bins the best results are achieved when this is not applied. Given the time savings offered by these initial queries we will implement them in this work unless otherwise specified. We note that there is room for determining the optimal implementation of magnitude pseudo-priors for various photo-$z$ science goals, but we leave this for future work. At this point, we find it sufficient to show that our initial queries to reduce computation time produces unbiased photo-$z$ results that are appropriate for our analysis.

\textbf{Percent Point Function (PPF) Value --} As described in Section~\ref{ssec:photoz}, the PPF defines the boundaries of the color-matched subset of training galaxies: a lower value is a more restrictive match in color-space, and a higher value is less restrictive. In Figure \ref{fig:ppfcsq} we compare the impact of using a PPF value $0.68$ or $0.95$ (purple and orange lines), and find that as expected, the robust standard deviation is improved with a lower value of the PPF. The robust bias is unchanged over the full redshift range $0.3 \leq z_{\rm phot} \leq 3.0$, but an improvement at higher redshifts is seen with a more restrictive PPF value. We furthermore confirm that while a lower PPF value does increase the number of test galaxies for which no photo-$z$ is returned (i.e., their color-matched subset based on $D_M$ is empty), it remains $<4\%$, which is an acceptable level.

\textbf{Method of Photo-$z$ Selection --} At the point where we have identified the color-matched sample of training set galaxies for a given test galaxy, we can choose the photo-$z$ with one of three methods: a random selection, a random selection weighted by $D_M^{-1}$, or the best color match (i.e., the nearest neighbor in color-space). We found that weighting by $D_M^{-1}$ gives only a moderate improvement to the statistical measures, and so in Figure \ref{fig:ppfcsq} we only compare the random and best selection methods (red and purple lines). We find that choosing the best-matched training set galaxy instead of using a random selection results in a reduction of the robust standard deviation from $\sigma_{\rm IQR} = 0.019$ to $0.017$ over $0.3 \leq z_{\rm phot} \leq 3.0$. In higher redshift bins, we can see that the improvement is greater (e.g., from $\sigma_{\rm IQR} = 0.030$ to $0.025$ at $z_{\rm phot}\approx2.0$). Choosing the ``best" training set galaxy does not affect the bias over $0.3 \leq z_{\rm phot} \leq 3.0$, which is well within the SRD target value, but does worsen the bias at $z_{\rm phot}\approx2.0$. We also find that choosing the nearest color neighbor from the training set also decreases the number of outliers with $\Delta z > \pm1$ (galaxies with truly high-$z$ that are assigned a low-$z_{\rm phot}$). Ultimately we think that the improvement to the standard deviation outweighs the minor degradation in bias and will use the ``best" selection option for our experiment, but note that this might not be appropriate for all scientific uses of this photo-$z$ estimator.

\subsection{Size of the Test Set}\label{ssec:N}

\begin{figure}
\begin{center}
\includegraphics[width=8.5cm,trim={1cm 4.5cm 1cm 4.5cm},clip]{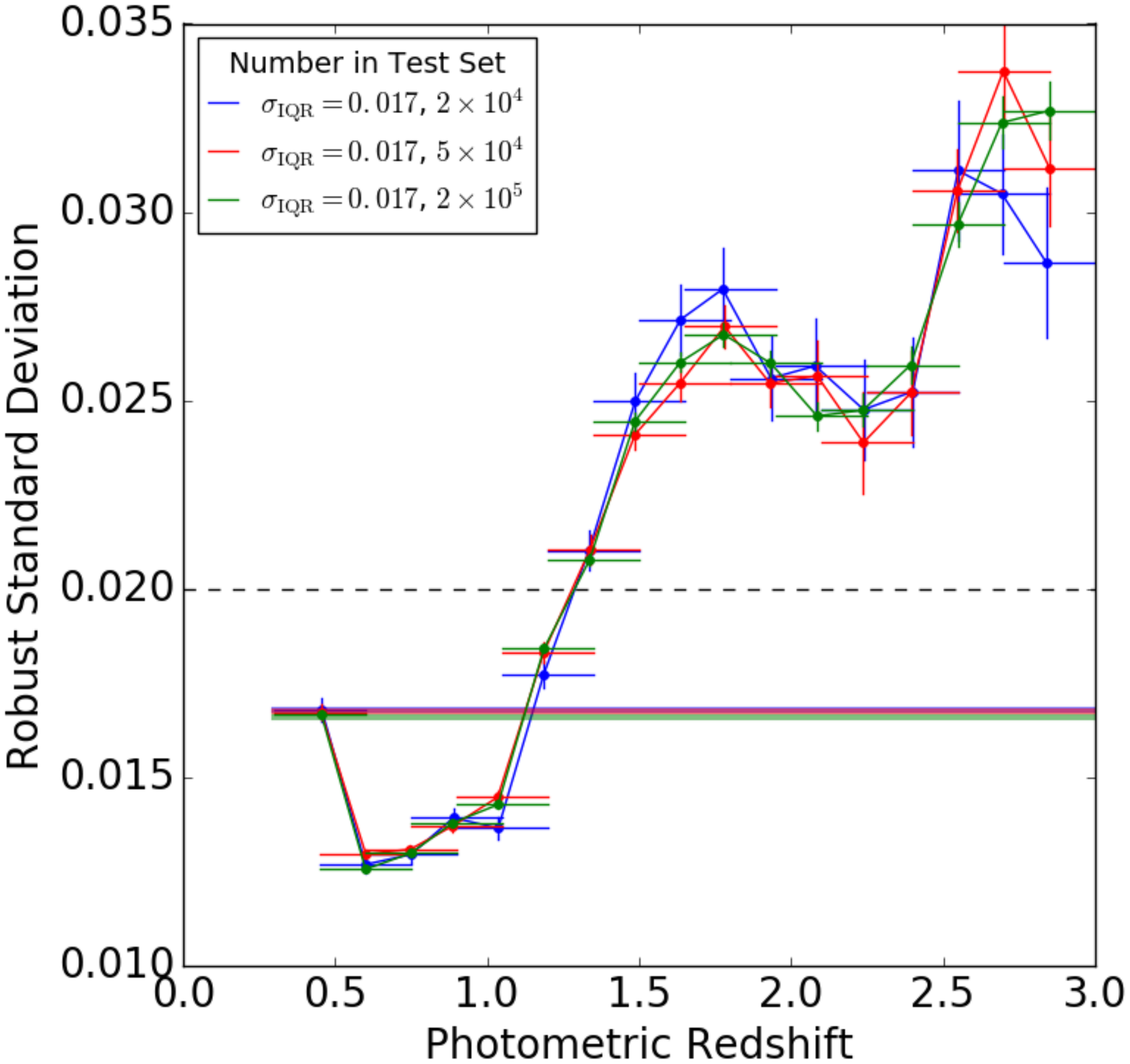}
\includegraphics[width=8.5cm,trim={1cm 4.5cm 1cm 4.5cm},clip]{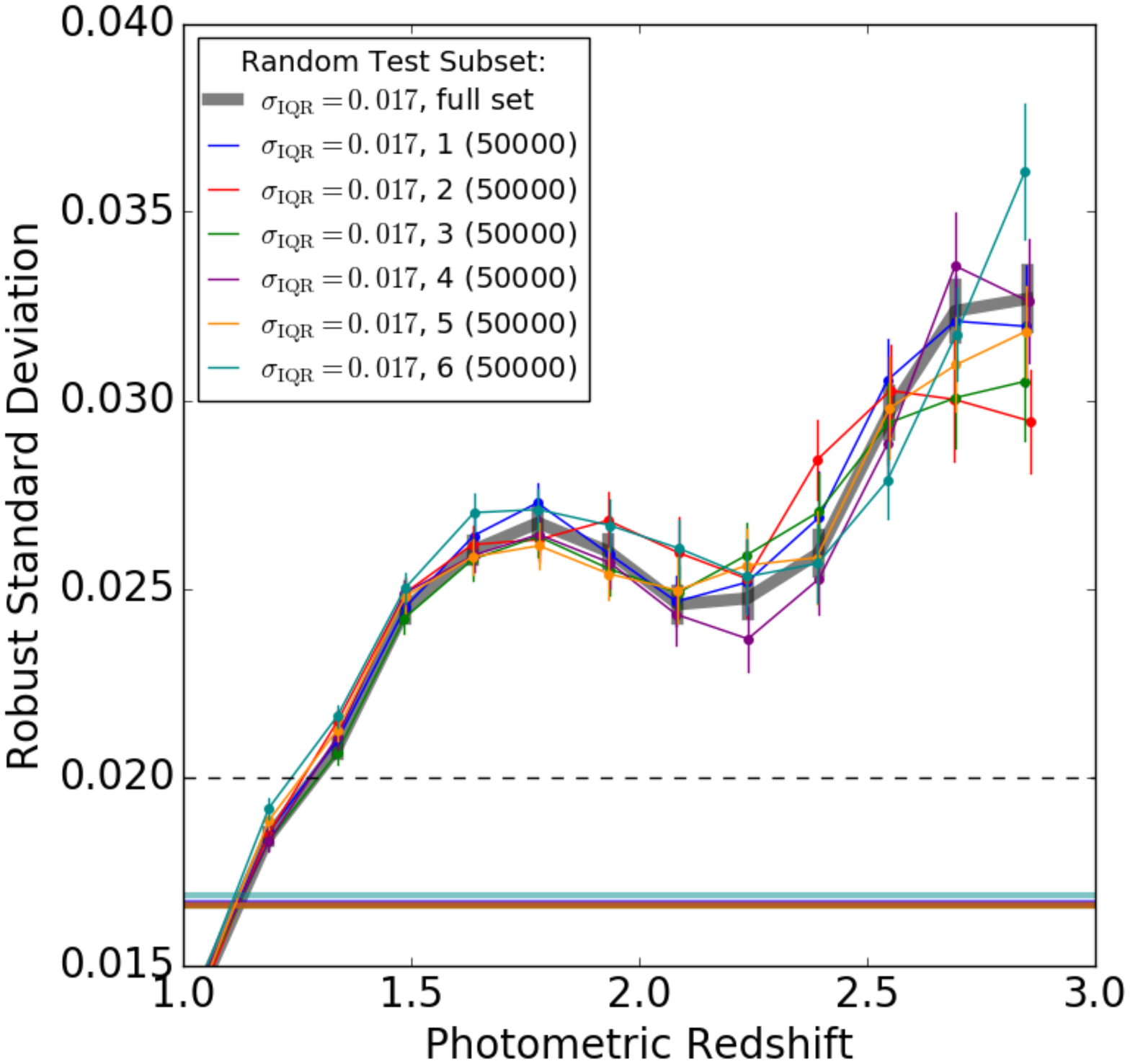}
\caption{{\it Top:} The robust standard deviation in $\Delta z_{(1+z)}$ as a function of $z_{\rm phot}$ for three runs of our photo-$z$ estimator in which we vary the number of galaxies in the test set. {\it Bottom:} We make six random draws of $5\times10^4$ test galaxies from the run with a total of $2\times10^5$ test set galaxies, and compare the dispersion in robust standard deviation of $\Delta z_{(1+z)}$ between these six subsets and the original in the highest redshift bins. Horizontal colored lines mark the value of the statistic over the full redshift range $0.3 \leq z_{\rm phot} \leq 3.0$, which are also listed in the legend. Horizontal dashed lines mark the SRD's target value for each statistic. \label{fig:N}} 
\end{center}
\end{figure}

In order to confirm that our analysis is not influenced by random fluctuations in, e.g., the magnitude or redshift distributions of test set that we choose from the larger catalog, we compare the results our photo-$z$ estimator when we vary the number of galaxies in the test set. For these runs we simulate magnitude uncertainties predicted for a 10-year LSST survey. We use a PPF value of $0.68$, the ``best" color-match selection method, and initial queries in both magnitude and color as discussed in Section~\ref{ssec:ppfcsq}. In the top panel of Figure \ref{fig:N} we show the robust standard deviation as a function of $z_{\rm phot}$, and over the redshift range of $0.3 \leq z_{\rm phot} \leq 3.0$, for runs in which we use a different number of test set galaxies. 

We find that the robust standard deviation in the photo-$z$ results over the redshift range of $0.3 \leq z_{\rm phot} \leq 3.0$ is not significantly affected by the size of the test set, as shown by the values of $\sigma_{\rm IQR}$ in the legend of the top panel of Figure \ref{fig:N}. However, in the highest redshift bins -- which contain a small fraction of the total number of galaxies (Figure \ref{fig:cats}) -- the standard deviation is affected by the number of test galaxies. Specifically, we can also see how the error in standard deviation increases when a smaller sized test set is used; e.g., the size of the error bar in the highest redshift bin is visibly larger for a smaller test set. Furthermore, we see a misleading turn-over in the $\sigma_{\rm IQR}$-$z_{\rm phot}$ relation in the final redshift bin in two of our runs that use $2\times10^4$ and $5\times10^4$ test set galaxies (blue and red in the top panel of Figure \ref{fig:N}). However we can also see that the values of $\sigma_{\rm IQR}$ for the two final bins agree within their error bars, indicating that this turn-over is not statistically significant, and raising the question of whether we need to include a systematic component in the error bar for $\sigma_{\rm IQR}$ when we use a test set size of $5\times10^4$ galaxies or less. 

To test for the presence of such a systematic, we repeat our statistical assessment using six random draws of $5\times10^4$ test galaxies chosen from the larger set of $2\times10^5$ test galaxies (represented by the green line in Figure \ref{fig:N}). In the bottom panel of Figure \ref{fig:N}, we compare the robust standard deviation in high redshift bins for these six subsets. If we take the final redshift bin as an example, we can see that the error bars on $\sigma_{\rm IQR}$ for runs 1, 3, 4, and 5 include the value of $\sigma_{\rm IQR}$ for the full test set. Four out of six is $67\%$, and since we are using $1\sigma$ from the bootstrapped statistical error, this is consistent with expectations. We find that our use of $1\sigma$ bootstrapped statistical errors adequately represents the uncertainty on $\sigma_{\rm IQR}$ in a bin, and therefore we know that an additional systematic to compensate for under-sampling the test galaxy set is not required when $5\times10^4$ galaxies are included. In subsequent sections we will assess the affects of altering various LSST survey parameters on the photo-$z$ results, and some of that analysis will rely small changes to the $\sigma_{\rm IQR}$ in high redshift bins. Based on what we have learned here, we will use identical test sets of at least $5\times10^4$ galaxies to ensure that any variation we observe in the standard deviation is due to the survey parameters being altered, and not due to fluctuations in the test set composition.

\subsection{Size and Depth of the Training Set}\label{ssec:train}

Our photometric redshift estimator will fail to return a photo-$z$ for a test galaxy if there are no training set galaxies well match in color-space, and so we also investigate how the failure rate is dependent on training set size to ensure we are using a large enough sample. We find that as we increase the size of the training set from $5\times10^5$ to $10^6$ to $2\times10^6$ galaxies, the fraction of test galaxies that fail to obtain a photo-$z$ decreases from $6.6\%$ to $4.5\%$ to $3\%$, respectively, over the redshift range $0.3 \leq z_{\rm phot} \leq 3.0$. For high redshift, $z\approx2.5$, the attrition rates are closer to $15\%$, $10\%$, and $7\%$. While these numbers are acceptable, we do not want to compromise our ability to robustly assess the statistical measures at high redshift and so will not use less than $10^6$ training set galaxies in the experiments of this work.

A training set composed of a million galaxies is also a realistic simulation of the future sample of galaxies with spectroscopic redshifts. Since obtaining the spectra for such a large, deep set of galaxies requires a considerable investment of observing time, it is reasonable to assume that the corresponding photometry would be obtained with LSST as soon as possible. For this work we assume that the training set has photometric uncertainties equivalent to the full 10-year depth of LSST. In reality, a spectroscopic training set imaged to a 10-year equivalent depth could have even lower magnitude errors than a 10-year test set because the spectra can further be used to refine the photometric corrections, and/or have LSST imaging that is deeper than the full 10-year depth due to e.g., a mini-survey or deep drilling field during commissioning or the first year of survey operations. Spectroscopic training sets might also have different magnitude and redshift distributions from the test set, and these would affect the performance of our photo-$z$ estimator. In order to focus on survey strategy and its impact on LSST photometric redshifts, we use the same training set of $10^6$ galaxies with photometry equivalent to the 10-year depth of LSST for all of our experiments.


\section{The Effects of LSST Parameters}\label{sec:lsstparams}

In this section we use our photo-$z$ estimator to demonstrate how certain aspects of the LSST survey that can change the magnitude errors may diminish, or in some cases enhance, the resulting photometric redshifts. We investigate how the LSST photo-$z$ are impacted in four scenarios: in Section~\ref{ssec:years} we demonstrate how the photo-$z$ results improve over the years, assuming a uniform progression; in Section~\ref{ssec:visits} we evaluate the photo-$z$ results when we vary the relative number of visits allotted to the $u$- and $y$-band filters; in Section~\ref{ssec:year1gri} we consider the potential gains if the LSST only observed in filters $gri$ for the first year; and in Section~\ref{ssec:errors} we study generic scenarios in which systematic or random errors are added to the photometric uncertainties. For the experiments in this section we use the same test and training sets of $5\times10^4$ and $10^6$ galaxies respectively. We use a PPF value of $0.68$, select the best match from the training set, and apply initial queries to both magnitude and color.

\subsection{Survey Progression}\label{ssec:years}

\begin{table} 
\begin{center} 
\caption{Statistical measures with survey year.} 
\label{tab:years} 
\begin{tabular}{cccc} 
\hline 
\hline 
Years for     & Robust Standard & Robust & Fraction \\ 
Test Set      &  Deviation             & Bias     & of Outliers \\ 
\hline 
1  &  $ 0.0364 \pm 0.0002 $  &  $ -0.0039 \pm 0.0001 $  &  $ 0.15 $  \\  
2  &  $ 0.0262 \pm 0.0002 $  &  $ -0.0019 \pm 0.0 $  &  $ 0.12 $  \\  
5  &  $ 0.0197 \pm 0.0001 $  &  $ -0.0005 \pm 0.0 $  &  $ 0.08 $  \\  
10  &  $ 0.0165 \pm 0.0001 $  &  $ -0.0001 \pm 0.0 $  &  $ 0.04 $  \\  
\hline 
\end{tabular} 
\end{center} 
\end{table} 

\begin{figure*}
\begin{center}
\includegraphics[width=8cm,trim={1cm 4.5cm 1cm 4.5cm},clip]{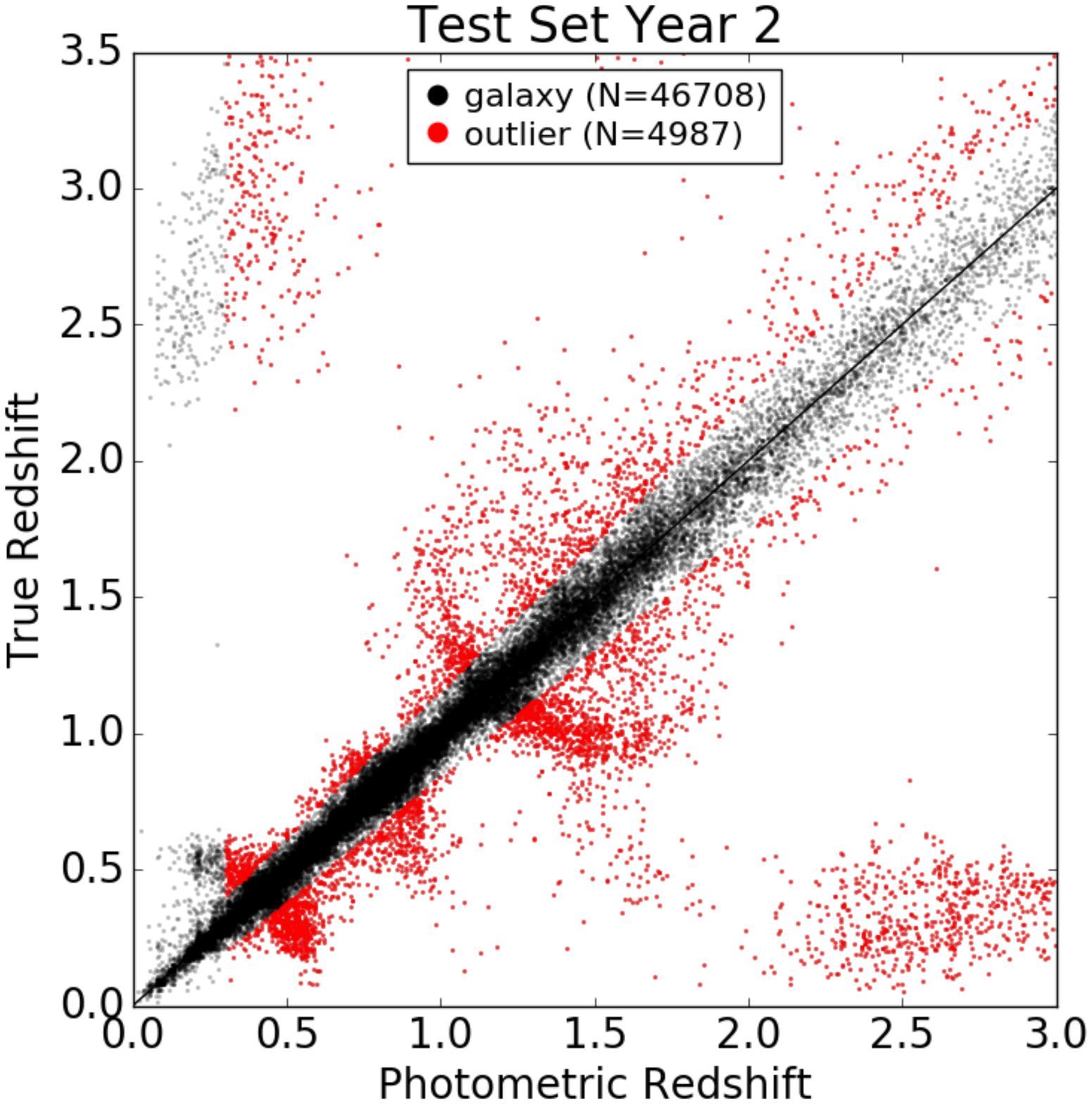}
\includegraphics[width=8cm,trim={1cm 4.5cm 1cm 4.5cm},clip]{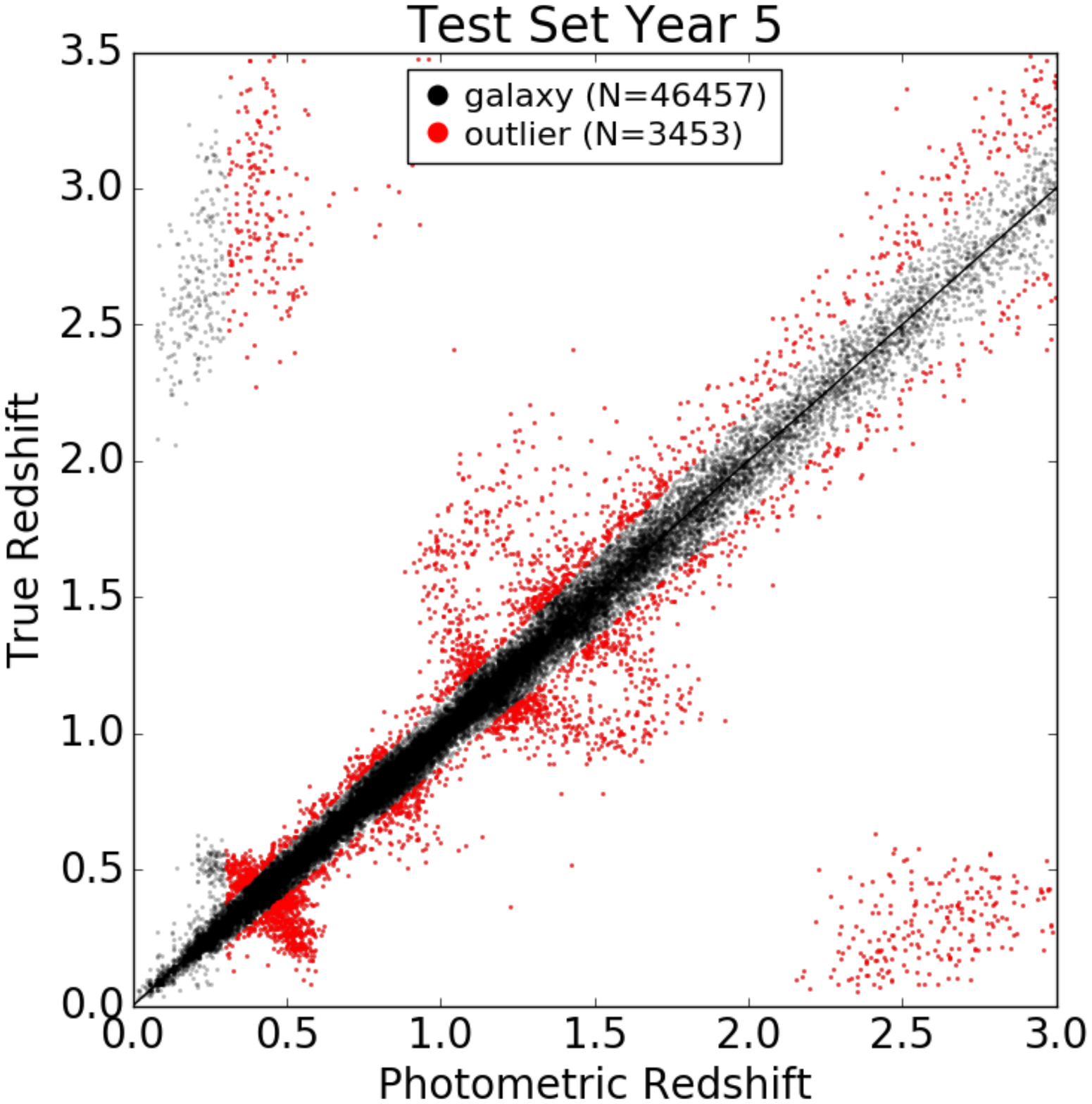}
\includegraphics[width=8cm,trim={1cm 4.5cm 1cm 4.5cm},clip]{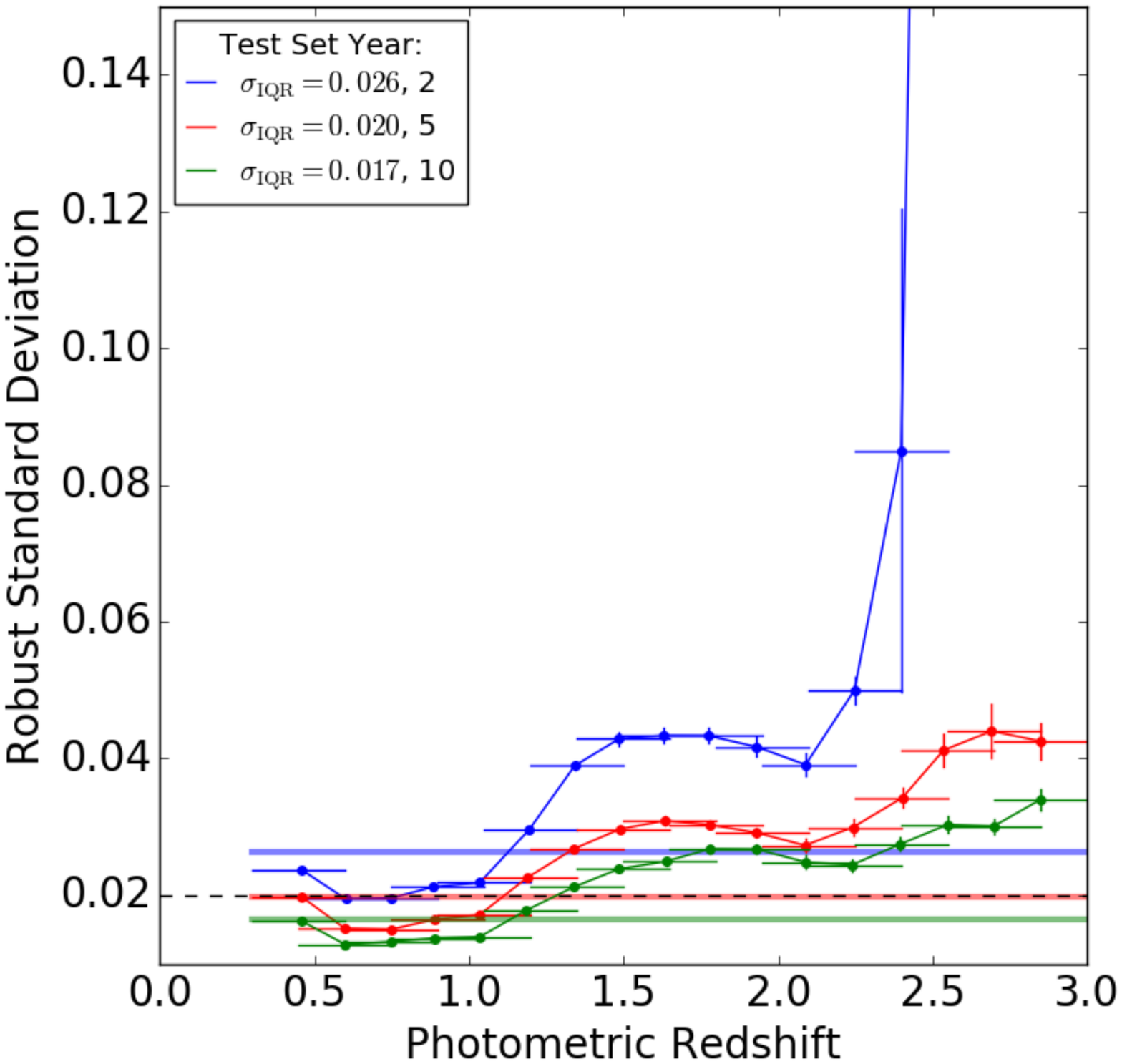}
\includegraphics[width=8cm,trim={0.5cm 4.5cm 0.5cm 4.5cm},clip]{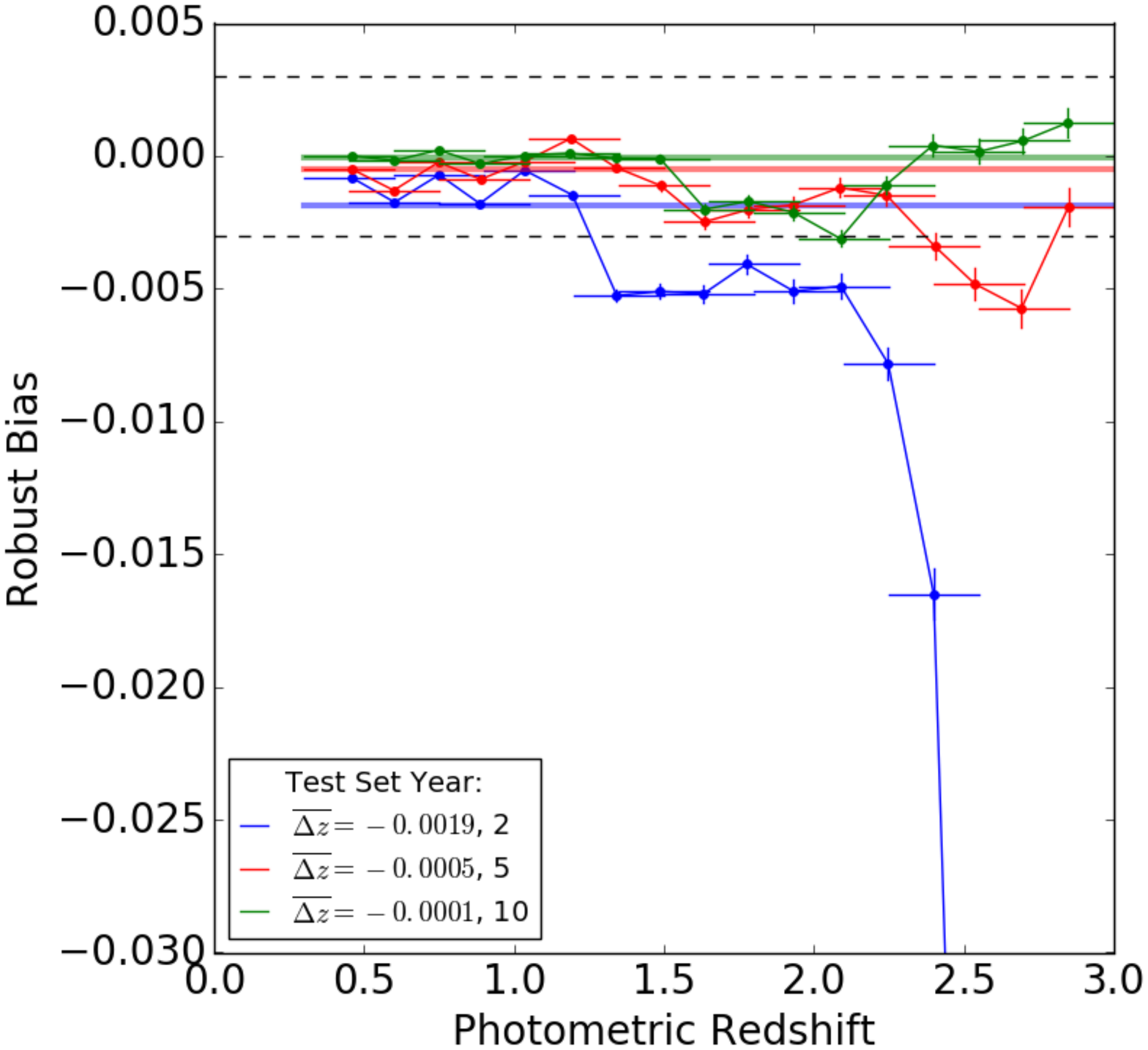}
\caption{{\it Top row:} True $vs.$ photometric redshifts when we simulate errors for the test set galaxies that are equivalent to the LSST data at 2 ({\it left}) and 5 ({\it right}) years of the survey. {\it Bottom row:} Statistical measures of robust standard deviation ({\it left}) and bias ({\it right}) in $\Delta z_{(1+z)}$ as the photometric errors improve from 2 to 10 years of survey time elapsed. Axes limits are necessarily set to cut off some points for clarity. Horizontal colored lines mark the value of the statistic over the full redshift range $0.3 \leq z_{\rm phot} \leq 3.0$, which are also listed in the legend. Horizontal dashed lines mark the SRD's target value for each statistic. \label{fig:results_year}}
\end{center}
\end{figure*}

With a uniform progression survey, the photometric uncertainties will improve steadily from the first year of the LSST through to the full 10 years of the survey, as shown in Figure \ref{fig:cats}. The photometric redshifts will likewise improve proportionally to the magnitude errors. At 2, 5, and 10 years we simulate the photometric uncertainties in all filters as described in Section~\ref{sec:cat}, apply them to the true apparent magnitudes for all test galaxies to simulate observed magnitudes with realistic errors for that stage of the survey, and run our photo-$z$ estimator for these data sets.

In the top two panels of Figure \ref{fig:results_year}, we plot $z_{\rm true}$ $vs.$ $z_{\rm phot}$ at 2 and 5 years, which can be compared to the results at 10 years shown in the lower left panel of Figure \ref{fig:magair1}\footnote{The lower left panel of Figure \ref{fig:magair1} is an appropriate representative of the nominal 10-year results despite the fact that it includes a prior from observed $\mathcal{M}_u$ because we showed the prior was ineffective.}. Between year 2 and 5, we can see a visible improvement in the fraction of outliers and in the spread of the scatter around $z_{\rm phot}=z_{\rm true}$. In the bottom two panels of Figure \ref{fig:results_year} we compare the robust standard deviation and bias in bins of $z_{\rm phot}$ for our simulated results at 2, 5, and 10 years. Although the improvement between years 2 and 5 is greater than that between 5 and 10, we still see a significant amount of improvement in the second half of the survey -- especially in the higher redshift bins. We also list the values for each of our statistical for measures with year of survey in Table \ref{tab:years}.

\subsection{Visits per Filter}\label{ssec:visits}

\begin{figure*}
\begin{center}
\includegraphics[width=8cm,trim={1cm 4.5cm 1cm 4.5cm},clip]{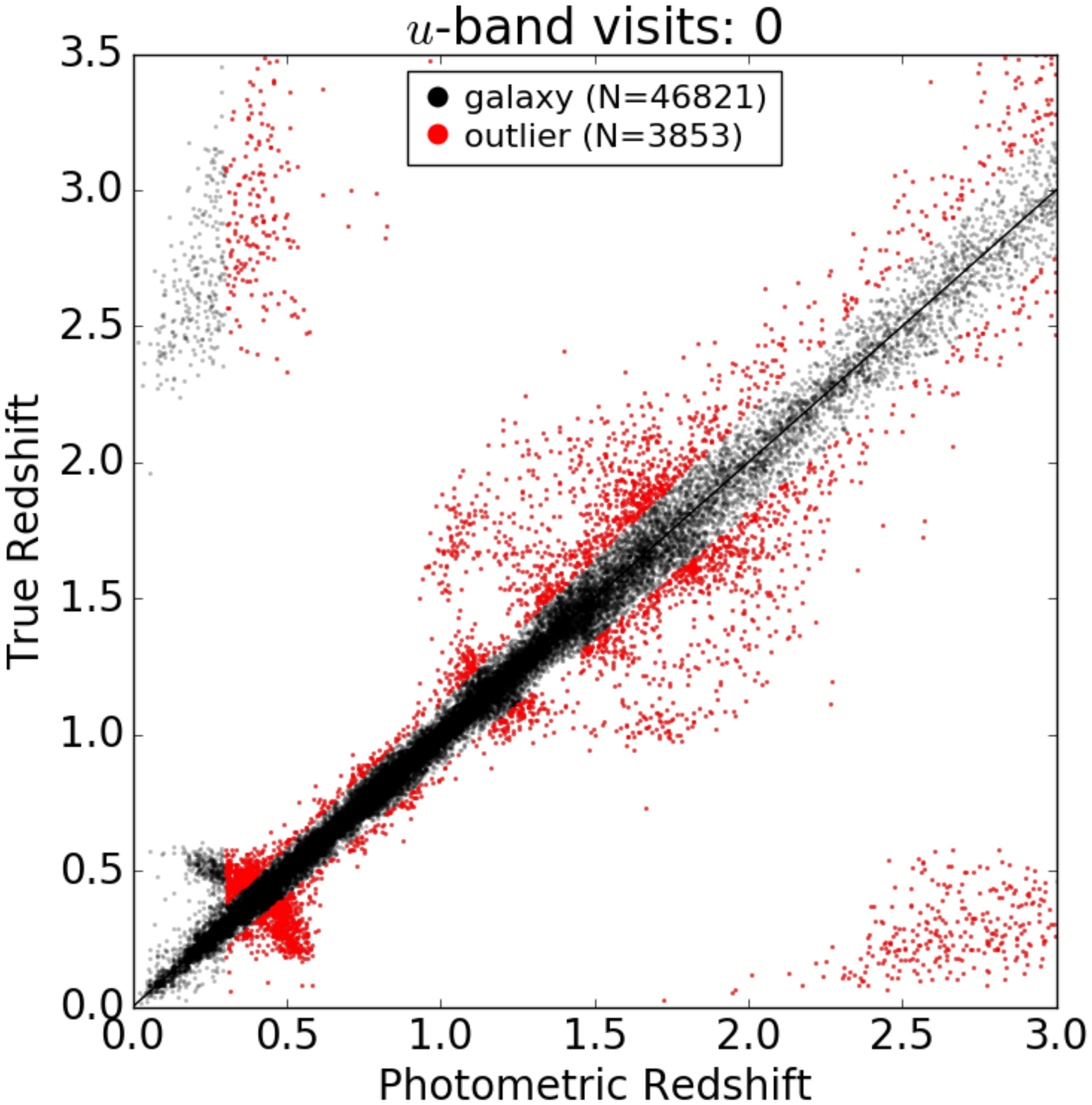}
\includegraphics[width=8cm,trim={1cm 4.5cm 1cm 4.5cm},clip]{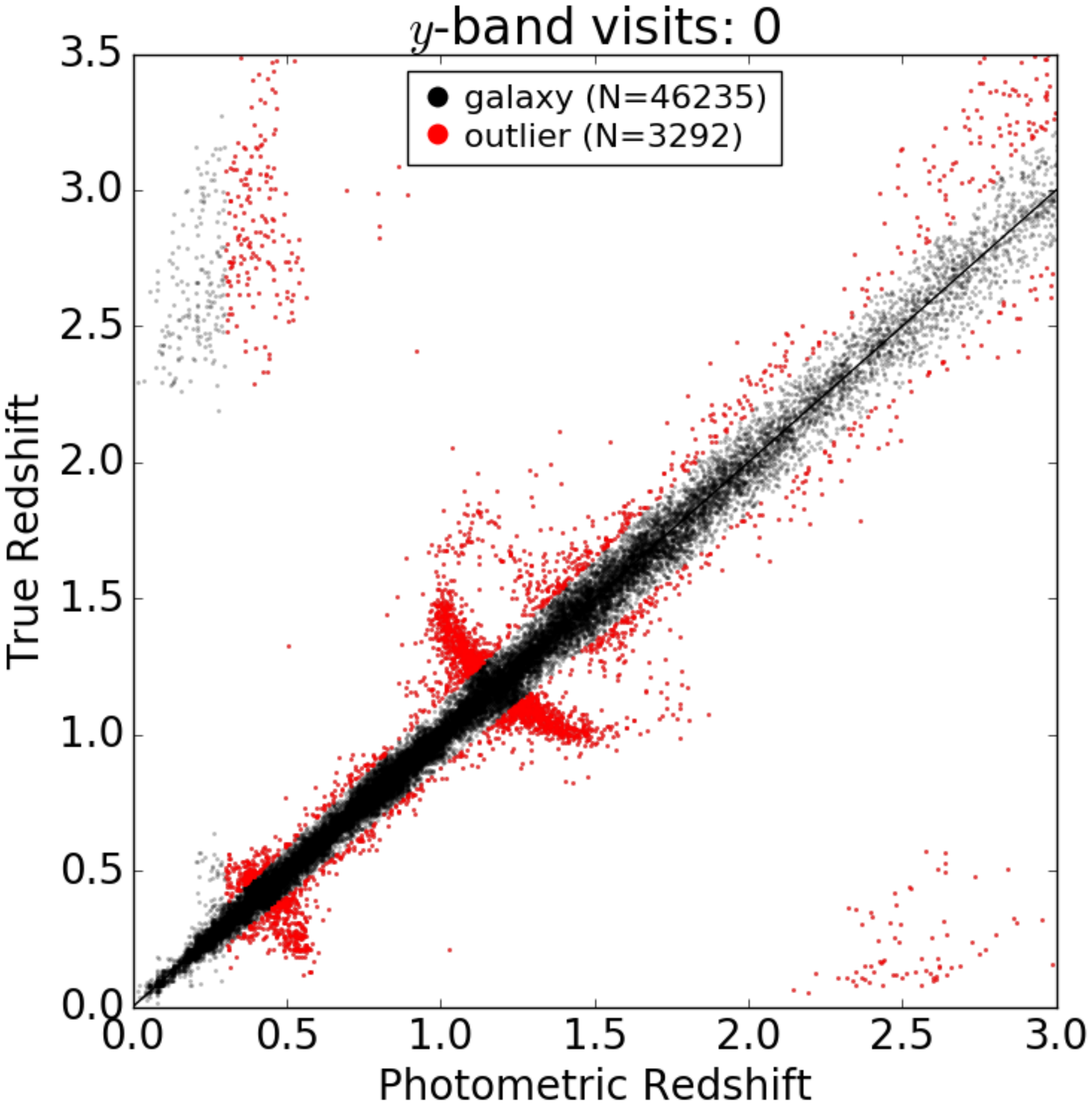}
\includegraphics[width=8cm,trim={1cm 4.5cm 1cm 4.5cm},clip]{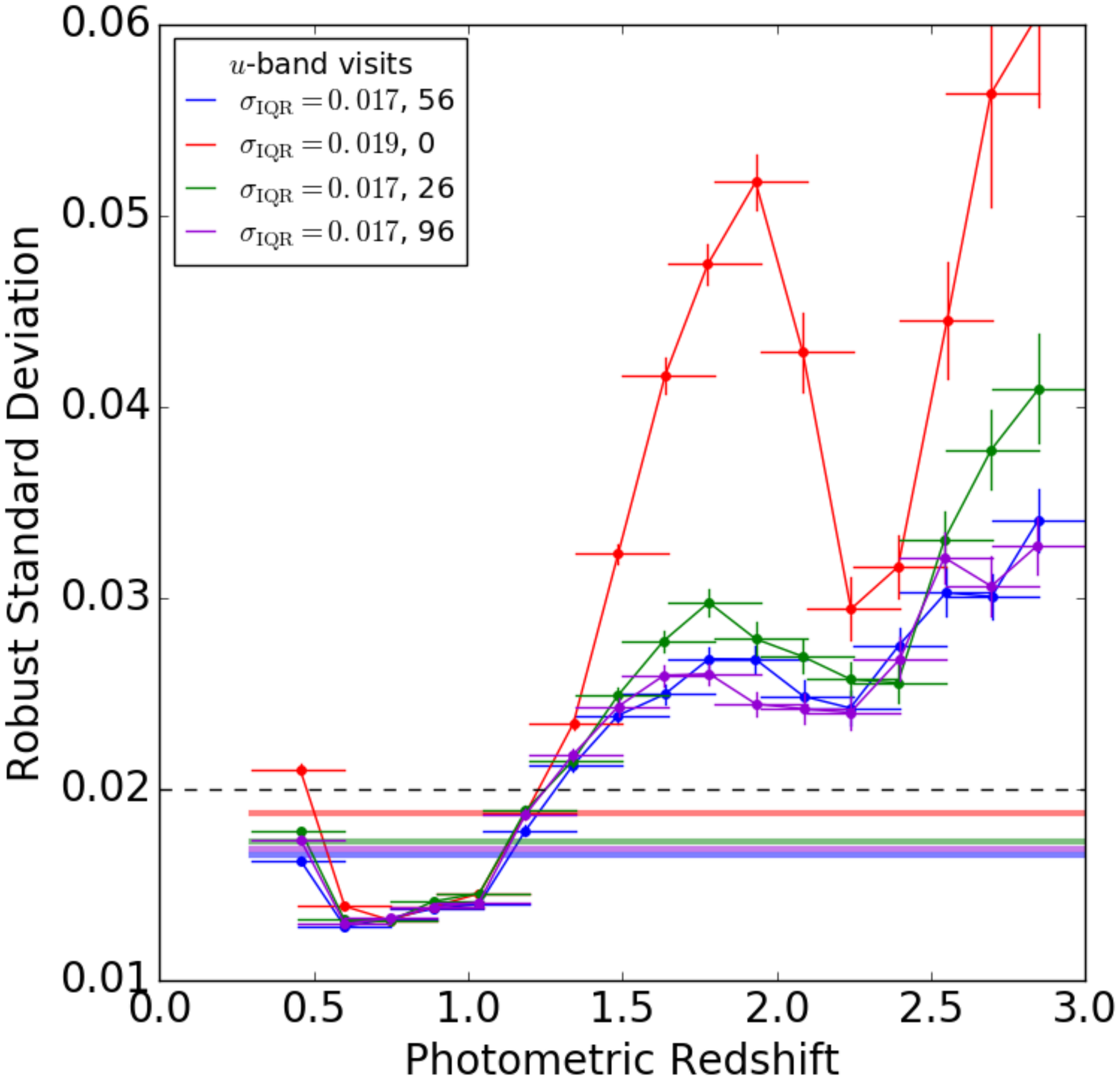}
\includegraphics[width=8cm,trim={1cm 4.5cm 1cm 4.5cm},clip]{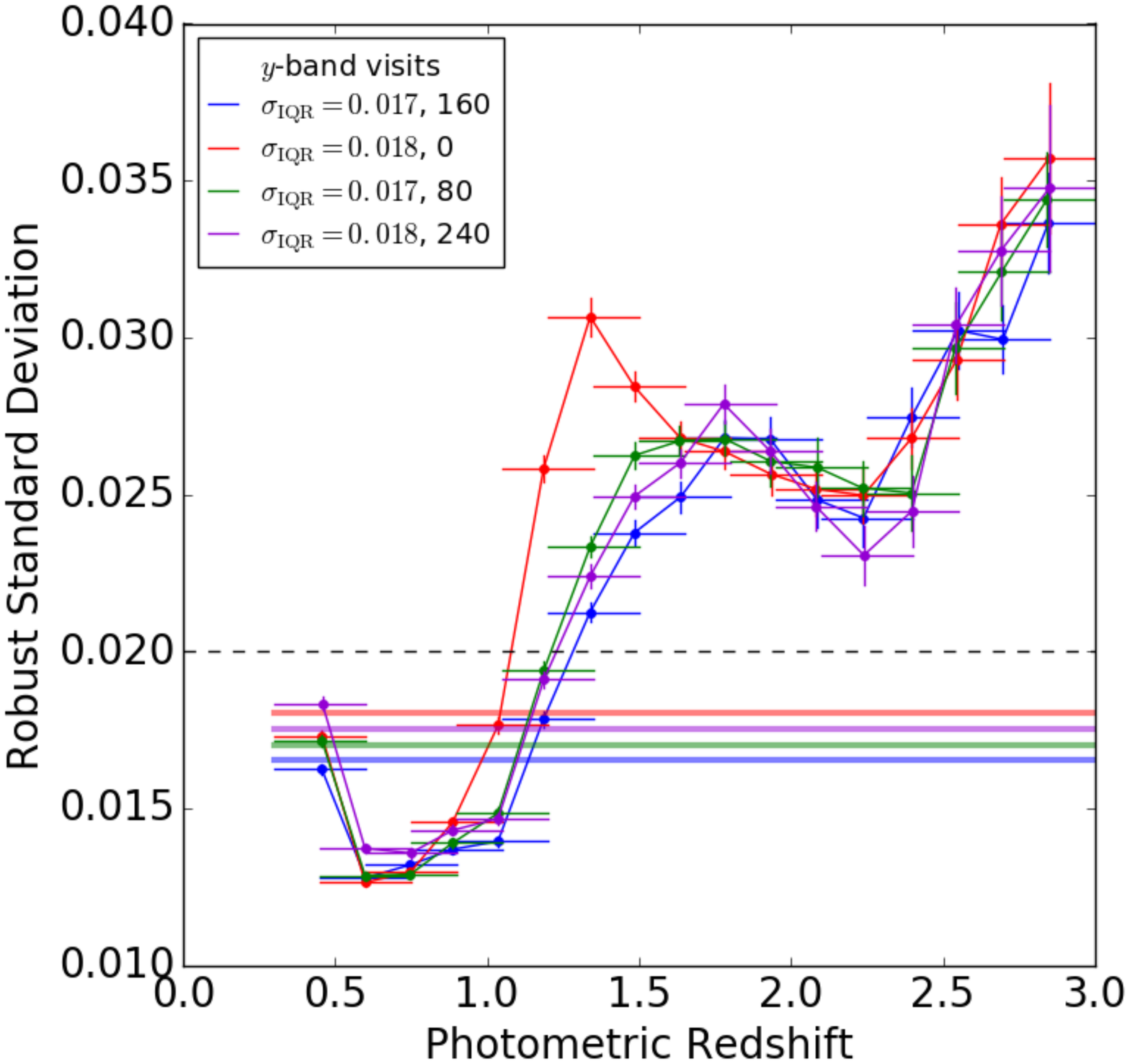}
\caption{{\it Top row:} True $vs.$ photometric redshifts when we simulate errors for the test set in an extreme case where $0$ visits are allotted to the $u$-band filter ({\it left}; the default is $56$ visits) or the $y$-band filter ({\it right}; the default is $160$ visits). A 10-year equivalent training set with the default number of visits per filter is used. {\it Bottom row:} The robust standard deviation in $\Delta z_{(1+z)}$ as we vary the number of visits allotted to the $u$- and $y$-band filters ({\it left} and {\it right}, respectively). The default number of visits is shown with a blue line for both filters. Horizontal colored lines mark the value of the statistic over the full redshift range $0.3 \leq z_{\rm phot} \leq 3.0$, which are also listed in the legend. Horizontal dashed lines mark the SRD's target value for each statistic. \label{fig:results_visit}}
\end{center}
\end{figure*}

\begin{figure*}
\begin{center}
\includegraphics[width=8cm,trim={1cm 4.5cm 1cm 4.5cm},clip]{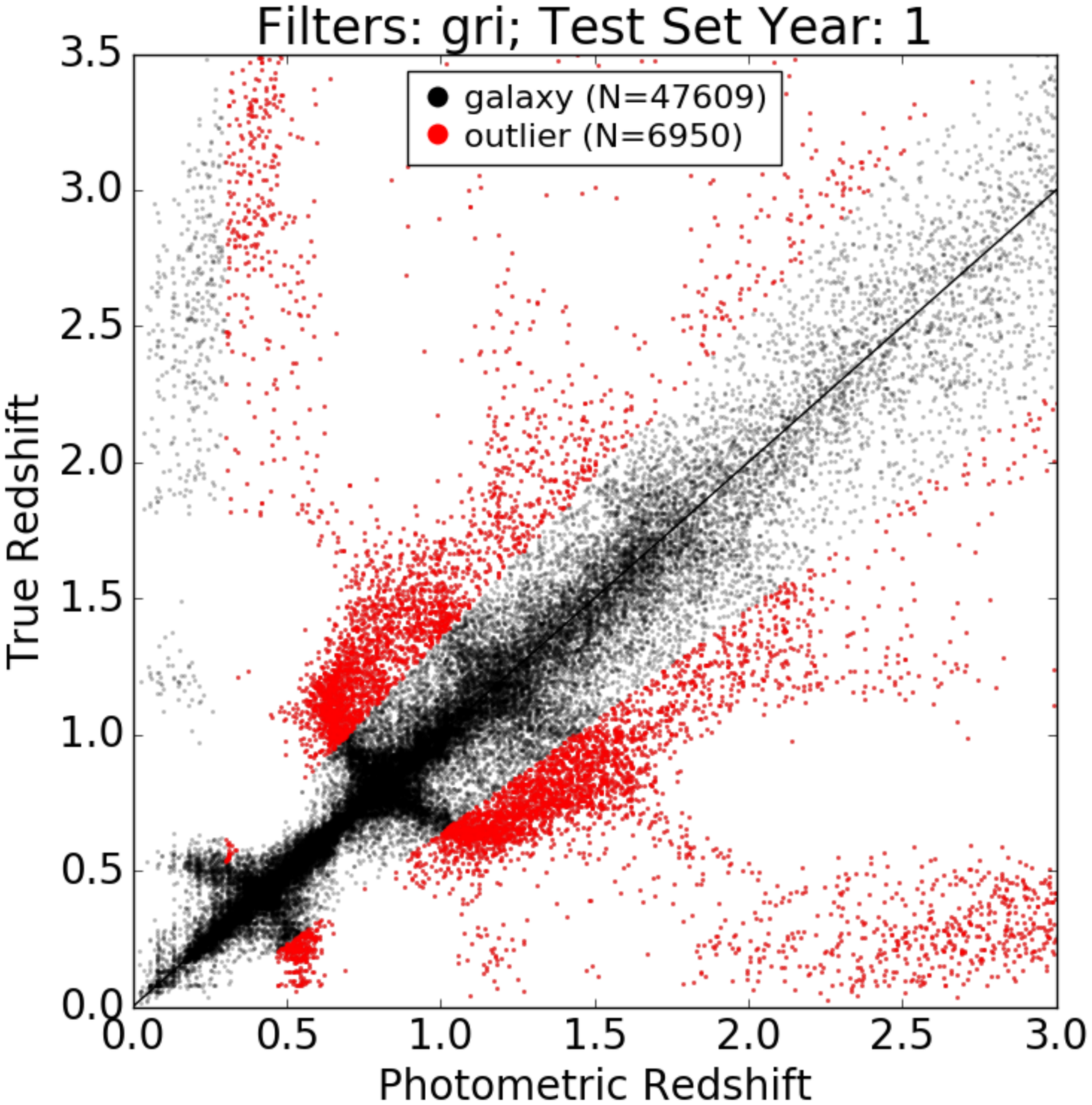}
\includegraphics[width=8cm,trim={1cm 4.5cm 1cm 4.5cm},clip]{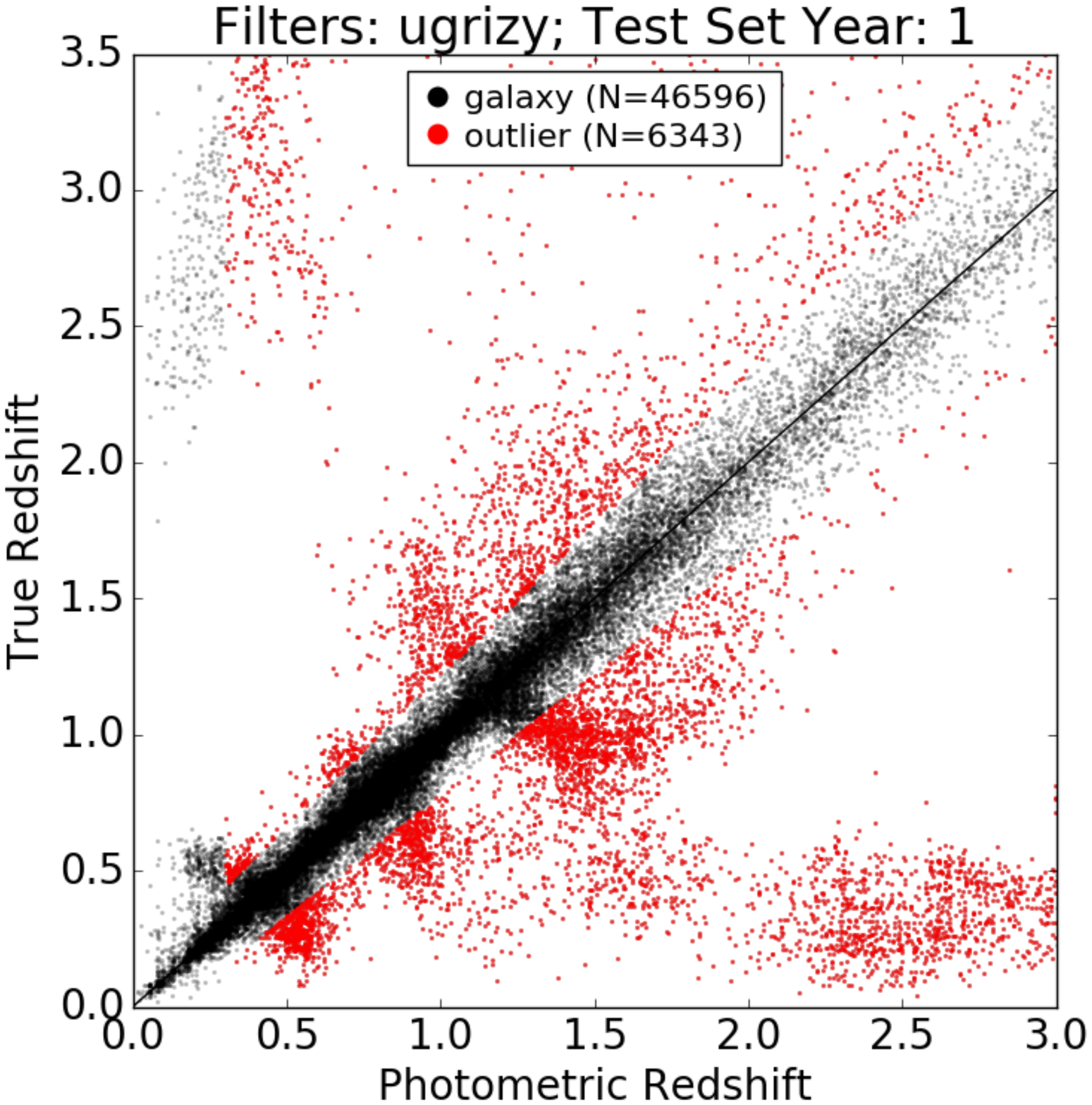}
\includegraphics[width=8cm,trim={1cm 4.5cm 1cm 4.5cm},clip]{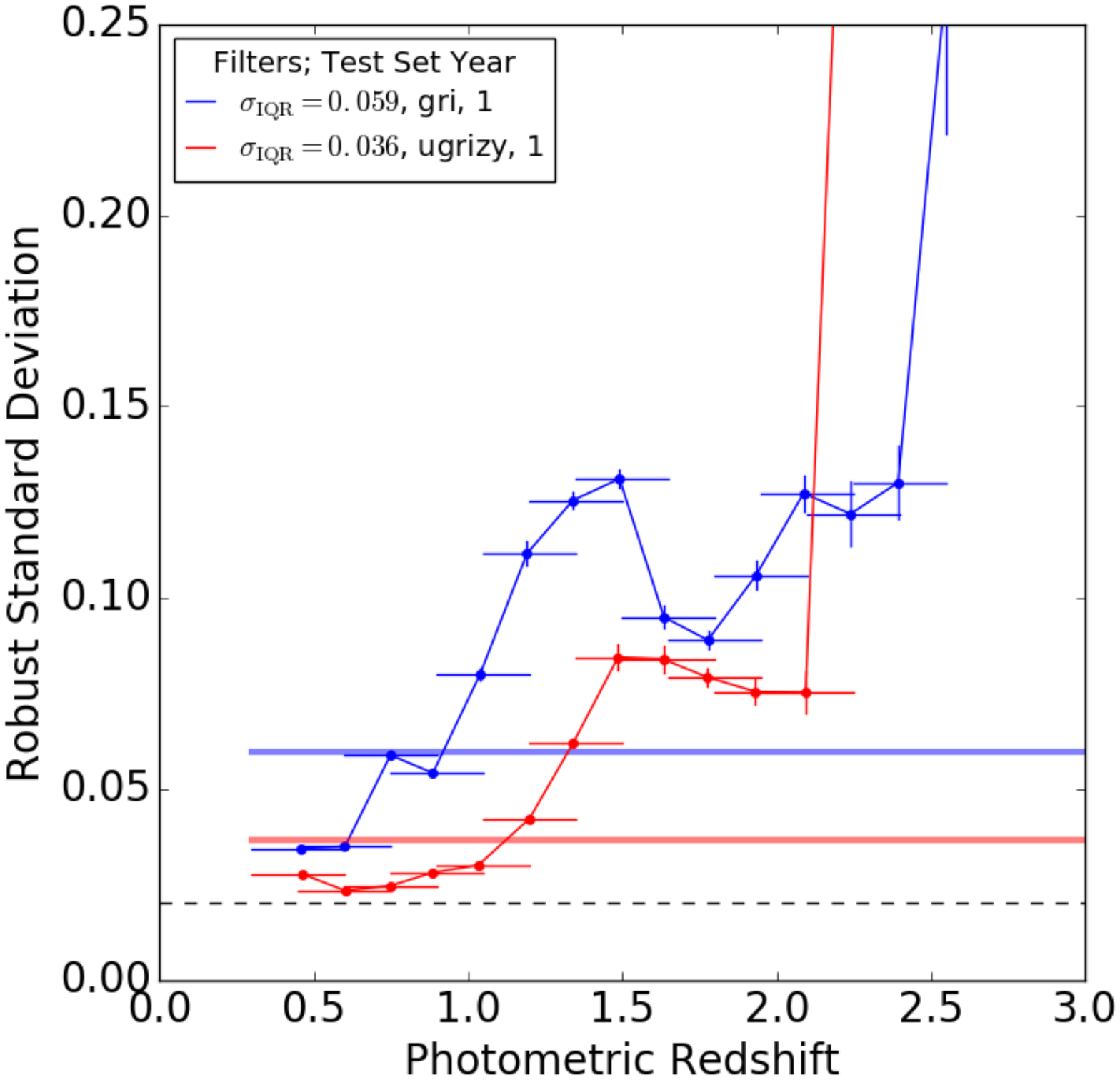}
\includegraphics[width=8cm,trim={1cm 4.5cm 1cm 4.5cm},clip]{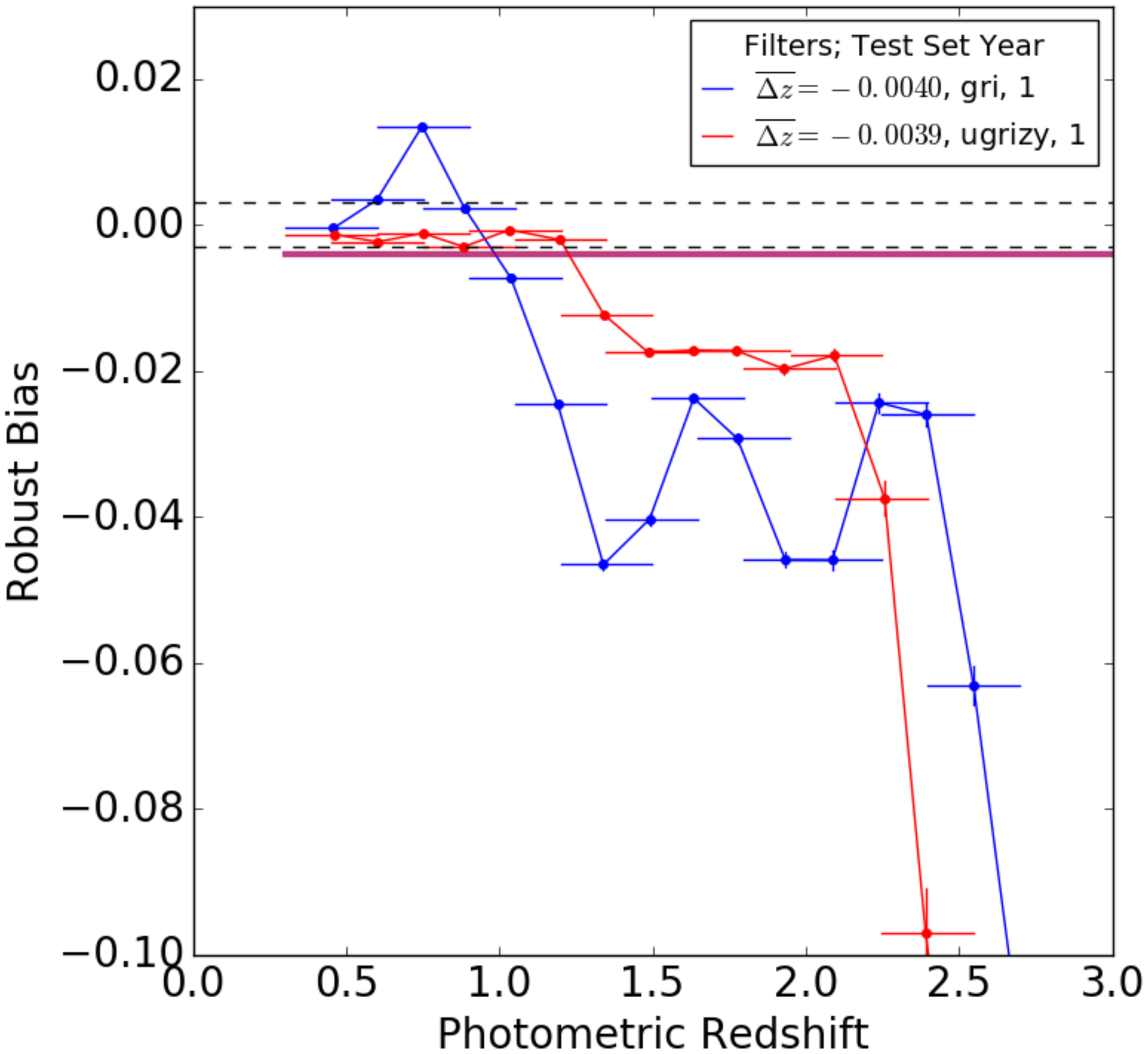}
\caption{{\it Top row:} True $vs.$ photometric redshifts at 1 year if we limit to filters $gri$ ({\it left}) or if we use all six filters $ugrizy$ ({\it right}). {\it Bottom row:} The robust standard deviation ({\it left}) and bias ({\it right}) in $\Delta z_{(1+z)}$ after 1 year of LSST if we limit to filters $gri$ only (blue), or use all six filters $ugrizy$ (red). Horizontal colored lines mark the value of the statistic over the full redshift range $0.3 \leq z_{\rm phot} \leq 3.0$, which are also listed in the legend. Horizontal dashed lines mark the SRD's target value for each statistic. \label{fig:results_year1gri}}
\end{center}
\end{figure*}

\begin{figure*}
\begin{center}
\includegraphics[width=8cm,trim={1cm 4.5cm 1cm 4.5cm},clip]{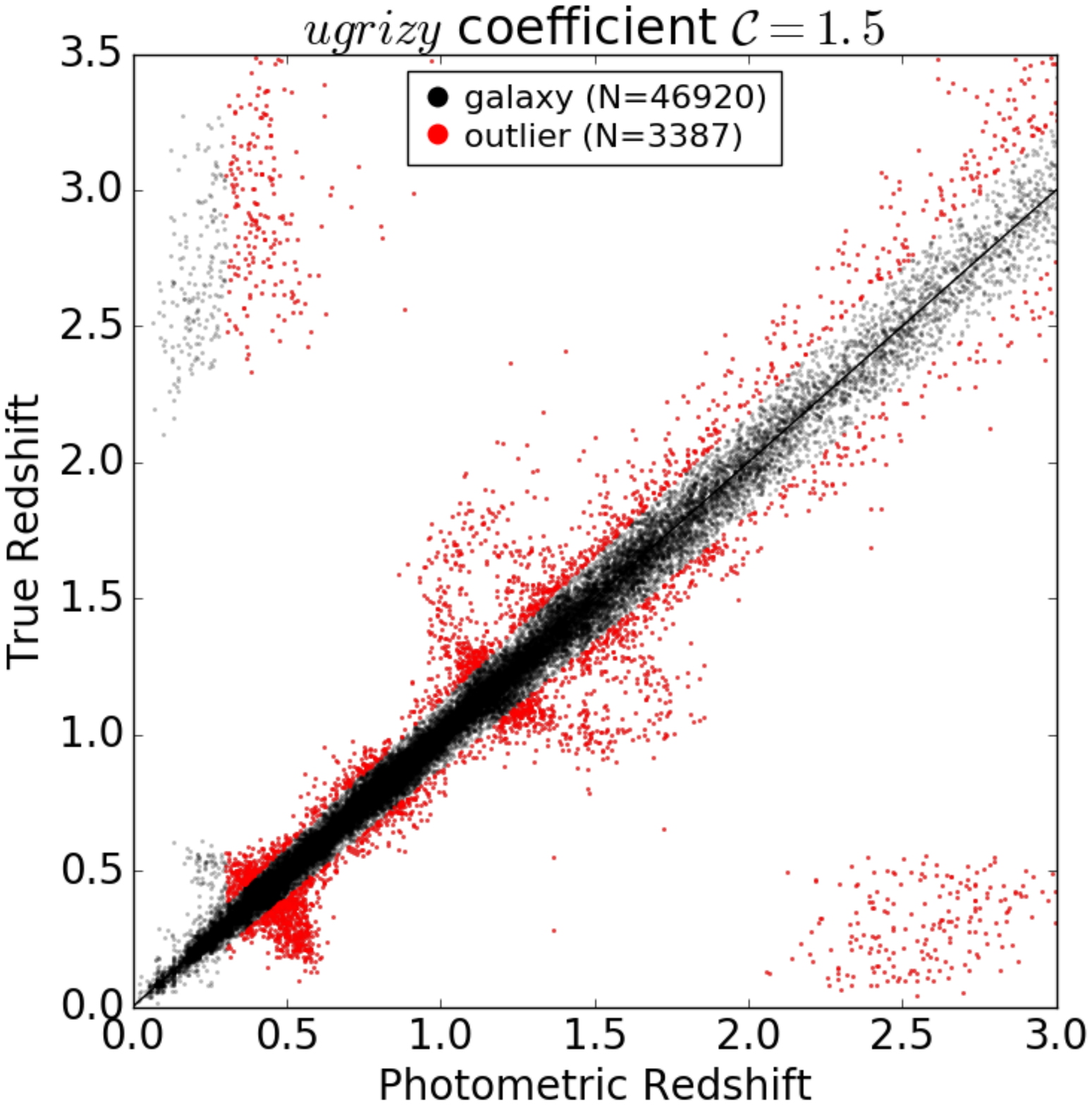}
\includegraphics[width=8cm,trim={1cm 4.5cm 1cm 4.5cm},clip]{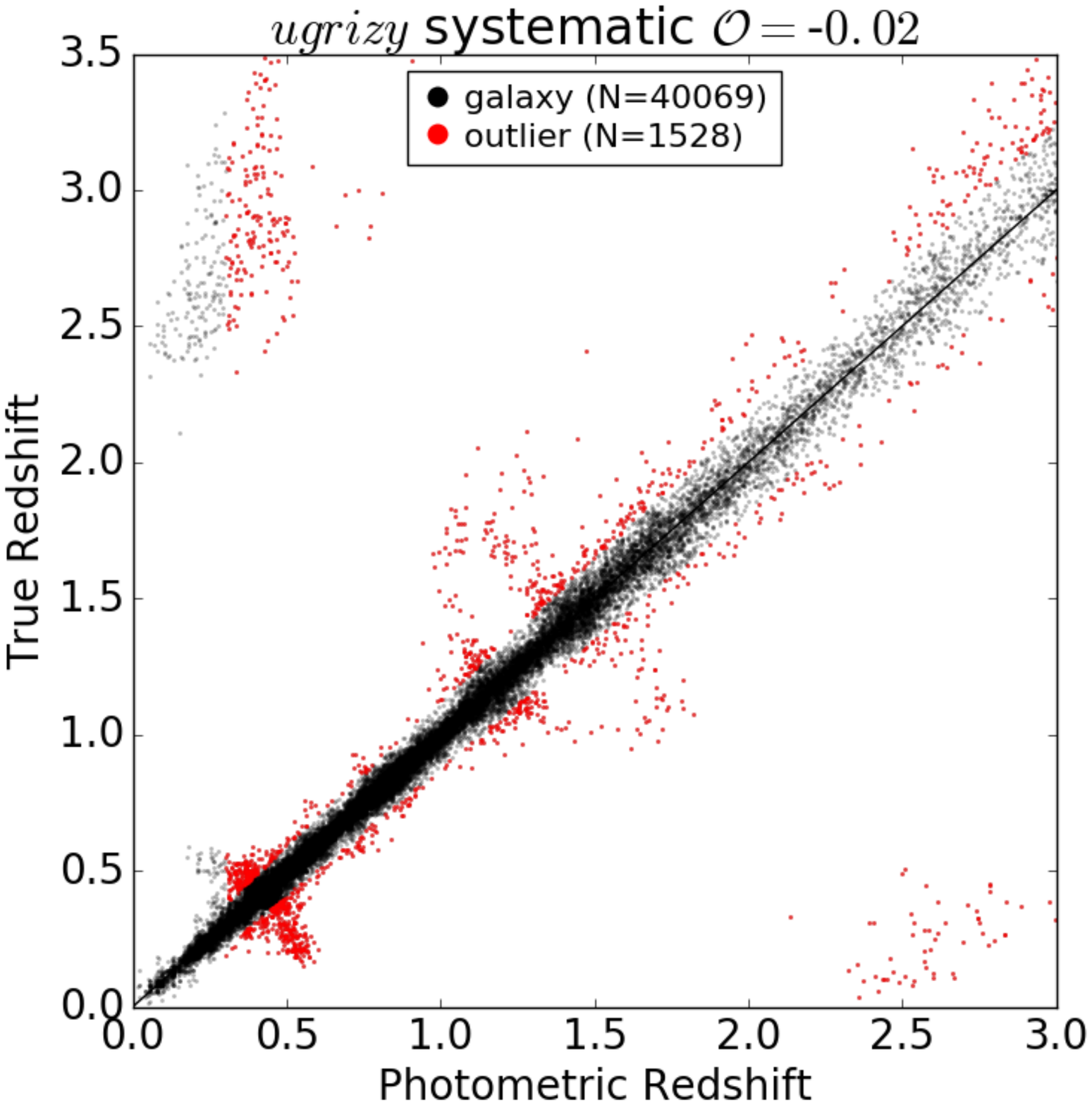}
\includegraphics[width=8cm,trim={1cm 4.5cm 1cm 4.5cm},clip]{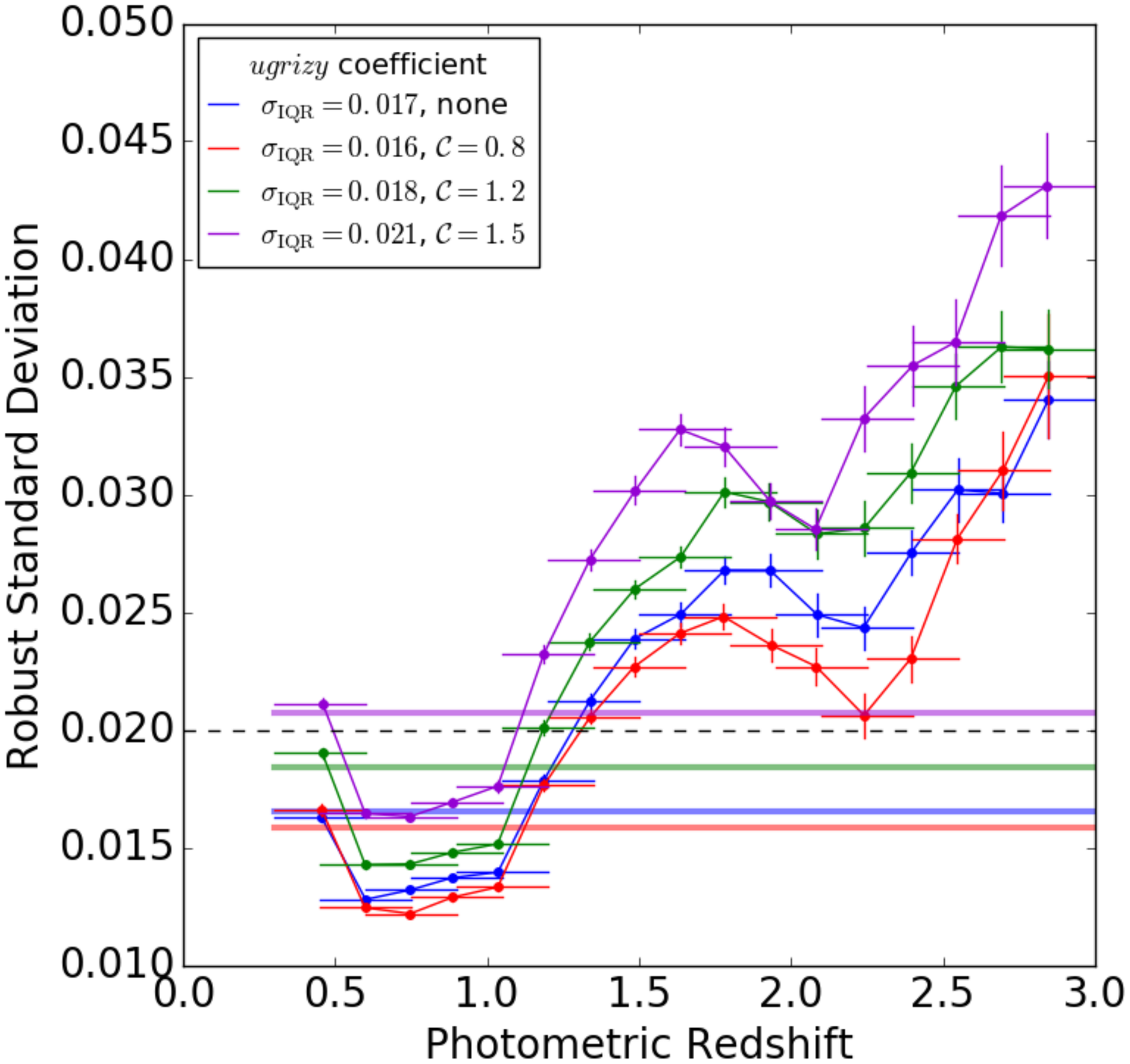}
\includegraphics[width=8cm,trim={1cm 4.5cm 1cm 4.5cm},clip]{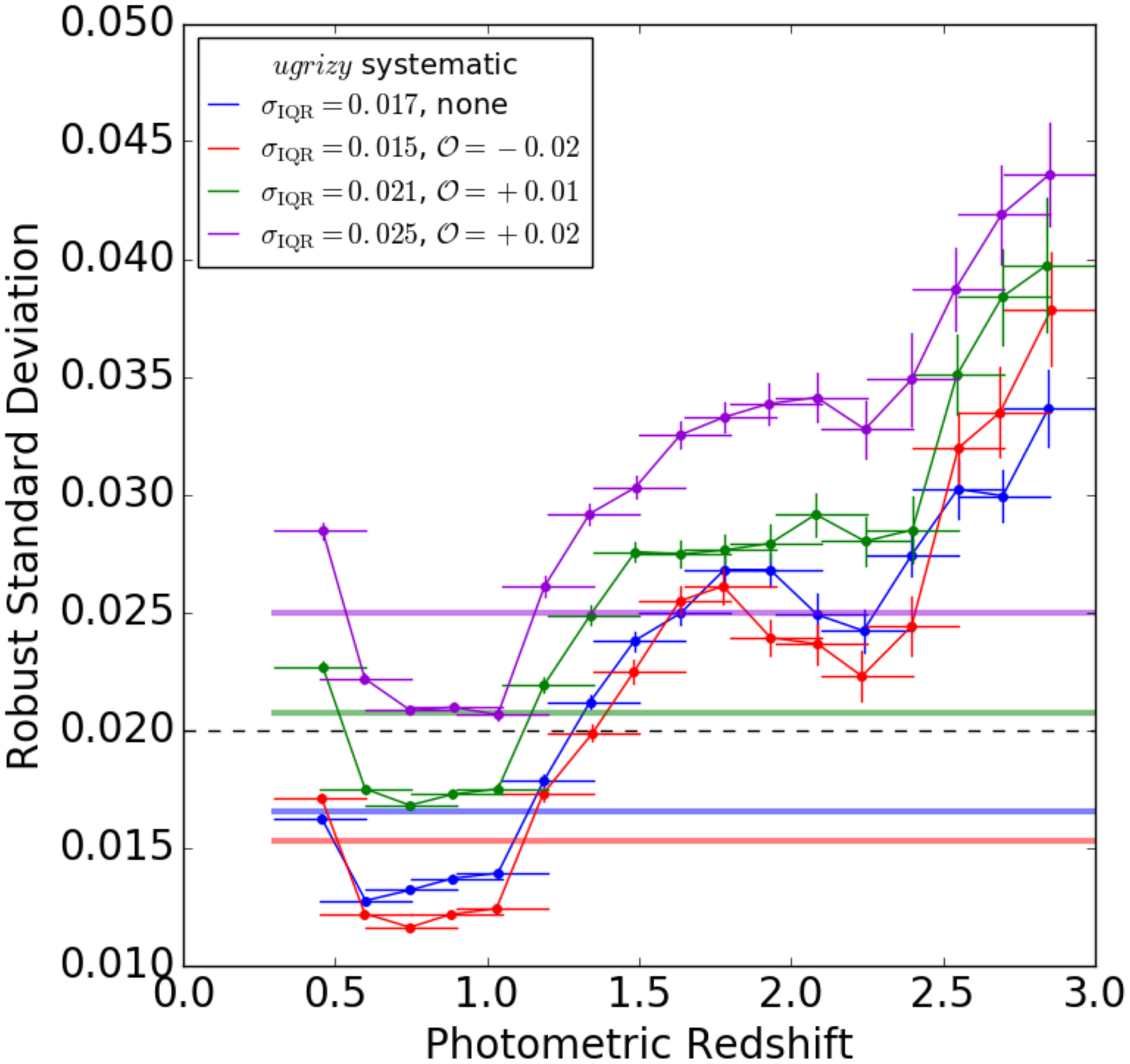}
\caption{{\it Top row:} True $vs.$ photometric redshifts if we multiply the magnitude errors in all filters by a coefficient, $\mathcal{C}$ ({\it left}) or if we add a systematic offset to the magnitude errors in all filters, $\mathcal{O}$ ({\it right}). In both scenarios a training set with no alterations to the photometry was used. {\it Bottom row:} The robust standard deviation in $\Delta z_{(1+z)}$ when we apply an increasingly large coefficient to the magnitude errors ({\it left}), and when we change the value of the systematic ({\it right}). Results for a standard test set with no changes to the photometry are represented by the blue line in each plot. Horizontal colored lines mark the value of the statistic over the full redshift range $0.3 \leq z_{\rm phot} \leq 3.0$, which are also listed in the legend. Horizontal dashed lines mark the SRD's target value for each statistic. \label{fig:results_error}}
\end{center}
\end{figure*}

The final 10-year photometric errors for the LSST are predicted based on a total number of visits equal to 56, 80, 184, 184, 160, and 160 in each of filters $ugrizy$, respectively. Here we examine how the photo-$z$ results are altered if we remove or add visits to the two end filters, $u$ and $y$. Visits that are removed or added to these filters are redistributed evenly to, or taken evenly from, the other five filters to maintain a constant total number of visits across all filters. As in Section~\ref{ssec:years}, for a given number of visits per filter we re-calculate the photometric errors and observed magnitudes in order to simulate a realistic catalog of test galaxy photometry, and then re-run our photo-$z$ estimator using a training set with a depth equivalent to a full 10-year survey with the default number of visits per filter (i.e., the same training set as used in the other experiments).

In the top two panels of Figure \ref{fig:results_visit} we show the $z_{\rm true}$ $vs.$ $z_{\rm phot}$ results in the extreme case where no visits are allotted to filter $u$ or $y$. We can see that the overall scatter around $z_{\rm true} = z_{\rm phot}$ is larger than with the default visit distribution shown in the bottom left panel of Figure \ref{fig:magair1}. We can furthermore see the filters' influence on the positions and population levels of the features perpendicular to $z_{\rm true} = z_{\rm phot}$ that contain outlier galaxies. For example, without the $y$-band filter outlier spurs appear at $1.0<z_{\rm phot}<1.5$, and without the $u$-band filter such spurs appear at $z_{\rm phot} \sim 0.5$.

In the bottom two panels of Figure \ref{fig:results_visit} we show the changes in the robust standard deviation of the photometric redshift results as we vary the number of visits allotted to the $u$- or $y$-band filters (with the default number represented by the blue line, and zero visits by the red line). We find that zero visits with the $u$-band filter results in a large increase in the robust standard deviation at low-$z_{\rm phot}$ and $z_{\rm phot} > 1.5$, but due to the redshift distribution of the galaxy catalog the minimum target value for $\sigma$ over $0.3 \leq z_{\rm phot} \leq 3.0$, as described in Section~\ref{ssec:stats}, is still met. Zero visits with the $y$-band filter also delivers poorer quality results, but the difference is less dire. Although we do not show the robust bias in Figure \ref{fig:results_visit} we report that the SRD's targets over $0.3 \leq z_{\rm phot} \leq 3.0$ are met for all the visit distributions considered, but that zero $u$-band visits results in a significant bias towards higher photometric redshifts in bins with $z_{\rm phot}>1.5$. We also report that the SRD's targets for the fraction of outliers over $0.3 \leq z_{\rm phot} \leq 3.0$ are met for all the visit distributions considered, but that zero $u$- or $y$-band visits results in a large increase in $f_{\rm out}$ at $z_{\rm phot}>1$.

\subsection{Photo-$z$ Results at 1 Year}\label{ssec:year1gri}

In this section we investigate whether concentrating the LSST's observing program on building up the signal-to-noise ratio a limited filter set could improve the photometric redshifts and possibly unlock science goals earlier than anticipated. For this experiment, we re-simulate the photometry of the test galaxy sample if all of the available survey time in the first year was dedicated to filters $g$, $r$, and $i$, which amounts to 21, 31, and 31 visits respectively. As in previous sections, we then re-run our photo-$z$ estimator using a training set with a depth equivalent to a full 10-year survey with the default number of visits per filter. We compare to the 1 year photometric redshifts with the full $ugrizy$ filter set.

In the top two panels of Figure \ref{fig:results_year1gri} we plot the $z_{\rm true}$ $vs.$ $z_{\rm phot}$ at 1 year when only $gri$ filters are used (left) and when the full set of $ugrizy$ are used (right). The impact of the loss of filters $u$, $z$, and $y$ can be seen in the increased scatter around $z_{\rm true} = z_{\rm phot}$ and the population of the spur-like perpendicular features, similar to the effects of dropping the visits for filters $u$ and $y$ as seen in Section~\ref{ssec:visits}. From these two plots it does not appear that building up the signal-to-noise ratio in $gri$ at the expense of color information is beneficial to the photometric redshift results at 1 year.

In the bottom two panels of Figure \ref{fig:results_year1gri} we compare the robust standard deviation and bias in bins of $z_{\rm phot}$ of these photometric redshift results. Comparing the blue and red lines, we find that the standard deviation across $0.3 \leq z_{\rm phot} \leq 3.0$ is lower by a factor of $0.6$ when the full filter set is used during year 1, and that the SRD's target value for standard deviation is not met in year 1 (as expected). We also find that the robust bias over $0.3 \leq z_{\rm phot} \leq 3.0$ is not as significantly affected by changes to the filter set, and that the SRD's target values are not met in either case.

In these experiments we have considered all test galaxies with observed apparent $i<25$ magnitudes, but early science goals might focus on a brighter sample. If we instead limit to $i<24$ and recalculate the statistical measures in the range $0.3 \leq z_{\rm phot} \leq 3.0$ when only $gri$ filters are used in the first year, we find that the results improve to $\sigma_{\rm IQR} = 0.036$ and $\overline{\Delta z} = -0.0016$. However the results for $i<24$ with all six filters are still better, at $\sigma_{\rm IQR} = 0.019$ and $\overline{\Delta z} = -0.0013$. Furthermore, the fraction of outliers is $<10\%$ after 1 year if all six filters are used. We therefore caution that any future proposal to alter the relative fraction of visits per filter must consider the magnitude limits of the desired galaxy catalogs as well as the potential impacts to other early science goals, which may depend on brighter or fainter galaxies.

\subsection{The Direct Impact of Photometric Uncertainty}\label{ssec:errors}

In the previous experiments we investigated how the distribution of LSST visits in time and across the available filters affects the photometric redshifts, due to the relationship between the number of visits per filter and the magnitude errors. In Section~\ref{sec:method} we described how we developed this photo-$z$ method because it has a direct relation between the photometric uncertainties and photo-$z$ results without intervening steps, such as the selection of a spectral template. In this experiment we exploit this direct aspect of our method and apply artificial deterioration to the photometric errors of the test galaxies and evaluate the consequences for the photometric redshift results.  Any change to the value of the photometric errors influences the photo-$z$ in two ways. First, as described in Section~\ref{sec:cat}, the observed magnitudes are calculated from the true catalog magnitudes by adding a normal random scatter based on the magnitude error. Any increase in the error will pull the observed magnitudes further from the true magnitudes and change the composition of the color-matched subset of training galaxies, thereby introducing more scatter and perhaps biases into the photometric redshifts. Second, as described in Section~\ref{ssec:photoz}, the photometric error of the test galaxy is used in the denominator of the Mahalanobis distance in Equation \ref{eq:DM}. Any increase in the error will increase $D_M$ and include more training galaxies in the color-matched subset. When the option to choose randomly from this subset is used to estimate the photo-$z$, an increase in the error would introduce more scatter and perhaps more bias into the photometric redshifts. When the option to choose the nearest-neighbor from this subset is used, an increase in the photometric error would alter the error that we estimate for the photo-$z$, $\delta z_{\rm phot}$.

As described in Section 3.2.1 of \cite{2008arXiv0805.2366I}, the total magnitude error is the sum of a systematic and a random component. To test changes to either of these factors individually, we perform two experiments to simulate deteriorated photometry for the test set: (1) we multiply the magnitude errors in all filters by a coefficient $\mathcal{C}=0.08,1.2,1.5$ or (2) we add a systematic offset to the errors in all filters of $\mathcal{O}=-0.02,0.01,0.02$ mag. These values are larger than any expected deviations from the predicted magnitude errors for LSST data, but we have exaggerated for the sake of demonstration. As in previous experiments, we keep the photometry for the training set galaxies at the 10-year level and only alter the test set galaxy photometry.

In the top two panels of Figure \ref{fig:results_error} we plot the $z_{\rm true}$ $vs.$ $z_{\rm phot}$ when we apply the largest value of the coefficient, $\mathcal{C}=1.5$ and the smallest value of the systematic, $\mathcal{O}=-0.02$ (which are quite similar looking to the results for small values of the coefficient or large values of the systematic). The dominant effect of increasing the photometric errors is an increase in the scatter around $z_{\rm true} = z_{\rm phot}$ and not, for example, the population of perpendicular features as seen when the number of visits in certain filters are limited (Section~\ref{ssec:visits}). In the bottom two panels of Figure \ref{fig:results_error} we compare the robust standard deviation in redshift bins when we apply a range of values for $\mathcal{C}$ and $\mathcal{O}$. When the photometric errors are improved we see a corresponding improvement in the value of $\sigma_{\rm IQR}$ over the redshift range $0.3 \leq z_{\rm phot} \leq 3.0$, and vice versa. These two bottom panels clearly show that $\mathcal{C}\gtrsim1.5$ or $\mathcal{O}\gtrsim0.01$ will lead to a failure to reach the SRD's target value for standard deviation.

Although we do not show it, we find that the robust bias over $0.3 \leq z_{\rm phot} \leq 3.0$ is relatively unaffected by these changes to the photometric errors, but that in high redshift bins $z_{\rm phot}>2$, increasing the photometric errors induces a negative bias and the photo-$z$ overestimates the true redshift by $\overline{\Delta z}\lesssim-0.01$. Furthermore, we find that when the errors are improved for the test galaxies relative to the training set ($\mathcal{C}=0.8$ or $\mathcal{O}=-0.02$), the fraction of galaxies which fail to be assigned a photo-$z$ increases from $5$--$10$\% up to $15$--$30$\%. As a final note, we reiterate that our simulated systematics are applied to the magnitude {\it errors} and not the magnitudes themselves; simulating the effect on photo-$z$ results from a systematic offset to the magnitudes is something we leave for other work, except for the particular case where that magnitude offset is the result of the airmass effect, which we discuss in Section~\ref{sec:air}.

\section{The Airmass Effect}\label{sec:air}

\begin{figure*}
\begin{center}
\includegraphics[width=13cm,trim={0cm 9.75cm 0cm 9.75cm},clip]{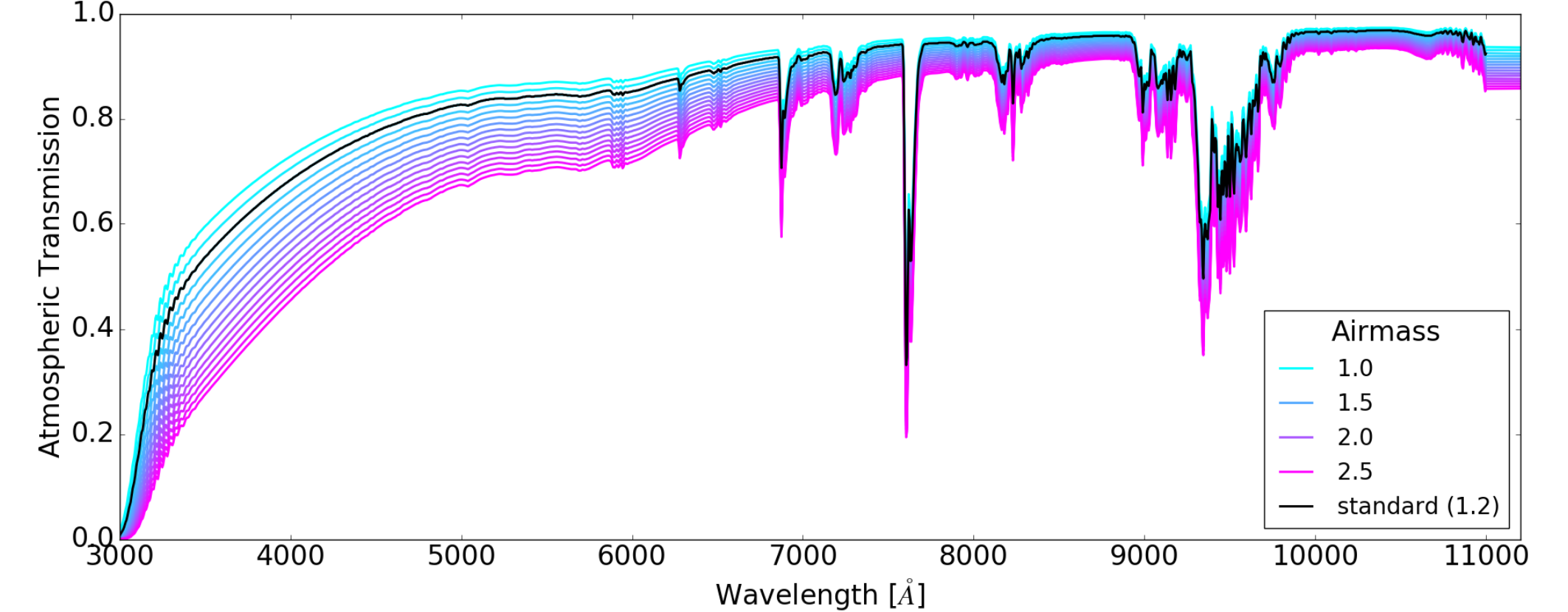}
\includegraphics[width=13cm,trim={0cm 9.75cm 0cm 9.75cm},clip]{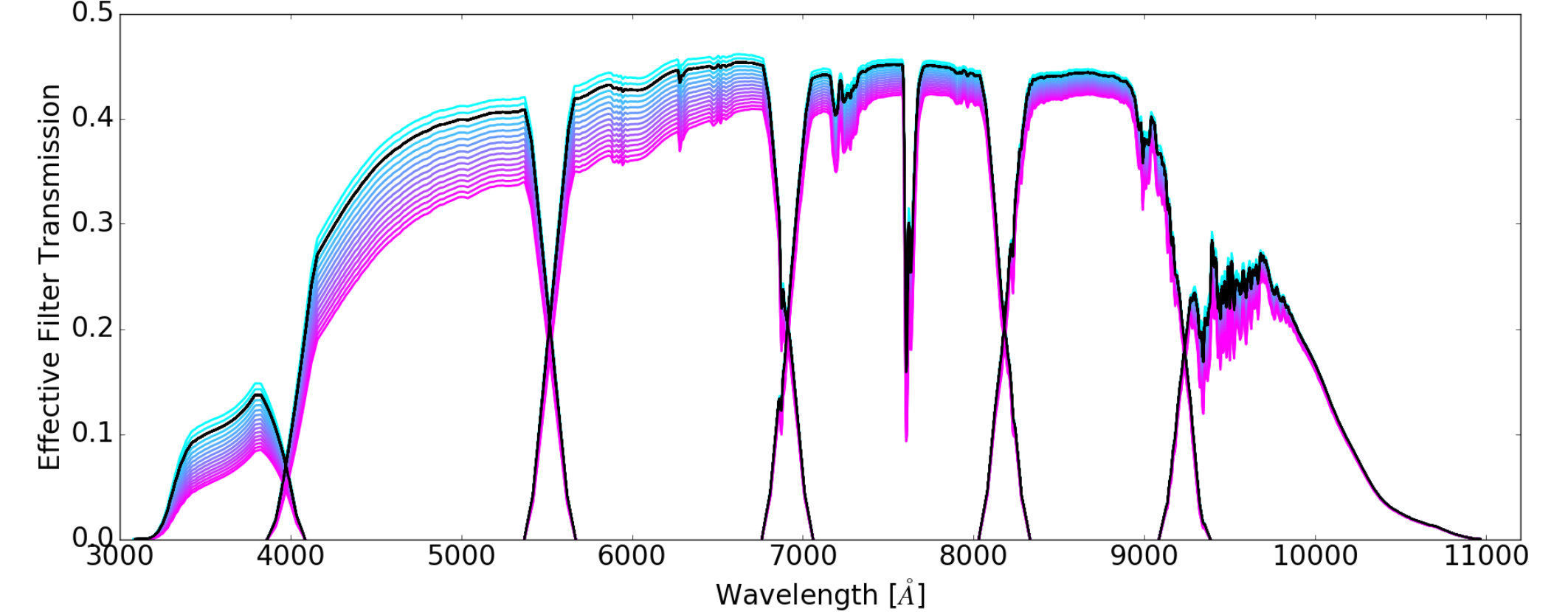}
\hspace*{-0.4cm}\includegraphics[width=13.4cm,trim={0cm 9.75cm 0cm 9.75cm},clip]{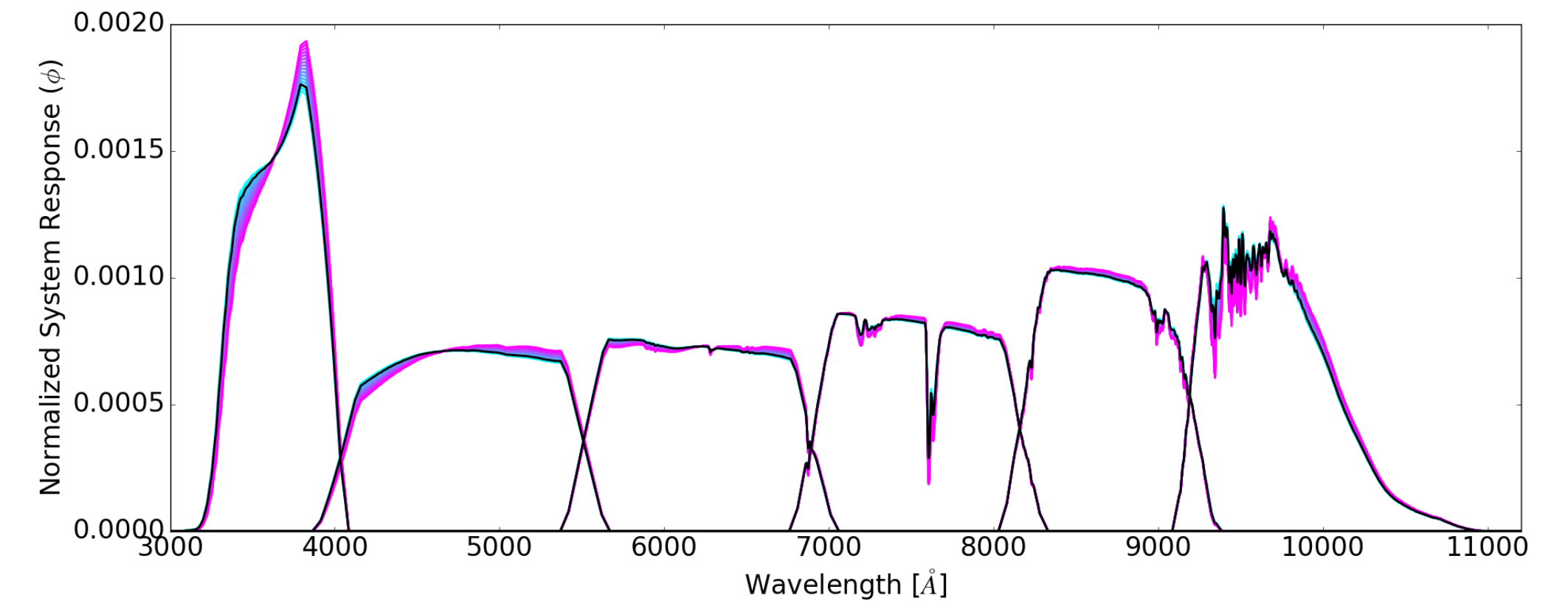}
\includegraphics[width=13cm,trim={0.5cm 9.2cm 0cm 9cm},clip]{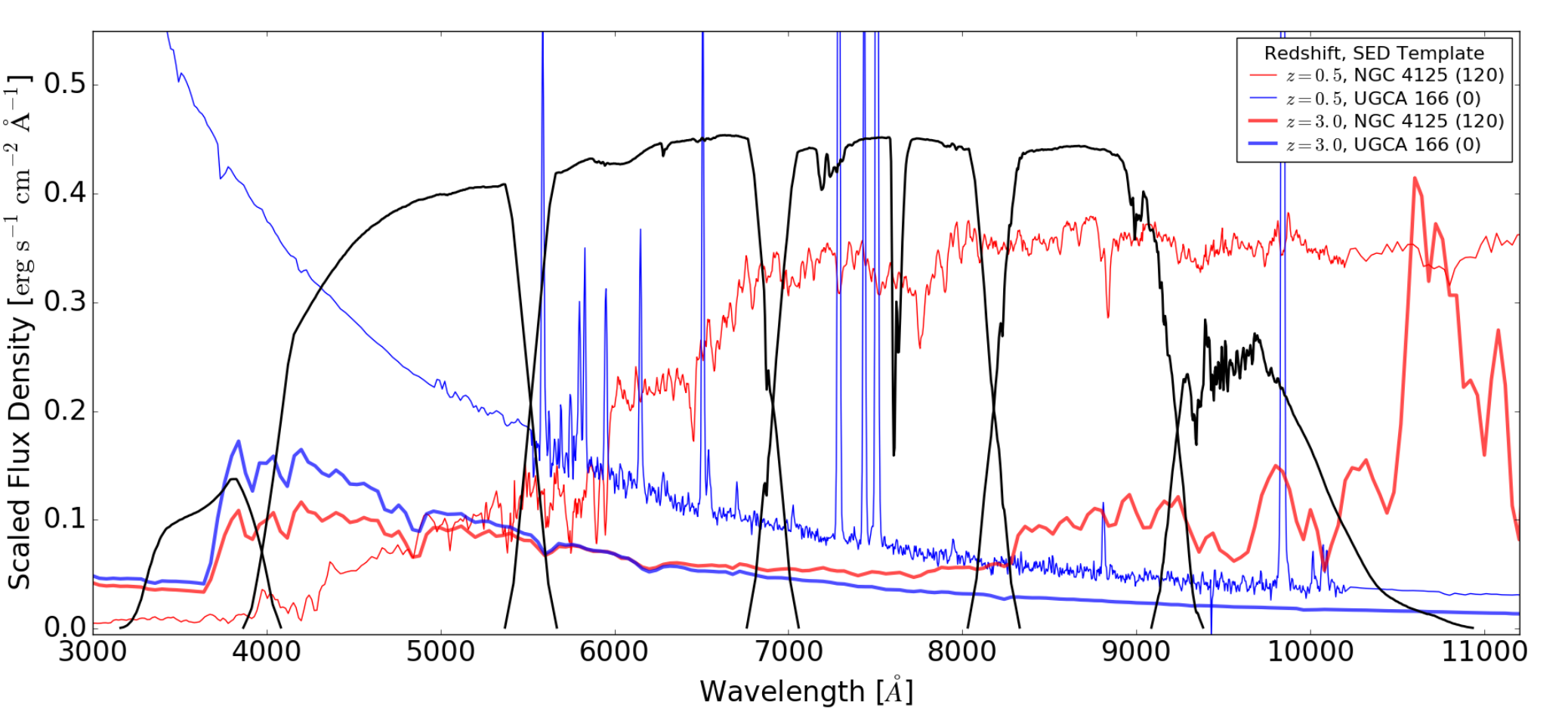}
\caption{{\it Top:} The atmospheric transmission curves with aerosols for different airmass (blue to pink is low to high airmass; note that airmass $X=1.2$ is used as the default standard for the full system throughput, and is represented in each panel as a black line). {\it Second:} The effective filter transmission function for the six LSST filters, $ugrizy$ (left to right), including the total system throughputs and the atmospheric transmissions for different airmass, with the same color convention as the top panel. {\it Third:} The normalized system response for different airmass, with the same color convention as the top panel. {\it Bottom:} Examples of spectra for one late-type (blue) and one early-type (red) galaxy SED templates redshifted to $z=0.5$ (thin) and $z=3$ (thick), with the nominal filters over-plotted in black. \label{fig:filters}}
\end{center}
\end{figure*}

The wide-area survey style of LSST, which must achieve $>800$ visits of each field over ten years, necessitates that observations be obtained over a range of airmass. This affords us the possibility of the unique opportunity of using an SED- and $z$-dependent atmospheric affect on the photometry as a redshift indicator. In this section we test whether we can extract SED information for galaxies that is independent of photometric color and use it to improve the photometric redshifts for LSST -- and furthermore, whether this technique is powerful enough that it should influence the LSST survey strategy (i.e., the airmass distributions of LSST visits to a field in each filter).

The Earth's atmosphere has two main effects on incident light from a celestial object: atmospheric refraction causes astrometric distortion, and atmospheric absorption causes extinction and reddening. The amplitude of these effects depends on the amount of atmosphere traversed by the photons, commonly represented by the zenith angle, $\Phi$, or airmass $X = \sec{\Phi}$, and on the spectral energy distribution of the object. In this way, information about the object's SED is encoded in trends between airmass and its observed astrometry and photometry. Here, we focus on the photometric effects of airmass and leave astrometric effects for future work, but note that such astrometric techniques would have the advantage of being minimally affected by large-scale spatial variations in the photo-$z$ error due to non-uniform coverage and foreground dust extinction, which we have not included in the present analysis. (For a demonstration of incorporating astrometric effects into photometric classification schemes, see e.g., \citealt{2015ApJ...811...95P}). 

Aside from airmass, atmospheric temperature and water content also a change to the filters' normalized effective transmission curves and their cut-off edges, and this can be a larger photometric effect than airmass. The LSST auxiliary telescope will constantly measure the water vapor column in the atmosphere above LSST and photometry will be corrected for this effect, but it could still be used to glean additional information about an object's SED. We also note that small alterations to the effective filter transmission introduced by hardware, e.g., between the center and edge of the focal plane, or from the two types of CCD sensors in the LSST camera, might also be used as an SED indicator. Since these effects are not as directly associated with the LSST survey strategy, we leave them for future analyses. 

\subsection{The Magnitude-Airmass Slope, $\mathcal{M}$}\label{ssec:airM}

In Figure \ref{fig:filters} we demonstrate how airmass can influence changes to the effective filter transmission functions that alter the observed magnitude of an object in a manner that is dependent on the object's SED. In the top panel we show the atmospheric transmission as a function of wavelength, for airmass $1<X<2.5$ (blue to pink). Note that airmass $X=1.2$ is used as the standard for the full system throughput, and is represented as a black line in each panel. In the second panel we show the effective filter transmission functions for the six LSST filters, $ugrizy$. These curves include the full system throughput -- the filters, all reflective and refractive surfaces, and the CCD's quantum efficiency -- as well as the nominal atmosphere for a variety of airmass values. In these top two panels the extinction and reddening from atmospheric absorption are seen as an overall decrease in transmission with airmass, with a the relatively larger decrease in bluer filters. These effects are relatively well understood, and the corrections for atmospheric extinction and reddening that are typically applied to the photometry are quite robust. 

The effect that we are interested in for this work is the subtle warping of the normalized system response, $\int\phi\ d\lambda = 1$, for different airmass values, as shown in the third panel of Figure \ref{fig:filters}. The effect exists for all filters, but is more obvious in the $u$-band, where the low- and high-airmass transmission curves exhibit the strongest differences (i.e., the blue and pink lines are the most separated for the left-most bandpass in the third panel of Figure \ref{fig:filters}). Even after the apparent magnitude in a given filter has been corrected for the appropriate amount of extinction and reddening based on the airmass at the time of observation, there will be a residual relationship between magnitude and airmass, and that relationship depends on the object's SED through that filter's bandpass. For example, a galaxy with a SED that is decreasing in flux through the $u$-band (i.e., a negative slope in flux $vs.$ wavelength) will experience slightly more total extinction over that bandpass at high airmass than an a galaxy with a SED that is increasing in flux through the $u$-band. Therefore, measuring this residual slope between magnitude and airmass in the LSST photometry could provide an independent indication of the SED of an object without the need to obtain spectra. In the bottom panel of Figure \ref{fig:filters}, we illustrate how the spectra of early- and late-types of galaxies at low- and high-redshifts have different shapes through the LSST passbands, especially in filters $u$ and $g$. 

\begin{figure*}
\begin{center}
\includegraphics[width=15cm,trim={0cm 5cm 0cm 5cm},clip]{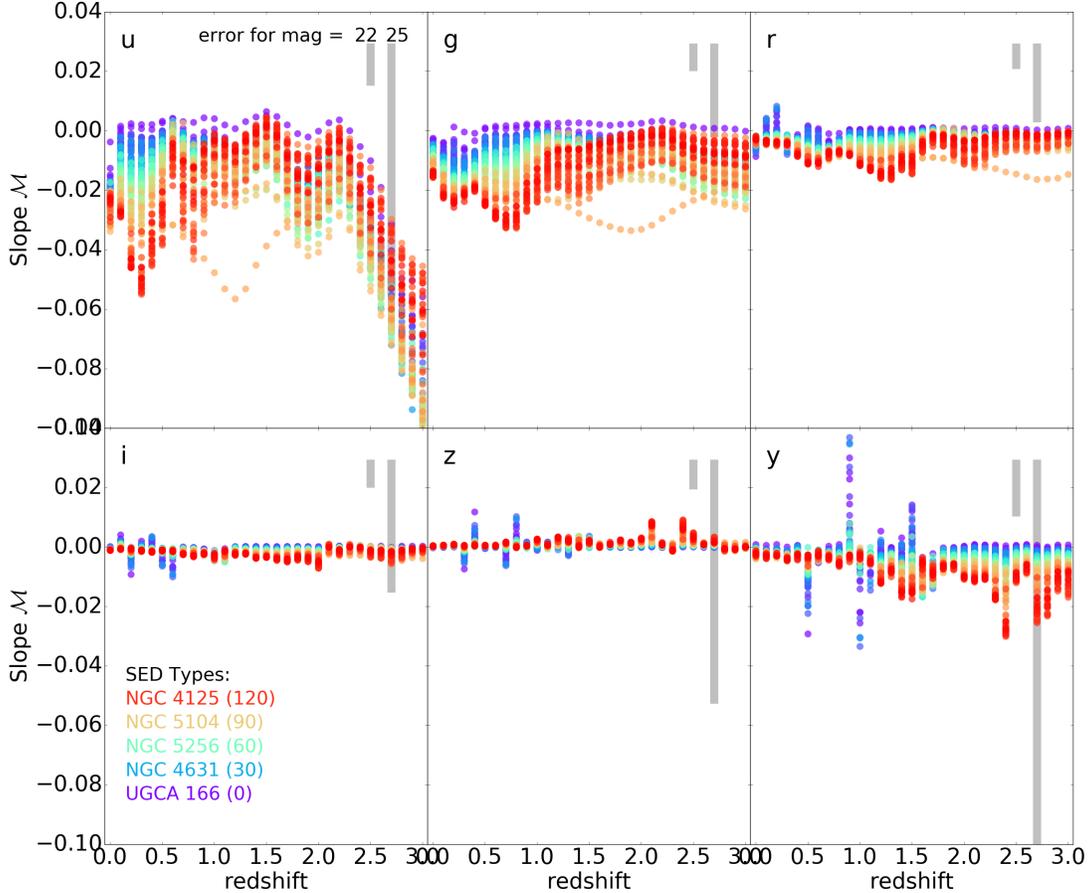}
\caption{The slope of the magnitude-airmass relation, $\mathcal{M}$, as a function of redshift for the six LSST filters $ugrizy$ (top left to bottom right panels) for 128 galaxy SED templates from early-type (red, index 120 in the SED catalog) to late-type (purple, index 0 in the SED catalog). Grey bars in the background show the typical magnitude error for galaxies of magnitude 22 and 25 in the filter of each panel. \label{fig:slope_vs_z}}
\end{center}
\end{figure*}

To simulate the potential magnitude-airmass slope values for a range of SED types, redshifts, and filters we start with the spectral templates from \cite{2014ApJS..212...18B}, two of which are illustrated in the bottom panel of Figure \ref{fig:filters}. For a given SED type and redshift, we calculate the magnitudes for a range of airmass values by convolving the SED with the effective filter transmission function for each airmass, and then determine the slope of the relationship between magnitude and airmass, $\mathcal{M}$. We do this for all SED types, for a grid of redshifts, for each LSST filter. In Figure \ref{fig:slope_vs_z}, we plot the value of the magnitude-airmass slope $\mathcal{M}$ as a function of redshift for each of the six LSST filters (top left to bottom right panels) for all of the 128 SED templates from \cite{2014ApJS..212...18B} (represented by the color of the point). These plots demonstrate how the magnitude-airmass slope $\mathcal{M}$ is correlated with both SED type and with redshift, which indicates it is potentially useful as an independent piece of information to include in our photo-$z$ estimator. For example, the absolute value of $\mathcal{M}$ is largest for early-type galaxies (red points) except for some specific filter/redshift combinations (e.g., $y$-band and $z\approx0.9$--$1.0$) where the emission lines of late-type galaxies can cause large variations in $\mathcal{M}$. We can also see that the $u$-band appears to yield the strongest correlation between $\mathcal{M}$ and $z$, and also offers the largest spread in $\mathcal{M}$ over SED type (i.e., the red and purple points are further apart). In Figure \ref{fig:slope_vs_z} we also plot vertical grey bars that have a length equal to the magnitude errors for galaxies of $22$ or $25$ mag in each panel's filter. In many cases these errors are larger than the expected value of $\mathcal{M}$, which indicates that the potential use of $\mathcal{M}$ may be limited to only filters $ugr$ and only to brighter galaxies.

The next steps are to figure out how well we could measure $\mathcal{M}$ with LSST-like photometry, and for what fraction of galaxies is it a useful redshift indicator, given our catalog has a realistic mix of SED types, redshifts, and magnitudes. To begin, we match all the galaxies in our simulated catalog to an SED template from \cite{2014ApJS..212...18B} using the rest-frame $g-r$ color from the catalog, and impose a random uniform scatter in SED type to avoid discretization. We use this associated template to calculate both true and observed values of $\mathcal{M}$ for each of the galaxies in our training and test subsets. The true slope is calculated directly from the SED with no observational errors; the true slope would be appropriate to use for the training set, which is meant to represent a spectroscopic subset in the photo-$z$ estimator. To simulate the observed slopes we assume the LSST 10-year photometric errors and the airmass distribution presented in Table \ref{tab:amdistr}, which we apply to all filters\footnote{For readers familiar with LSST simulations this distribution is similar to the {\sc minion 1016} OpSim run in which most of the observations are at low airmass. A preliminary analysis found that there is no obvious distribution that significantly improves our ability to measure $\mathcal{M}$ with LSST, and so we are not including a study of airmass distribution optimization in this work.}. For each catalog galaxy we randomly choose an airmass from this distribution for each visit, assume the catalog's true magnitude to represent $X=1$, and use the true slope $\mathcal{M}$ to calculate a true apparent magnitude for each visit. We then add observational error to every visit measurement by scattering the true magnitudes by an amount proportional to their uncertainty, and perform a linear regression between airmass and simulated observed magnitudes for all visits to determine the observed slope and its uncertainty\footnote{We have used the {\sc python} package {\sc scipy.stats.linregress}, and use the returned standard error on slope as the uncertainty}.

\begin{table}
\begin{center}
\caption{Assumed Distribution of Airmass}
\label{tab:amdistr}
\begin{tabular}{cc}
\hline
\hline
Airmass & Cumulative Fraction \\
 & of Visits (all filters) \\
 \hline
1.0 & 0.1 \\
1.1 & 0.2 \\
1.2 & 0.3 \\
1.3 & 0.4 \\
1.4 & 0.5 \\
1.5 & 0.6 \\
1.6 & 0.65 \\
1.7 & 0.7 \\
1.8 & 0.75 \\
1.9 & 0.8 \\
2.0 & 0.85 \\
2.1 & 0.9 \\
2.2 & 0.94 \\
2.3 & 0.96 \\
2.4 & 0.98 \\
2.5 & 1.0 \\
\hline
\end{tabular}
\end{center}
\end{table}

In Appendix \ref{sec:ap} we present additional analysis based on these true and observed values of $\mathcal{M}$ that shows the magnitude-airmass slope $\mathcal{M}$ is a unique redshift indicator which is independent of color, and could help to mitigate color-redshift degeneracies. We find that this is true especially in filters $u$, $g$, and $y$, and show that it is a measurable property for galaxies with LSST-like photometric errors, especially in filters $u$, $g$ and $r$, but that this may be limited to bright galaxies only. In the next section, we explore how the values of $\mathcal{M}$ that we have added to our simulated catalog could be incorporated into our photo-$z$ estimator.

\subsection{Incorporating $\mathcal{M}$ Into Photometric Redshifts}\label{ssec:airInc}

\begin{figure*}
\begin{center}
\includegraphics[width=8cm,trim={1cm 4.5cm 1cm 4.5cm},clip]{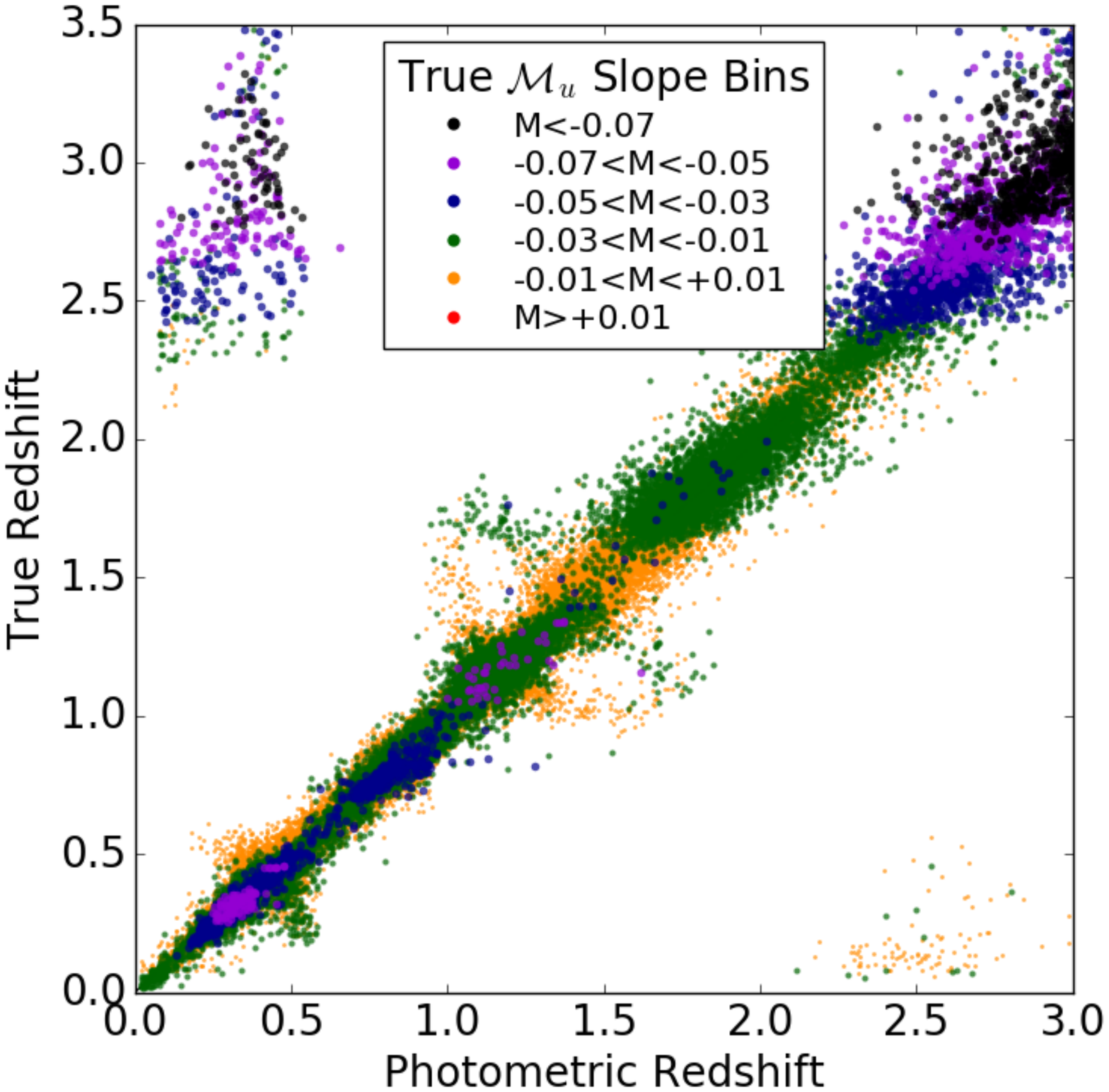}
\includegraphics[width=8cm,trim={1cm 4.5cm 1cm 4.5cm},clip]{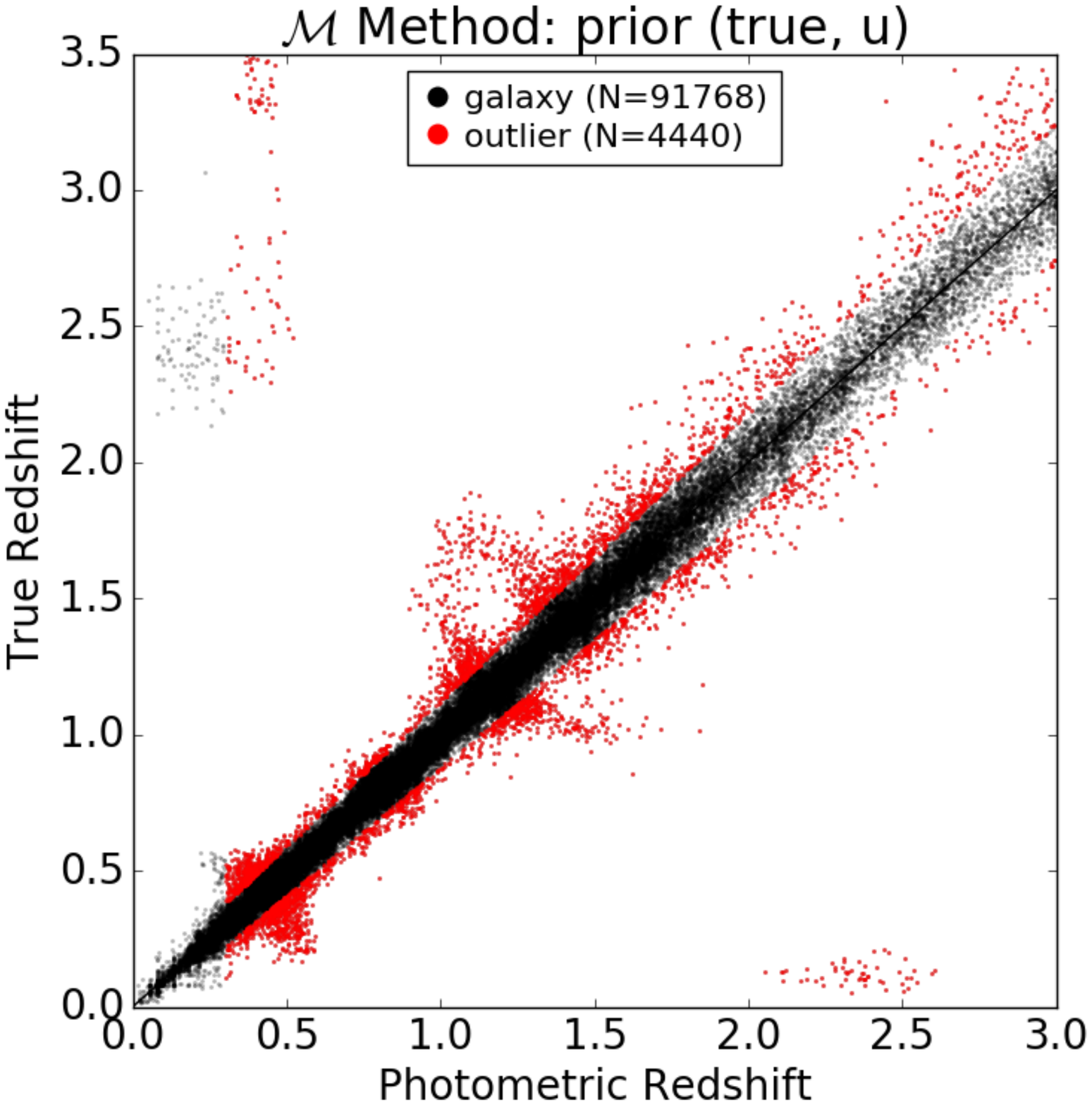}
\includegraphics[width=8cm,trim={1cm 4.5cm 1cm 4.5cm},clip]{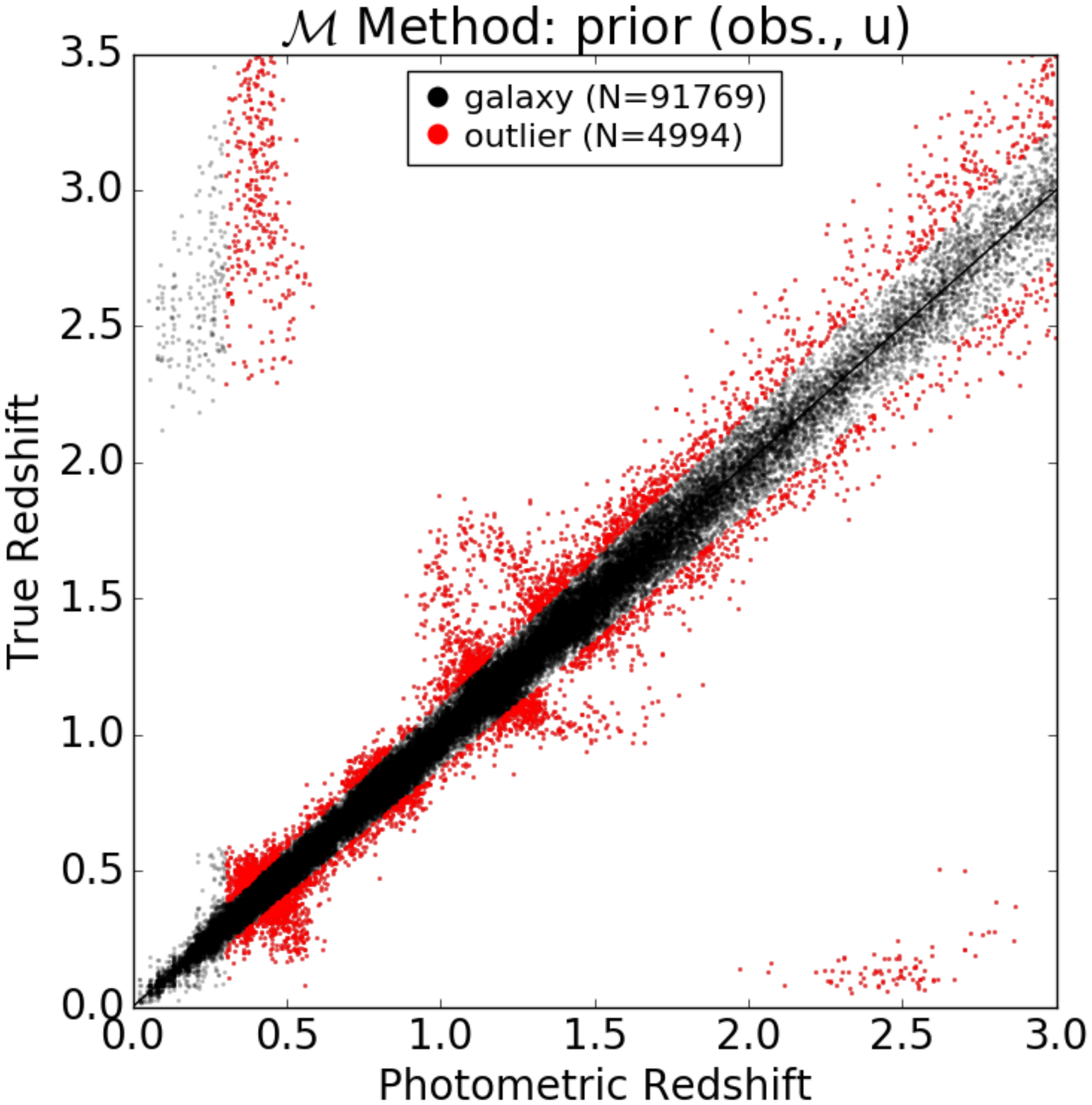}
\includegraphics[width=8cm,trim={1cm 4.5cm 1cm 4.5cm},clip]{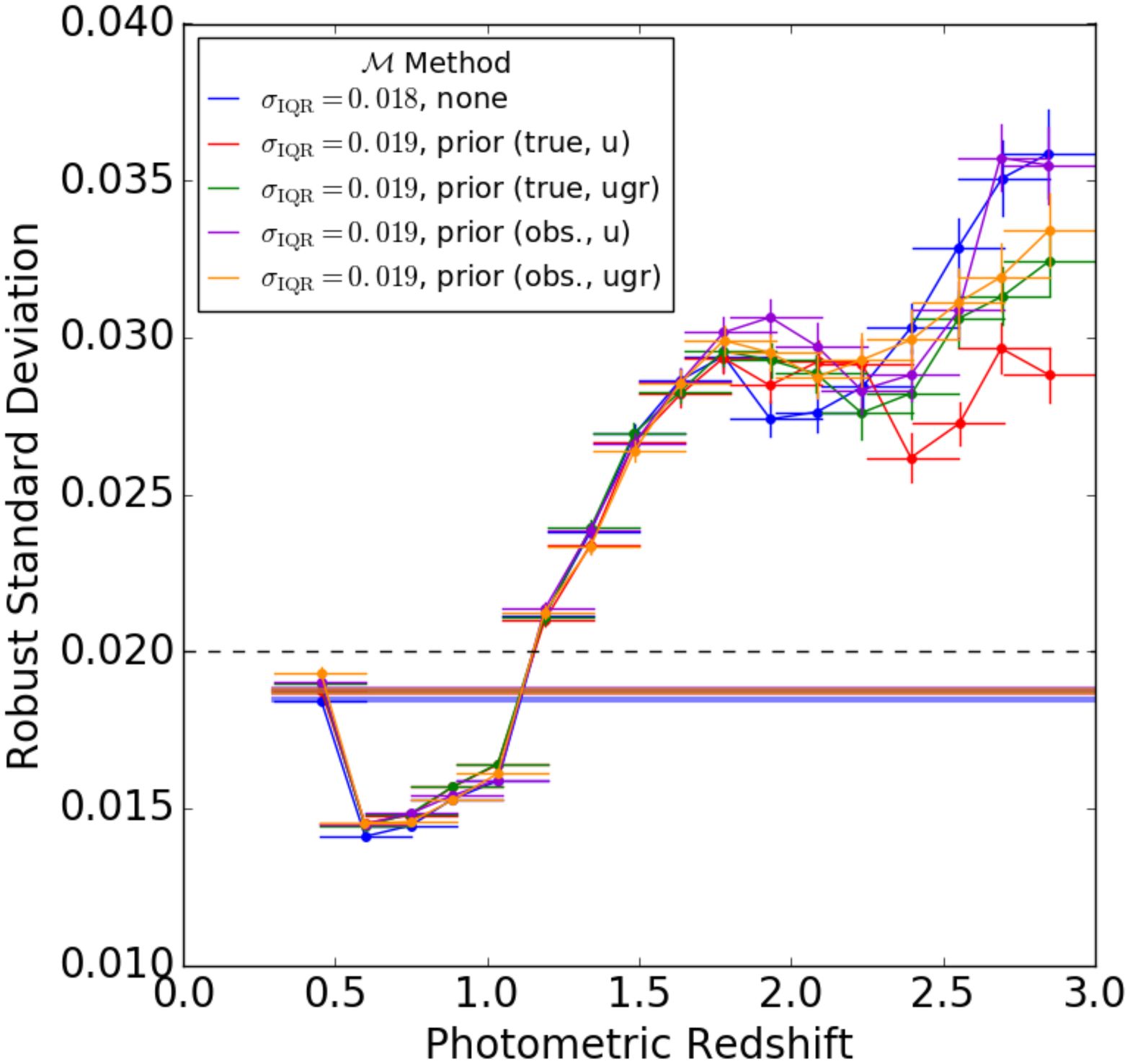}
\caption{{\it Top Left:} The $z_{\rm true}$ $vs.$ $z_{\rm phot}$ results when $\mathcal{M}$ is not included in the photo-$z$ estimate, and we choose randomly from the color-matched subset of training galaxies. Points are colored by the galaxy's true $\mathcal{M}_u$ value. {\it Top Right:} The photo-$z$ results if we incorporate the true $\mathcal{M}_u$ as a prior on photometric redshift. {\it Bottom Left:} The photo-$z$ results if we incorporate the observed $\mathcal{M}_u$ as a prior on photometric redshift. {\it Bottom right:} The robust standard deviation in $\Delta z_{(1+z)}$ when we apply a prior based on the true or observed magnitude-airmass slope in $u$-band only, or in $u$, $g$, and $r$ bands. Horizontal colored lines mark the value of the statistic over the full redshift range $0.3 \leq z_{\rm phot} \leq 3.0$, which are also listed in the legend. Horizontal dashed lines mark the SRD's target value for each statistic. \label{fig:magair1}}
\end{center}
\end{figure*}

In this section we explore a method for incorporating the magnitude-airmass slope into our photometric redshift estimator by generating and applying a prior based on the redshift distribution of training set galaxies with a similar $\mathcal{M}$ as a given test set galaxy. To do this, for each test galaxy we first identify a set of slope-matched training galaxies using the Mahalanobis distance. This is similar to the method used to identify a color-matched subset in Section~\ref{ssec:photoz} and Equation \ref{eq:DM}, but uses $\mathcal{M}$ in any desired set of filters instead of color:

\begin{equation}\label{eq:DMslp}
D_\mathcal{M} = \sum_{\rm 1}^{N_{\rm filters}} \frac{( \mathcal{M}_{\rm train, true} - \mathcal{M}_{\rm test, obs} )^2}{ (\delta \mathcal{M}_{\rm test, obs})^2},
\end{equation}

\noindent We use a PPF $=0.95$ to identify a slope-matched subset of training galaxies and create a redshift distribution based on their true redshifts, similar to the ones shown in Figure \ref{fig:slope_prior} in Appendix \ref{sec:ap}. This distribution is applied to the color-matched subset of training galaxies as a weight, and then a random training set galaxy is chosen to provide the photometric redshift for the test galaxy. As a test of our process we also run our photo-$z$ estimator using the true slope $\mathcal{M}_{\rm test, true}$ of the test galaxy, with a nominal error of $\delta \mathcal{M}_{\rm test, obs} = 0.01$. This is a completely unrealistic scenario, but is useful as a simulation of the maximum possible impact of including $\mathcal{M}$ as a prior.

We also considered a second method that applies a cut based on $\mathcal{M}$ to the color-matched subset of training galaxies. To use the slope $\mathcal{M}$ as a cut, we calculate $D_\mathcal{M}$ for only the training galaxies in the color-matched subset. We then use a PPF $=0.95$ to identify a slope-matched sub-subset of these training galaxies, from which a random training galaxy is chosen to provide the photometric redshift for the test galaxy. However, we found that this had no measurable impact on the photo-$z$ results, and so have not included it in the rest of this analysis.

To test whether incorporating $\mathcal{M}$ as a prior leads to improved results, we run our photo-$z$ estimator using a training set of $10^6$ galaxies and a test set of $10^5$ galaxies. We include the initial queries based on magnitude and color, use a PPF $=0.68$ to identify the color-matched subset of training galaxies.

In the top left panel of Figure \ref{fig:magair1} we show the photo-$z$ results of a run in which $\mathcal{M}$ is not incorporated, and we simply choose randomly from the color-matched subset of training galaxies (as described in Section~\ref{ssec:ppfcsq}). In this plot, we color the points by each test galaxy's value of true $\mathcal{M}_u$, and overlay the plots in order from most to least common slope value. E.g., galaxies with $-0.01<\mathcal{M}_u<+0.01$ are plotted first, in orange, and the rare few galaxies with $\mathcal{M}_u<-0.07$ are overplotted last, in black. This allows us to clearly see clumping of galaxies with unique values of $\mathcal{M}_u$ in the $z_{\rm true}$--$z_{\rm phot}$ plane. We expect that it is these clumped, $\mathcal{M}_u<-0.03$ galaxies -- especially the outliers and high-$z$ galaxies -- that will most benefit from the inclusion of the magnitude-airmass slope in the photo-$z$ estimator.

In the top right panel of Figure \ref{fig:magair1} we show the photo-$z$ results of a run in which the true value of $\mathcal{M}_u$ is incorporated as a prior. This represents an unrealistic scenario -- the maximum possible impact of $\mathcal{M}$ on the photometric redshifts. We can see that the populations of outliers with $\Delta z > 1$ have been visibly reduced, and by comparing with the top left panel we can see that it is galaxies with $z_{\rm true}>2.5$ and $\mathcal{M}_u<-0.05$ which are no longer assigned low-$z_{\rm phot}$. Specifically, we find that the fraction of outliers is reduced by $\sim1\%$ in low-$z$ bins and by 10\% in high-$z$ bins. If we instead incorporate the true $\mathcal{M}$ in three filters $ugr$ into our photo-$z$ estimator we find that the $z_{\rm true}$--$z_{\rm phot}$ diagram appears similar, and that the the fraction of outliers is reduced to a slightly lesser extent, by $\sim8\%$ at high-$z$. 

In the bottom right panel of Figure \ref{fig:magair1} we show the robust standard deviation in the photometric redshifts when we include priors based on the true $\mathcal{M}$ in filters $u$ or $ugr$ (the red and green lines). We find that the overall value of $\sigma_{\rm IQR}$ is not significantly affected, but that the standard deviation in high-$z$ bins decreases compared to the default run in which $\mathcal{M}$ is not incorporated (the blue line). Although not shown, we find that a decrease in $\sigma_{\rm IQR}$ is seen when the true $\mathcal{M}$ is incorporated into the photo-$z$ estimator for all combinations of filters so long as $u$-band is included (e.g., no improvement in the photo-$z$ results are seen if we use only the true value of $\mathcal{M}_g$).

In reality we will not know the true value of $\mathcal{M}$ for all galaxies. In fact, we find it will be fairly difficult to measure because the observational errors in the photometry are about the same size as the slope we are trying to measure, as demonstrated in Figure \ref{fig:slope_vs_z} and in Appendix \ref{sec:ap}. In the bottom left panel of Figure \ref{fig:magair1} we show the photo-$z$ results of a more realistic run in which the observed value of $\mathcal{M}_u$ is incorporated as a prior. We find that in this case the outliers are not visibly reduced. The results look qualitatively similar if we incorporate the observed $\mathcal{M}$ in three filters $ugr$. In the bottom right panel of Figure \ref{fig:magair1} we show how the robust standard deviation in the photo-$z$ results are not improved by including priors based on the observed $\mathcal{M}_u$ (i.e., the purple line is consistent with the blue line). However, when we include the observed $\mathcal{M}$ in filters $g$ and $r$ (the orange line), which have lower magnitude errors than $u$, we see a potential improvement in $\sigma_{\rm IQR}$ in high redshift bins -- but this affect is barely at the $1\sigma$ level. Although we do not show it, we find that the robust bias is only minimally affected when we incorporate a prior on $\mathcal{M}$ into our photo-$z$ estimator: in the bins at $z_{\rm phot}>2.5$, the photometric redshifts are under-estimated by an average of $\sim0.01$ when true $\mathcal{M}$ is included as a prior, compared to no net bias when no prior is included.

As a final note, in Appendix \ref{sec:ap} we found that $\mathcal{M}_{\rm obs}$ is most accurately measured for galaxies with magnitudes $<23$. We attempted to implement an option of only applying the prior in $\mathcal{M}$ for a given filter if that test galaxy is $<23$ mag in the filter, but this yielded no discernible improvement in the photo-$z$ results.

\subsection{Photo-$z$ Bias from Airmass-Induced Systematics}\label{ssec:air_bias}

\begin{figure}
\begin{center}
\includegraphics[width=8cm,trim={0.5cm 4.5cm 0.5cm 4.5cm},clip]{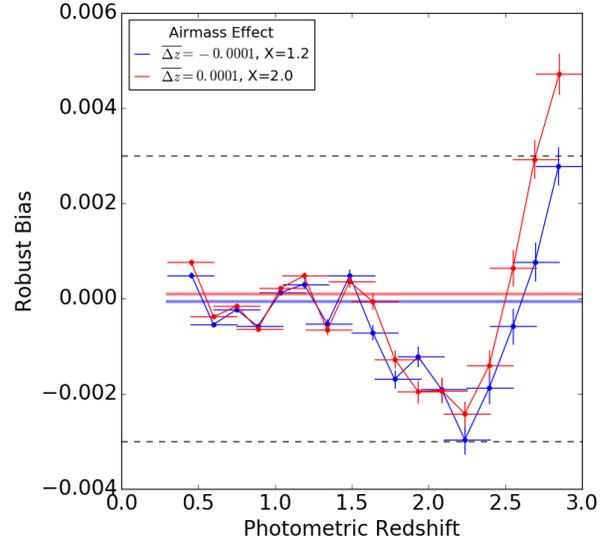}
\caption{The robust bias in $\Delta z_{(1+z)}$ when we have used LSST photometry that assumes all visits are done at (or corrected to) the nominal airmass of $X=1.2$ (blue), and when we apply a systematic offset that corresponds to change in the normalized effective filter transmission curve at an airmass of $X=2.0$ (red). The latter represents a worse-case scenario in which all observations are done at high airmass {\it and} this effect cannot be corrected for in the photometric calibrations. Horizontal colored lines mark the value of the statistic over the full redshift range $0.3 \leq z_{\rm phot} \leq 3.0$, which are also listed in the legend. Horizontal dashed lines mark the SRD's target value for each statistic. \label{fig:air_bias}}
\end{center}
\end{figure}

In most of this work we have focused on how changes to the photometric {\it errors} affect the photo-$z$ (i.e., Section~\ref{ssec:errors}), but since we have already covered the magnitude-airmass slope $\mathcal{M}$ in detail we take this opportunity to evaluate the effect of an airmass-induced systematic offset to the magnitudes. As described in Section~\ref{ssec:airM}, this systematic is the result of a change in the {\it normalized} effective filter transmission curve. As such, it is not simply corrected by applying corrections for atmospheric extinction and reddening, and so could lead to a bias in the photometric redshift results. Given that we did not find $\mathcal{M}$ to be significant enough to serve as a useful prior on photo-$z$, we do not expect to find a particularly detrimental impact -- but ultimately, this would depend on the final airmass distribution of the LSST survey. For demonstrative purposes we consider a worst-case scenario in which all observations for galaxies in the test set are obtained at airmass $X=2.0$, and add the corresponding systematic offset to their observed apparent magnitudes (but leave the photometric errors unchanged in order to isolate the change to a systematic on magnitude). In this way, we are simulating the maximum possible impact. Even so, this airmass-induced magnitude offset is greater than the error in magnitude for only a small fraction of galaxies in the test set ($\sim5\%$), most of which have a true redshift $<0.7$ (and a small number have $z_{\rm true}>2.0$). In almost all cases, this systematic causes the galaxy to appear brighter, especially in the $u$, $g$, and $r$-band filters, and so our prediction is that the photometric redshifts will be underestimated (a bias to lower redshifts). We then run our photo-$z$ estimator with the same parameters and training set as used above for our analysis of $\mathcal{M}$. In Figure \ref{fig:air_bias}, we show how the robust bias is affected when the test set is afflicted with a systematic magnitude offset that represents the results of all observations occurring at an airmass of $X=2.0$. Recall that bias is the mean value of $\Delta z = z_{\rm true} - z_{\rm phot}$. The small but significant increase in bias exhibited in Figure \ref{fig:air_bias}, which is larger in the high-$z$ bins, confirms that the photometric redshifts are systematically {\it lower}, as predicted. Although this photo-$z$ bias is unrealistically strong due to our adoption of a worse-case scenario airmass distribution, our point is that any future proposal to include -- or even prioritize -- high-airmass observations will have to consider the potential costs to the photo-$z$ bias.  

\subsection{Summary}\label{ssec:airsum}

In summary, we have shown that the slope of the magnitude-airmass correlation for galaxies due to a warping of the effective filter transmission curve by the atmosphere is a unique indicator of redshift. We have furthermore shown that when the value of this slope is accurately known, it can help to reduce the number of truly high-$z$ galaxies assigned an erroneously low $z_{\rm phot}$ and may reduce the overall standard deviation in high redshift bins. However, we have also shown that even in this ideal case, incorporating $\mathcal{M}$ into the photo-$z$ estimates does not lead to any significant improvement over $0.3 \leq z_{\rm phot} \leq 3.0$. Ultimately, we find that the predicted quality of the LSST photometric errors will allow the slope of the airmass-magnitude correlation to offer only a mild improvement on photometric redshifts. Based on this work, we do not think that the airmass distribution of the LSST should be constrained to impose high airmass visits on the survey strategy for the purpose of SED sampling to improve the photo-$z$ results, especially as we have also shown that this may induce a bias towards lower redshifts into the results.

\section{Conclusions}\label{sec:conc}

We have presented a nearest-neighbors color-matching photometric redshift estimator that avoids intermediate steps such as SED fitting, and for which the photo-$z$ precision and accuracy are directly related to the input magnitude uncertainties. These attributes make our method well-suited for testing the impact of changes to the LSST survey parameters, which affect the observed magnitude errors, on the quality of photometric redshifts. We characterized the performance of our photo-$z$ method by analyzing the results for a variety of input parameters using a simulated galaxy catalog. We established guidelines for the minimum number of test set galaxies, settings for the internal parameters that control the photo-$z$ assignment. We have shown that our broad interpretation of the SRD's minimum target values for standard deviation, bias, and the fraction of outliers over $0.3 \leq z_{\rm zphot} \leq 3.0$ can be met with our photo-$z$ estimator. However, meeting the LSST SRD's target values was not the main goal of this work and we are not claiming to have developed the best photo-$z$ estimator -- only that it is an appropriate tool for our particular experiments. To inform future applications of this photo-$z$ estimator, we have demonstrated how the choice between variance and sample size will have to be made based on the science goals of that application.

We have used our photo-$z$ estimator to explore the nominal survey strategy of the LSST: a uniform progression of all filters towards a total number of visits over 10 years. We determined that limiting the relative fraction of visits allotted to the $u$-band filter can significantly deteriorate the photo-$z$ quality, but that limiting the number of $y$-band visits has a smaller effect. We also showed that there would not be a significant, overall improvement to the photo-$z$ quality if the LSST concentrated all of its survey time on building depth in only filters $gri$ during the first year. We've also provided a list of our statistical measures of photo-$z$ at 1, 2, 5 and 10 years for readers who may be interested to use this relative progression to evaluate the feasibility of LSST science goals as a function of survey parameters (Table \ref{tab:years}). Finally, we exemplified the main motivation for developing this photo-$z$ estimator -- its direct correlation between photometric quality and photo-$z$ results -- by demonstrating how artificial deterioration to the magnitude errors impacts the photometric redshifts. In particular, we found that if the magnitude errors increase by 50\% or suffer a systematic of $+0.01$ mag the target value for the standard deviation in photo-$z$ error will not be met. We hope that this work will serve an informative guideline for the LSST science community in regards to the relative quality and quantity of photometric redshifts that will be available as the survey progresses.

The nominal survey strategy of the LSST will include observations over a distribution of airmass values. Due to a residual atmospheric component in the normalized effective filter transformation function, we have shown that there is a subtle relationship between the apparent magnitude and airmass that is uniquely correlated with the SED and redshift of each observed object. In this work we have sought to ascertain whether the slope of this magnitude-airmass relation can be employed as an additional prior in our photometric redshift estimator. We concluded that if it could be perfectly measured this effect would be useful in mitigating the number of high-$z$ galaxy interlopers with low-$z{\rm phot}$, but that it is generally too subtle to detect in the LSST photometry. We therefore do not recommend that any special effort is undertaken to increase the number of high-airmass observations during the LSST survey. This experiment with the magnitude-airmass relation has also demonstrated the ease with which alternative parameters can be incorporated into our photo-$z$ estimator, and could be used in future assessments of potential priors or survey strategies.

\section*{Acknowledgements}

We thank Carlton Baugh and the The Institute for Computational Cosmology for access to their simulated mock catalogs.

This material is based upon work supported in part by the National Science Foundation through Cooperative Agreement 1258333 managed by the Association of Universities for Research in Astronomy (AURA), and the Department of Energy under Contract No. DE-AC02-76SF00515 with the SLAC National Accelerator Laboratory. Additional LSST funding comes from private donations, grants to universities, and in-kind support from LSSTC Institutional Members.

AJC acknowledges support by the U.S. Department of Energy, Office of Science, under Award Number DE-SC-0011635, and partial support from the Washington Research Foundation and the DIRAC Institute. SJS was partially supported by the Heising-Simons Foundation. MJ acknowledges the support of the Washington Research Foundation Data Science Term Chair fund, and the University of Washington Provost's Initiative in Data-Intensive Discovery.

\bibliography{ms}

\appendix
\section{The Airmass-Magnitude Slope as a Redshift Indicator}\label{sec:ap}

In Section~\ref{sec:air} we defined the magnitude-airmass slope, $\mathcal{M}$, and proposed to incorporate it into our photometric redshift estimator. In order to do this we must first ensure that $\mathcal{M}$ can be well measured, and secondly verify that it holds some unique information about the redshift that is independent of a galaxy's photometric color. 

\begin{figure}[h]
\begin{center}
\includegraphics[width=8.2cm,trim={1cm 4.5cm 1cm 4.5cm},clip]{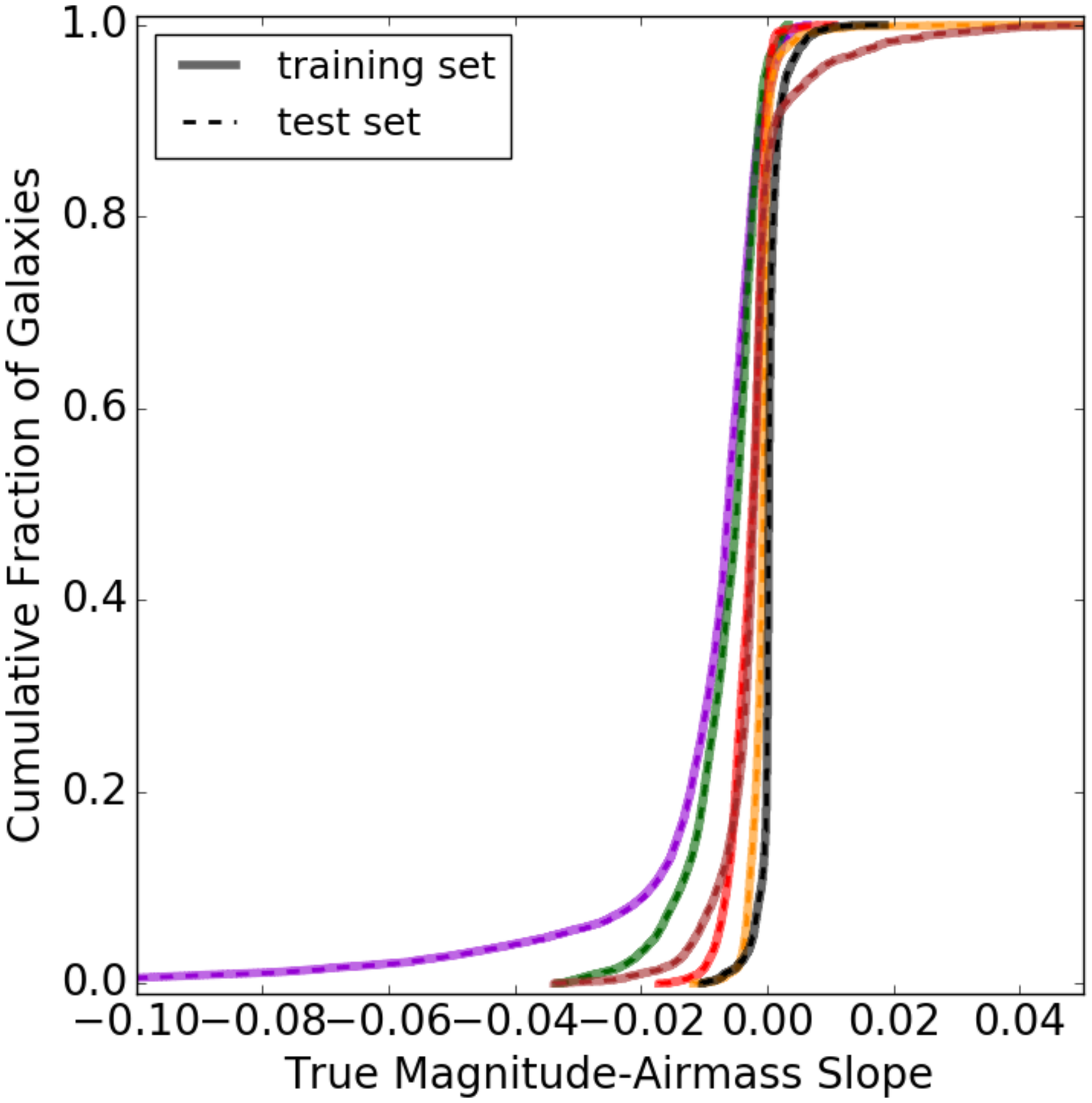}
\includegraphics[width=8.2cm,trim={1cm 4.5cm 1cm 4.5cm},clip]{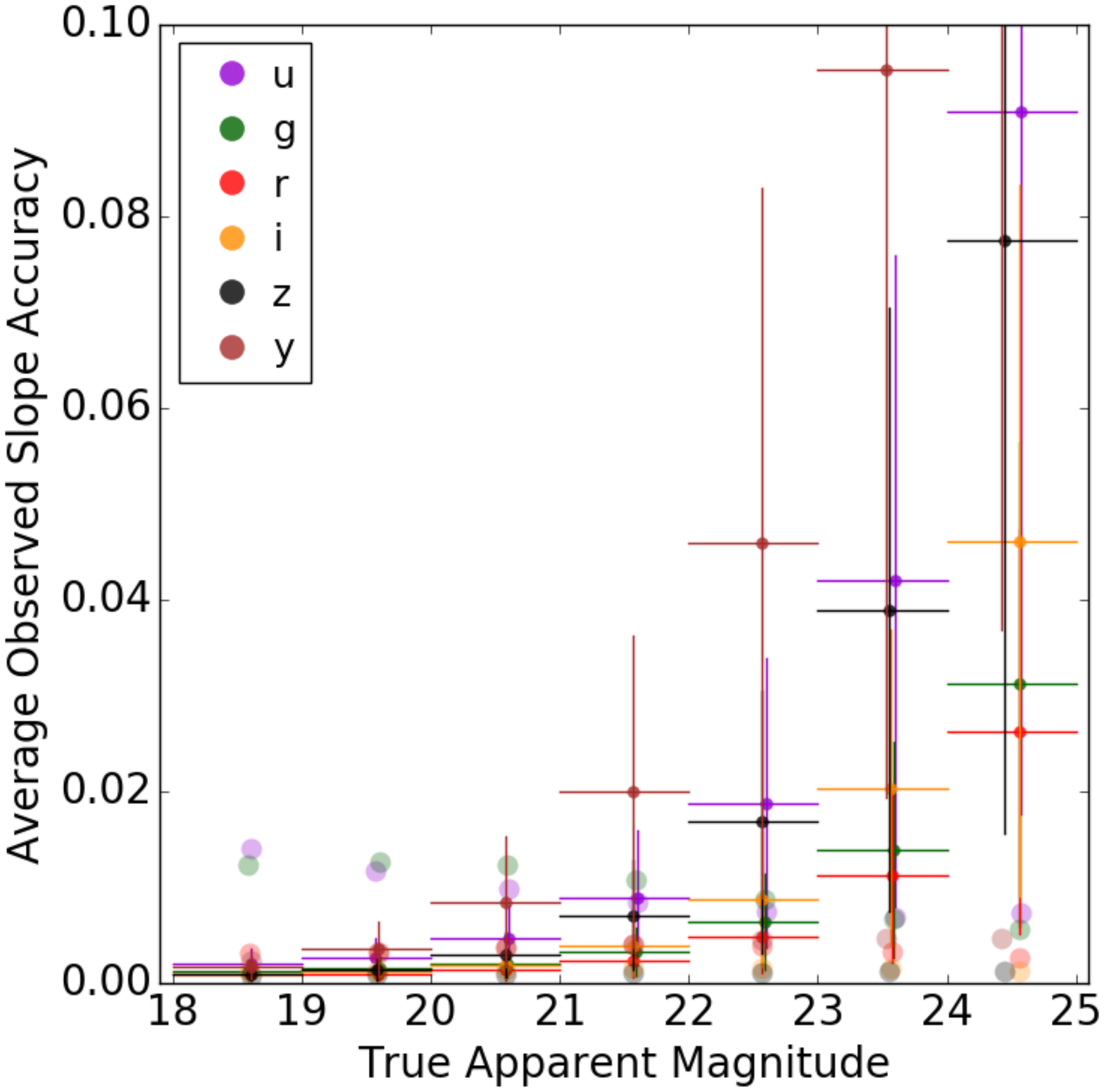}
\caption{{\bf Left:} Distribution of true slope values in each filter, for the training subset (solid) and the test subset (dashed). {\bf Right:} In bins of true apparent magnitude we plot the average of $\Delta \mathcal{M} = |\mathcal{M}_{\rm true}-\mathcal{M}_{\rm obs}|$. The error bars represent the standard deviation in this difference. Transparent circles underlying these data represent the mean absolute value of $\mathcal{M}_{\rm true}$ in that bin. A confident measure of $\mathcal{M}$ would result in $\Delta \mathcal{M} \lesssim \overline{\mathcal{M}}$, so, based on this plot we find that $\mathcal{M}$ cannot generally be confidently measured with LSST data for galaxies fainter than $23$ magnitude. \label{fig:tm_vs_m}}
\end{center}
\end{figure}

In Figure \ref{fig:tm_vs_m} we plot the distribution of true airmass-slope values $\mathcal{M}_{\rm true}$ for each filter. This shows how most galaxies have $\mathcal{M}_{\rm true}\sim0$, and for the small fraction with a non-zero $\mathcal{M}_{\rm true}$, filters $u$, $g$, and $y$ offer distinctive slope values. At right, we plot the the difference between the true and observed values of slope for a given galaxy ($\mathcal{M}_{\rm true}-\mathcal{M}_{\rm obs}$), in bins of apparent magnitude. The magnitude at which our ability to measure $\mathcal{M}$ is compromised by photometric uncertainties can be estimated as the bin in which $\mathcal{M}_{\rm true}-\mathcal{M}_{\rm obs}$ exceeds the mean value of $\mathcal{M}_{\rm true}$. Figure \ref{fig:tm_vs_m} shows that this is the case for galaxies with $ugr > 23$ magnitude, which is going to limit the use of $\mathcal{M}$ in photometric redshift estimation to low-$z$ galaxies. 

\begin{figure}
\begin{center}
\includegraphics[width=5.9cm,trim={1cm 4.5cm 1cm 4.5cm},clip]{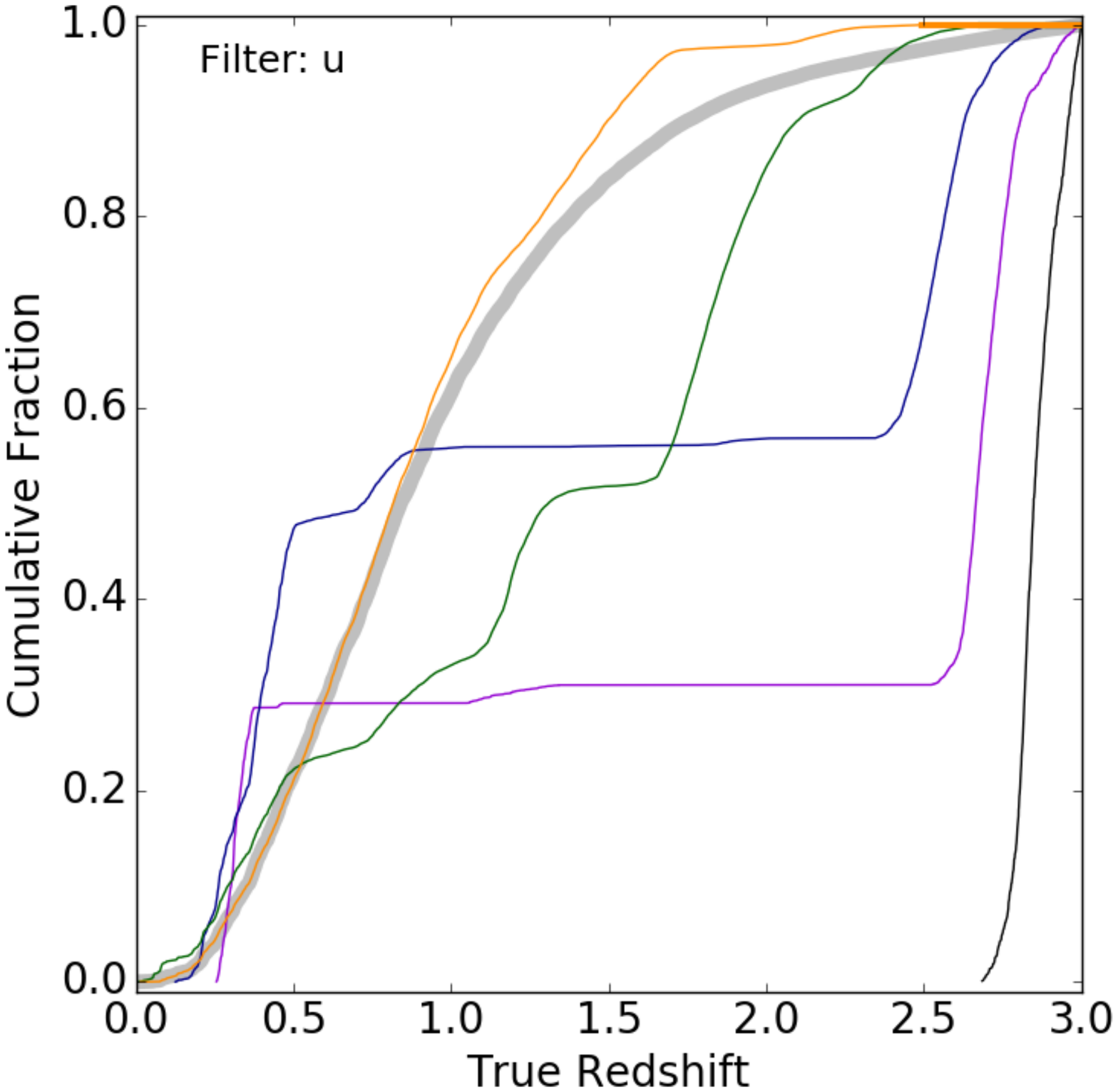}
\includegraphics[width=5.9cm,trim={1cm 4.5cm 1cm 4.5cm},clip]{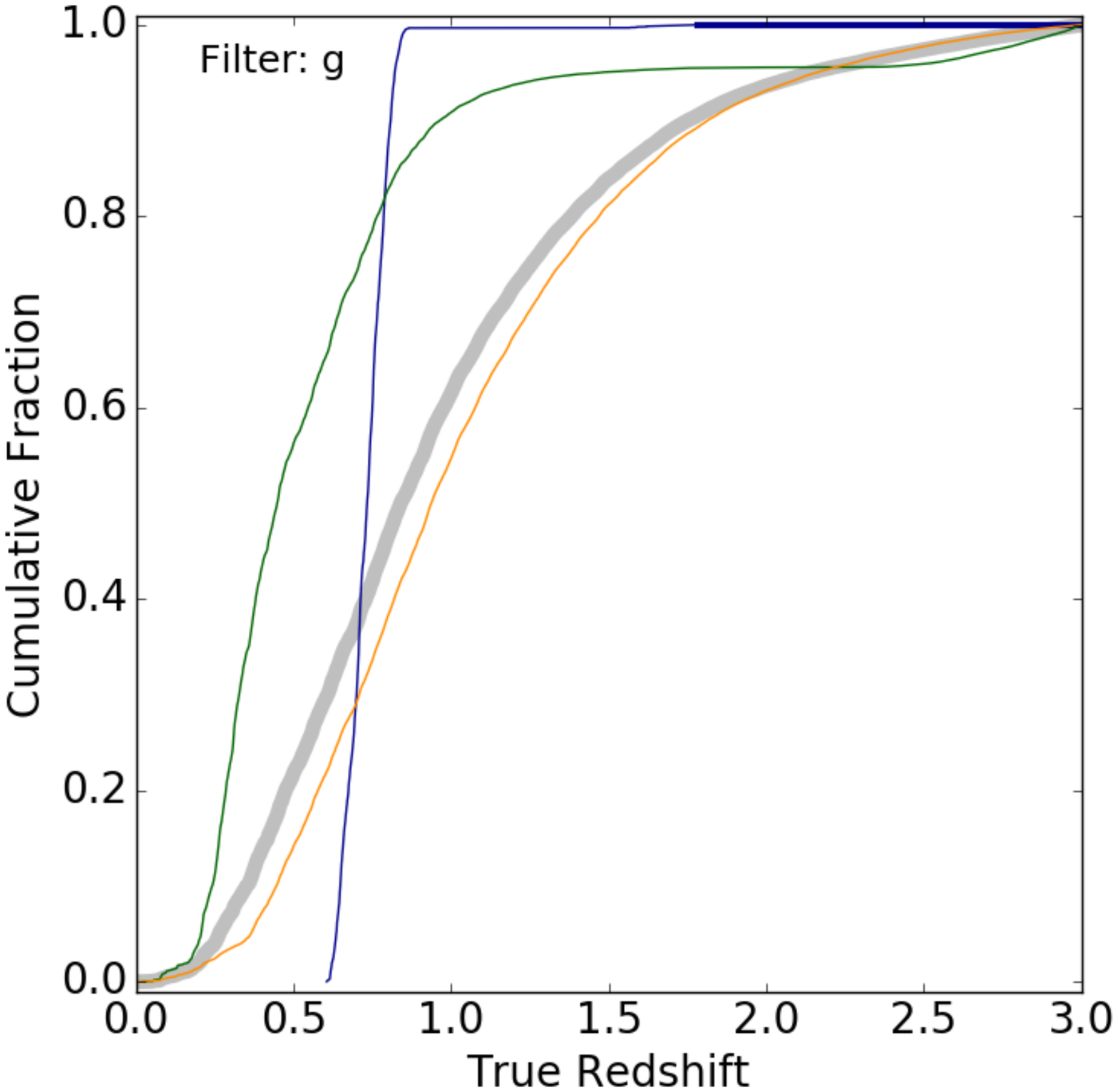}
\includegraphics[width=5.9cm,trim={1cm 4.5cm 1cm 4.5cm},clip]{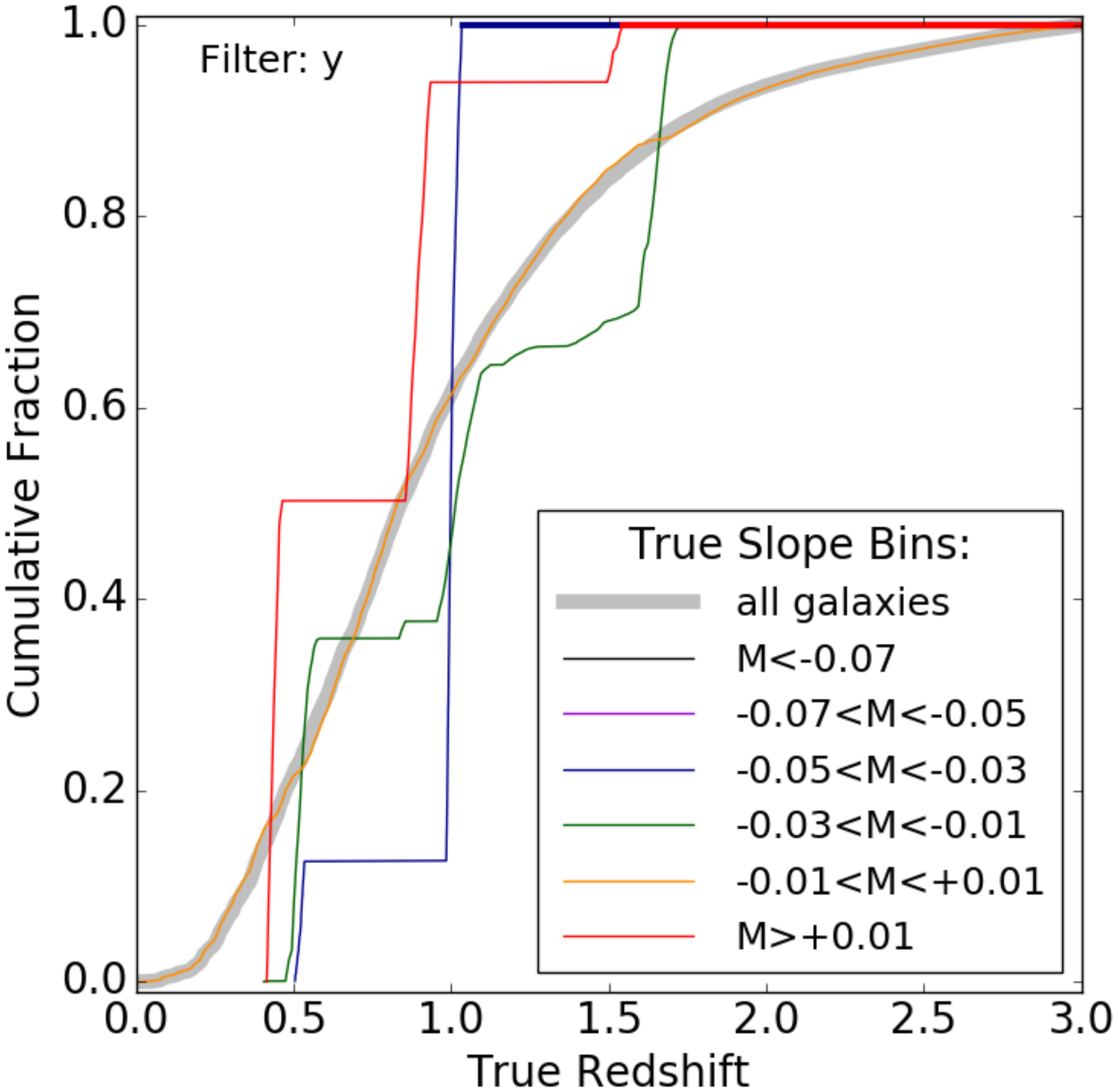}
\caption{The distributions in true catalog redshift for galaxies binned by the true magnitude-airmass slope $\mathcal{M}_{\rm true}$ in filters $u$, $g$, and $y$ (left to right). Similar plots for filters $r$, $i$, and $z$ were omitted because the distributions were similar for all slope bins. The legend in the right panel describes the bins of $\mathcal{M}_{\rm true}$ used for each plot. The redshifted distribution for all galaxies is shown as a thick grey line. \label{fig:slope_prior}}
\end{center}
\end{figure}

\begin{figure}
\begin{center}
\includegraphics[width=5.9cm,trim={1cm 4.5cm 1cm 4.5cm},clip]{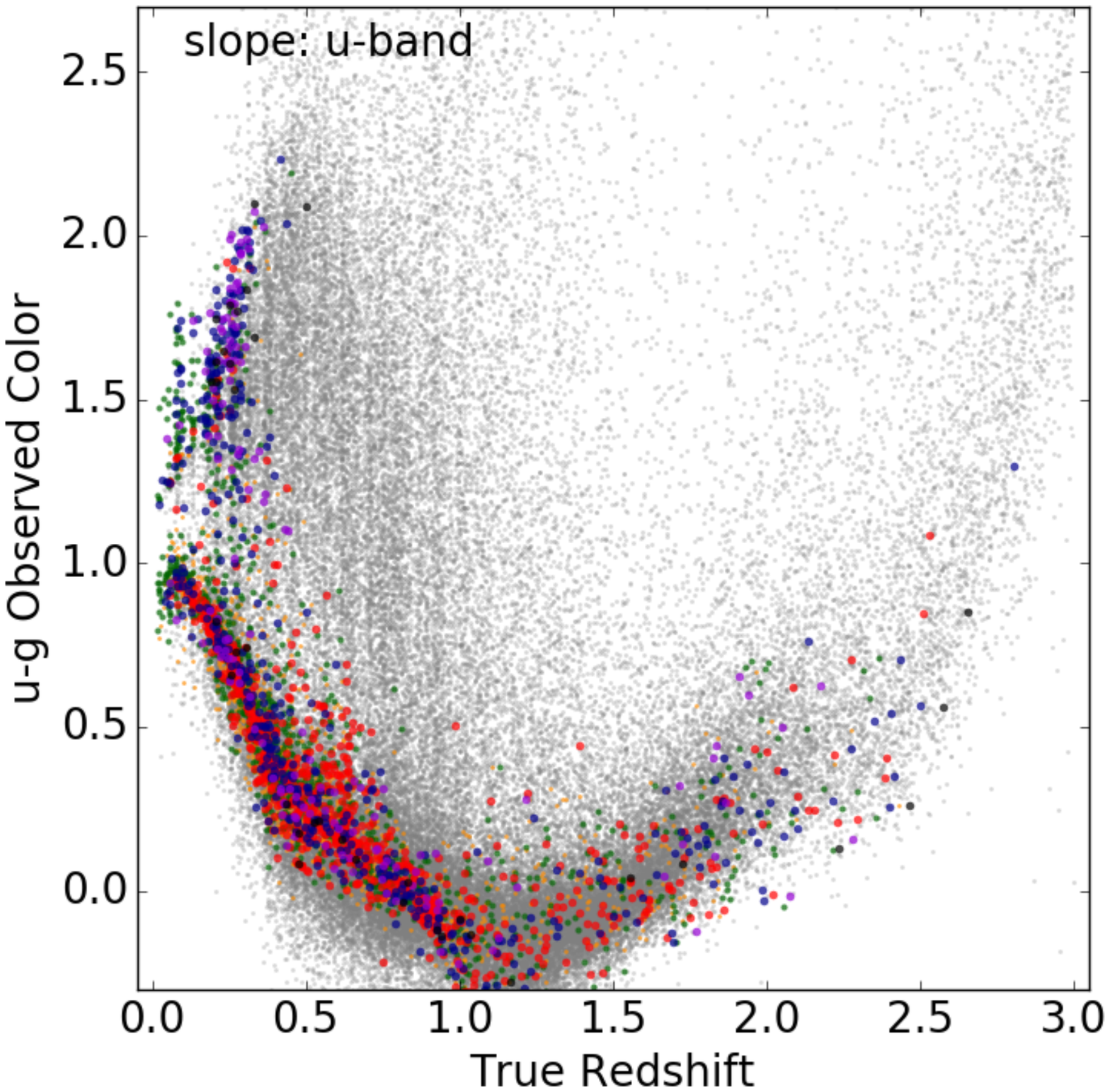}
\includegraphics[width=5.9cm,trim={1cm 4.5cm 1cm 4.5cm},clip]{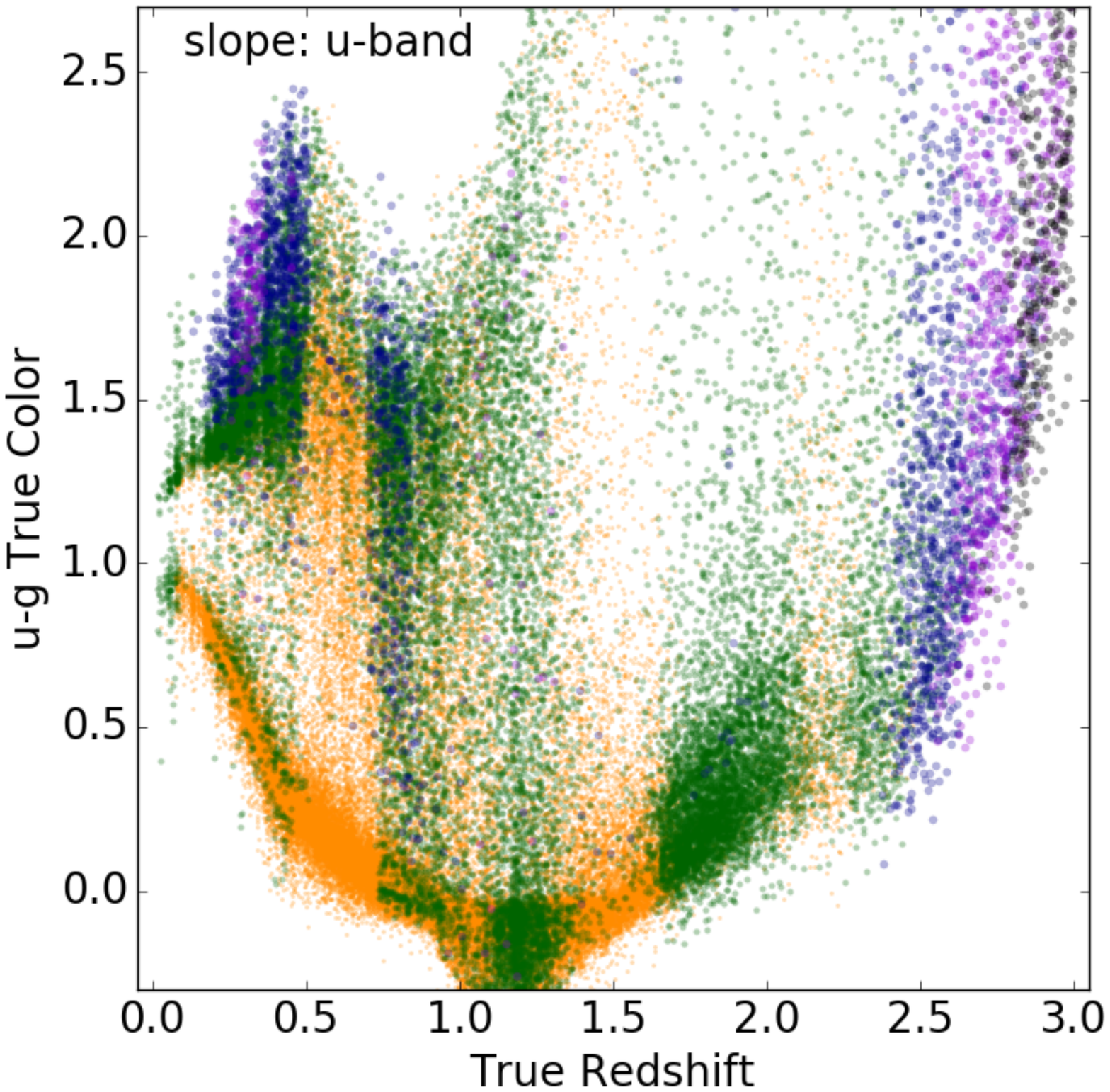}
\includegraphics[width=5.9cm,trim={1cm 4.5cm 1cm 4.5cm},clip]{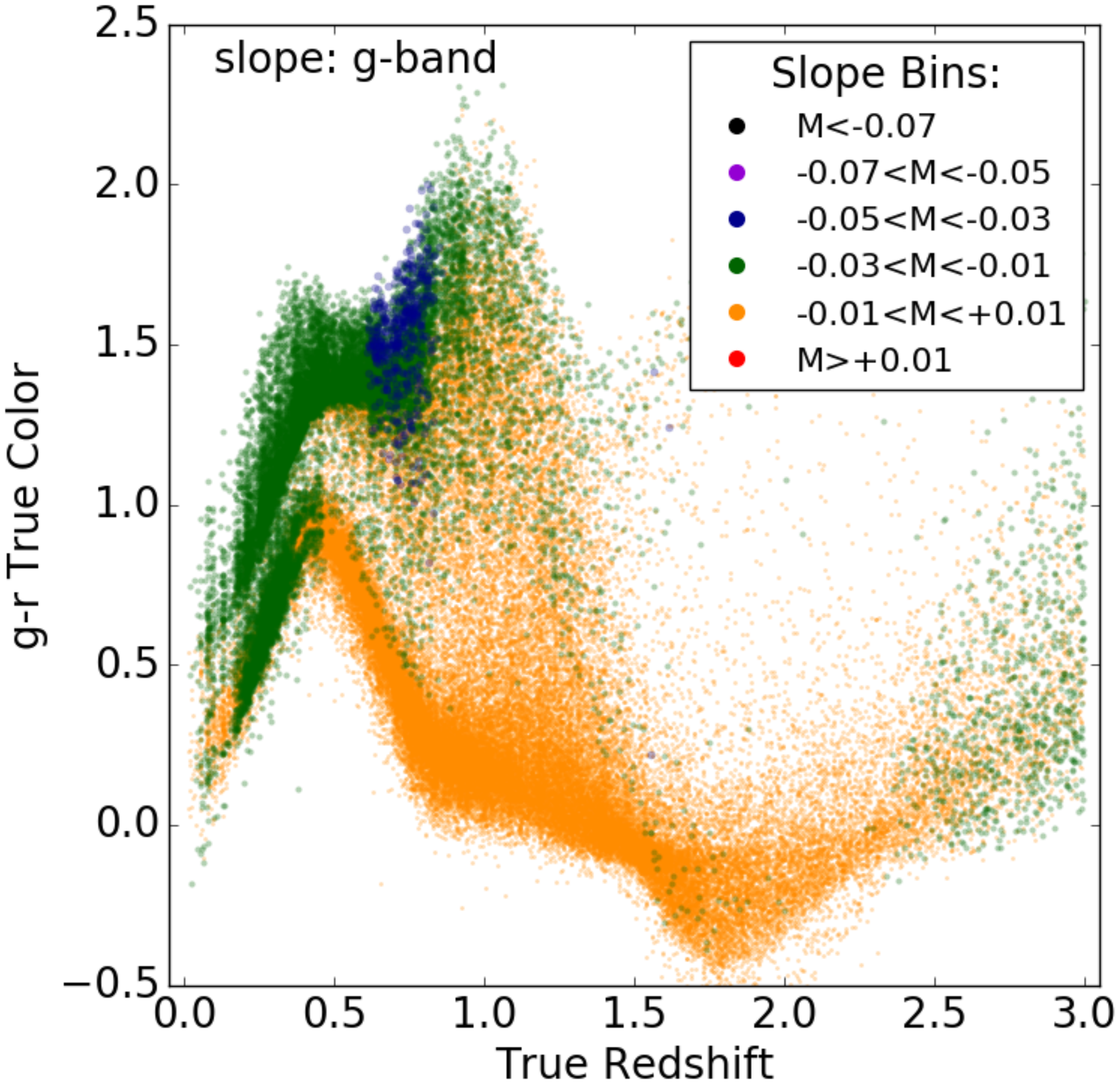}
\caption{Photometric color $vs.$ true redshift for our catalog galaxies. In the center and right panels we use the true $u-g$ and true $g-r$ colors on the $y$-axis, and in the left panel we use the observed $u-g$ color (i.e., including photometric errors that are predicted for the LSST after 10 years of survey time). The color of each point represents the galaxy's magnitude-airmass slope $\mathcal{M}$ as described in the legend in the right panel. From left to right we color based on the observed $\mathcal{M}_u$, the true $\mathcal{M}_u$, the true $\mathcal{M}_g$. In the leftmost plot, points are colored grey if that galaxy's $u$-band magnitude is $>23$ as a demonstration of how the predicted observational errors interfere with using $\mathcal{M}$ as an independent redshift indicator. \label{fig:slope_prior_2}}
\end{center}
\end{figure}

To assess whether $\mathcal{M}$ holds unique redshift information, in Figure \ref{fig:slope_prior} we bin all catalog galaxies by their true slope value and compare their normalized cumulative redshift distributions. We find that the redshift distributions are significantly different for galaxies of certain slopes in filters $u$, $g$, and $y$, and so we expect these filters to be the most effective when applying $\mathcal{M}$ as a prior on photometric redshift. As another visualization to assess whether $\mathcal{M}$ holds unique redshift information, in Figure \ref{fig:slope_prior_2} we plot the galaxy true color as a function of redshift and color the points by the value of the slope $\mathcal{M}$. We are looking for regions of the color-redshift space where points are clustered by the value of $\mathcal{M}$, especially in regions where photometric color is degenerate with redshift, as this indicates where $\mathcal{M}$ could be used to break degeneracies that plague the photo-$z$ estimator. For example, in the panel that plots $g-r$ color as a function of redshift, we can see that galaxies with a photometric color $1.3<g-r<1.7$ have redshifts $0.3<z<1.5$, but that if $-0.05<\mathcal{M}_{g}<-0.03$ (the clump of blue points), then the true redshift of that galaxy is $0.6<z<0.8$. In the other panels we see some other clumped points of similar color, especially in $\mathcal{M}_u$ (top left panel). However, once we incorporate observational errors -- as shown in the lower right plot for $u-g$ and $\mathcal{M}_u$ -- this clustering is severely reduced, which suggests that $\mathcal{M}$ may not be very effective as a prior on in the photo-$z$ estimator. 

\end{document}